\documentclass[11pt]{article}
\usepackage{amsfonts,amsmath,amssymb}
\usepackage{xcolor}
\usepackage{graphicx}
\usepackage{rotating}
\usepackage{multicol}
\usepackage{float}
\usepackage{multirow}
\usepackage{enumerate}
\usepackage{wasysym}
\usepackage{todonotes}
\usepackage{subfigure}
\usepackage[normalem]{ulem}
\usepackage[flushleft]{threeparttable} 
\usepackage[authoryear,round]{natbib}
\usepackage[colorlinks, citecolor=blue, linkcolor=blue]{hyperref}
\usepackage{xr}

\allowdisplaybreaks
\numberwithin{equation}{section}
\usepackage{ntheorem}
\theoremseparator{:}

\newtheorem{thm}{Theorem}[section]

\newtheorem{lm}{Lemma}

\newtheorem{as}{Assumption}[section]

\newtheorem{rem}{Remark}

\allowdisplaybreaks

\newcommand{\convP}{\stackrel{p}{\longrightarrow}}

\newcommand{\indep}{\perp\!\!\!\perp}

\begin{document}
\author{Nan Liu\footnote{Nan Liu, Wang Yanan Institute for Studies in Economics (WISE), Department of Statistics \& Data Science, School of Economics, Xiamen University.}
~~~~~~~~
Yanbo Liu\footnote{Yanbo Liu, School of Economics, Shandong University.}
~~~~~~~~
Yuya Sasaki\footnote{Yuya Sasaki, Department of Economics, Vanderbilt University.}
~~~~~~~~
Yichong Zhang\footnote{Yichong Zhang, School of Economics, Singapore Management University.}
}


\title{{\Large \textbf{Long-Term Quantile Treatment Effects with\\ Combined Experimental and Observational Data}}}

\date{}

\maketitle

\bigskip\bigskip
	
\begin{abstract}
This paper studies the identification, estimation, and inference of long-term quantile treatment effects (LTQTEs) when short-term experimental data are supplemented with long-term observational data. Under a latent unconfoundedness framework, we establish identification of the LTQTE curve and derive the corresponding efficient influence functions. Building on these influence functions, we propose a semiparametrically efficient estimation procedure and develop a multiplier bootstrap method for uniform inference on the LTQTE curve, yielding valid simultaneous confidence bands over a continuum of quantile indices. 

\vskip0.25cm
		
		
\vskip0.5cm
		
\baselineskip=15pt
\end{abstract}

\section{Introduction}\label{sec:intro}

Experimental data provide well-controlled treatment assignments, but long-term follow-up may be difficult or even infeasible due to attrition and the costs of conducting experiments. 
In contrast, observational data often allow for longer follow-up periods and hence the measurement of long-term outcomes, but lack experimental control over treatment assignment. 
A natural strategy is, therefore, to combine these two types of data to leverage their complementary strengths. 
Recent work has developed methods for doing so (e.g., \citealp*{athey2020combining}; \citealp*{athey2025surrogate_index_restud}). 
\citet{chen2023semiparametric} further develop semiparametric estimation methods for long-term treatment effects in this setting.

While this literature has developed rapidly, it has focused primarily on average long-term treatment effects. 
When researchers are interested in distributional effects, however, quantile treatment effects (QTEs) provide a natural and informative alternative. 
The goal of this paper is to complement the growing literature on combining experimental and observational data by studying the identification, estimation, and inference of long-term quantile treatment effects (LTQTEs) in this framework of data combination.
 
We consider the latent unconfoundedness (LU) framework proposed by \citet*{athey2020combining}, in which latent control variables facilitate the combination of short-term experimental data and long-term observational data to identify long-term treatment effects. Unlike the existing literature, which has focused on average effects, however, we study LTQTEs. We derive the efficient influence functions (EIFs) for the LTQTEs, which enable semiparametrically efficient estimation in a manner analogous to \citet{chen2023semiparametric}, with the key distinction that our analysis targets quantile rather than average treatment effects.

Our paper is also related to the recent work of \citet{efficientTransported2026} and \citet*{liu2026double}, who study QTEs when surrogate outcomes are available. In both papers, the primary outcome is missing for part of the sample, and surrogates are used as auxiliary information under a missing-at-random design to improve efficiency and transport treatment-specific outcome distributions. In contrast, our setting combines short-term experimental data with long-term observational data, where treatment assignment in the observational sample may be confounded. As in \citet*{athey2020combining}, we do not require short-term outcomes to be sufficient surrogates for long-term outcomes. Instead, the LU allows short-term latent outcomes to signal treatment selection. This distinction gives rise to a different identification strategy, different efficient influence functions, and a different estimation problem for the LTQTEs from the surrogacy approaches.
Table \ref{tab:surrogacy_vs_lu} summarizes the related literature and positions our contributions within it. See also Remark \ref{rem:lu_surrogate} below for a more detailed discussion.

In the conventional framework without data combination, there is a rich literature on semiparametric estimation and inference for unconditional QTEs \citep*{firpo2007efficient,firpo2009unconditional}. More recently, methods accommodating high-dimensional settings and modern machine learning techniques have been developed by \citet*{sasaki2022uqreg_hd}, \citet*{jiangetal2023}, and \citet*{kallusmaouehara2024}. Our goal is to provide counterparts to these recent advances in the framework of the data-combination setting, where experimental and observational data are combined through the LU structure to identify, estimate, and conduct inference on LTQTEs.

\begin{table}[tbp]
	\centering \caption{Summary of the Literature on Long-Term Treatment Effects}\label{tab:surrogacy_vs_lu}
	\renewcommand{\arraystretch}{0.8}
	\begin{tabular}{lccc} \hline\hline
		Approach & Stochastic Surrogacy && Latent Unconfoundedness \\
		\hline
		ATE & \citet{hotz2005predicting}, \citet{yang2020combining}, && \citet*{athey2020combining} \\
                             & \citet*{athey2025surrogate_index_restud}, && and \citet*{chen2023semiparametric}\\
                             & and \citet{zhao2025efficient} && \\ 
		\hline
		QTE & \citet{liu2026double} and \citet*{efficientTransported2026} && ---This Paper--- \\
		\hline\hline
	\end{tabular}
\end{table}


While the estimating-equation approaches based on the EIF often allow for modern machine learning methods \citep*[e.g., double machine learning;][]{chernozhukov2018double}, for the LTQTEs, these standard methods would require nonparametric estimation of the quantile density, which in turn requires additional tuning-parameter choices. Instead, we adapt the two-step gradient-based approach of \citet*{jiangetal2023} to reformulate the EIF-based estimating equation as an optimization problem and evaluate the nuisance components at a consistent preliminary estimator of the target quantile. This approach avoids direct estimation of the quantile density and its associated tuning parameters. Moreover, because the nuisance functions are evaluated at a consistent preliminary estimator of the target quantile, the resulting score preserves both Neyman orthogonality and convexity, yielding closed-form estimators in both the original and bootstrap samples. We establish the asymptotic normality of the proposed estimator and show that it attains the semiparametric efficiency bound.

QTEs allow researchers to study distributional heterogeneity in treatment effects. 
To facilitate such analysis, we develop a uniform inference method for the entire LTQTE curve. Specifically, we construct simultaneous confidence bands over a continuum of quantile indices using a multiplier bootstrap procedure. 
Extensive simulation studies support the validity of the proposed inference method, showing that the resulting uniform confidence sets provide reliable simultaneous coverage and control the Type I error uniformly over the quantile index.

The remainder of the paper is organized as follows. Section \ref{sec:model_setup} introduces the model setup. Sections \ref{sec:identification} and \ref{sec:estimation} establish the identification results, propose the estimation procedure, and develop the associated asymptotic theory. Section \ref{sec:bootstrap_uniform_inference} presents the multiplier bootstrap method and develops the uniform inference procedure. Section \ref{sec:simulation} reports the results of the simulation studies. Section \ref{sec:conclusion} concludes. Mathematical details are provided in the appendix.

\paragraph{Notation:}
Let $L_\infty(\mathcal X)$ denote the space of all bounded measurable functions $f:\mathcal X\to\mathbb R$, equipped with the uniform norm
$
\|f\|_\infty
:=
\sup_{x\in\mathcal X}|f(x)|.
$
For any measurable function $f:\mathcal X\to\mathbb R$ with finite second moment, let
$
\|f\|_{\mathbb P,2}
:=
\left\{
\int_{\mathcal X} f^2(x)\,d\mathbb P(x)
\right\}^{1/2}
$
denote its $L_2(\mathbb P)$ norm.
For any measurable function $f$, define the empirical expectation and empirical process, respectively, by
$
\mathbb E_n[f(W)]
:=
\frac{1}{n}\sum_{i=1}^n f(W_i)
$
and
$
\mathbb G_n[f(W)]
:=
\frac{1}{\sqrt n}
\sum_{i=1}^n
\left\{
f(W_i)-\mathbb E[f(W)]
\right\}
=
\sqrt n(\mathbb E_n-\mathbb E)f(W).
$

\section{Model Setup}\label{sec:model_setup}
Let $W_i$ denote the treatment indicator taking values in $\{0,1\}$. 
For each treatment status $w \in \{0,1\}$, individual $i$ has a potential short-term outcome $S_i(w)$ and a potential long-term outcome $Y_i(w)$. 
The realized short-term and long-term outcomes are given by
$S_i = (1 - W_i) S_i(0) + W_i S_i(1), $
and
$Y_i = (1 - W_i) Y_i(0) + W_i Y_i(1),$
respectively. 
Let $X_i$ denote a vector of observed covariates.

Researchers have access to two types of datasets. 
Let $G_i$ indicate the type of dataset to which individual $i$ belongs: $G_i=0$ and $G_i=1$ indicate the experimental and observational data, respectively.
The parameter of interest is the long-term quantile treatment effect (LTQTE) in the observational population.
Specifically, for a quantile level $\tau \in (0,1)$, the LTQTE is defined by
\begin{align}\label{eq:QTE}
\Delta_\tau := q_{1,\tau}-q_{0,\tau},
\end{align}
where 
$q_{w,\tau}:=Q_{Y_i(w)\mid G_i=1}(\tau):=\inf\left\{ y : \Pr\!\left(Y_i(w)\le y \mid G_i=1\right)\ge \tau \right\}$ denotes the $\tau$-th quantile of the potential outcome under the treatment assignment of $w$ among individuals in the observational population.
While the existing literature \citep*{athey2020combining,chen2023semiparametric} focuses primarily on long-term average treatment effects, we instead focus on the LTQTE defined in \eqref{eq:QTE}.

Our target population is the population from which the observational data ($G_i=1$) are collected.
However, observational data often suffer from endogenous treatment selection and attrition. 
This problem is exacerbated in long-term follow-ups.
We therefore exploit the complementary information provided by the experimental data ($G_i=0$). 
In particular, the experimental data provide a controlled treatment assignment mechanism, which we formally state as follows.

\begin{as}[Experimental Validity]\label{as:unconfoundedness} 
$
W_i \indep (Y_i(0), S_i(0), Y_i(1), S_i(1)) \mid X_i, \; G_i = 0 .
$
\end{as}

\noindent
This assumption requires that, in the experimental data, treatment assignment is independent of the short-term and long-term potential outcomes conditional on the pre-treatment covariates.

On the other hand, experiments are often costly, making it difficult, if not impossible, to observe the long-term outcome $Y_i$. 
Accordingly, the researcher observes only the subvector $(S_i, W_i, X_i')$ conditional on $G_i=0$, whereas the full vector $(Y_i, S_i, W_i, X_i')$ is observed conditional on $G_i=1$. 
To accommodate both cases, we collect the observed variables as {$B_i = (G_i,G_iY_i, S_i, W_i, X_i')$, where $Y_i$ is missing when $G_i=0$.}

To leverage the complementary features of the two datasets, \citet*{athey2020combining} take advantage of the following assumptions.

\begin{as}[External Validity]\label{as:external_validity}
$
G_i \indep (Y_i(1), Y_i(0), S_i(1), S_i(0)) \mid X_i.
$
\end{as}

\begin{as}[Latent Unconfoundedness, LU]\label{as:LUTM}
$
W_i \indep Y_i(w) \mid S_i(w), X_i, G_i = 1
$
for each $w \in \{0,1\}$.
\end{as}

Both of these assumptions can be demanding in some applications and therefore require careful consideration in the context of the specific empirical setting of interest.
Assumption \ref{as:external_validity} requires that the experimental data be externally valid with respect to the observational data, in the sense that the conditional distribution of the long-term and short-term potential outcomes given the covariates is the same across the two datasets.
This condition may be plausibly satisfied when the experimental and observational samples are drawn from the same underlying population conditional on observed characteristics.
Assumption \ref{as:LUTM} requires that the source of confounding be summarized by the short-term potential outcome $S_i(w)$. 
In other words, this latent variable (LU) serves as a control variable for evaluating long-term treatment effects. 
This assumption may be plausible in applications where the short-term outcome reflects innate ability conditional on $X_i$, which in turn also determines the long-term outcome up to random shocks.
We refer readers to \citet*{athey2020combining} for further discussion of these key assumptions.

\begin{rem}[{LU and Statistical Surrogacy}]\label{rem:lu_surrogate}
Two approaches are prevalent in the literature on combining experimental and observational data for long-term treatment evaluation: LU \citep*{athey2020combining} and statistical surrogacy \citep*{athey2025surrogate_index_restud}. Under LU, treatment is observed in the observational sample, and short-term potential outcomes are used to adjust for latent confounding. Importantly, this framework allows treatment to affect long-term outcomes through channels not captured by the short-term outcomes. By contrast, statistical surrogacy does not require treatment to be observed in the observational sample, but imposes the stronger restriction that the short-term realized outcomes fully capture the effect of treatment on long-term outcomes and that the relationship between short-term and long-term outcomes is stable across samples.
This explains the distinction between the two columns in Table \ref{tab:surrogacy_vs_lu}.
We adopt the LU framework, as formally stated in Assumption \ref{as:LUTM}.
\end{rem}

In addition, this literature imposes the following standard, albeit potentially restrictive, assumption commonly used in the causal inference literature.

\begin{as}[Overlap]\label{as:overlap} 
There exists a constant $c > 0$ such that for each $w,g \in \{0,1\}$,
\begin{align*}
c \leq \Pr(W_i = w \mid S_i, X_i, G_i = g) \leq 1-c \quad \text{and} \quad c \leq \Pr(G_i = g \mid S_i, X_i, W_i = w) \leq 1-c.
\end{align*}
\end{as}

\noindent
This assumption requires that the probabilities of treatment assignment and of belonging to the observational sample be bounded away from zero and one.

The four assumptions stated thus far (Assumptions \ref{as:unconfoundedness}--\ref{as:overlap}) are directly adopted from the existing literature \citep*[e.g.,][]{athey2020combining, chen2023semiparametric}. 
For the purposes of the present paper, which studies the LTQTE in \eqref{eq:QTE}, as opposed to mean effects, we further impose the following regularity condition considered in quantile analyses. 
Let $F_{Y_i(w)\mid G_i=g}$ and $f_{Y_i(w)\mid G_i=g}$ denote the cumulative distribution function and the corresponding density function, respectively, of the potential outcome $Y_i(w)$ for $w \in \{0,1\}$ and $g \in \{0,1\}$.

\begin{as}[Smoothness]\label{as:smoothness}
Let $\Upsilon\subset(0,1)$ be a compact set. For each $w\in\{0,1\}$, define the quantile set $\mathcal Q_w(\Upsilon)=\{q_{w,\tau}:\tau\in\Upsilon\}$ and its fixed $\delta$-enlargement
$
	\mathcal Q_w^\delta(\Upsilon)=\{q_{w,\tau}+v:\tau\in\Upsilon,\ |v|\leq\delta\},
$
where $\delta>0$ is a fixed constant value. There exist constants $0<c_f<C_f<\infty$ and $L_f<\infty$ such that the marginal density $f_{Y(w)\mid G=1}$ satisfies
\begin{align*}
	0<c_f\leq\inf_{w\in\{0,1\}}\inf_{y\in\mathcal Q_w^\delta(\Upsilon)}f_{Y(w)\mid G=1}(y)\leq\sup_{w\in\{0,1\}}\sup_{y\in\mathcal Q_w^\delta(\Upsilon)}f_{Y(w)\mid G=1}(y)\leq C_f.
\end{align*}
The marginal density is uniformly Lipschitz continuous on these enlarged sets:
\begin{align*}
	\sup_{w\in\{0,1\}}\sup_{\substack{y,y'\in\mathcal Q_w^\delta(\Upsilon)\\
			y\neq y'}}
	\frac{\left|f_{Y(w)\mid G=1}(y)-f_{Y(w)\mid G=1}(y')\right|}{|y-y'|}\leq L_f.
\end{align*}
Finally, the conditional density is uniformly bounded:
\begin{align*}
	\sup_{\substack{
			w\in\{0,1\},\,y\in\mathbb R\\
			x\in\operatorname{Supp}(X)}}
	f_{Y(w)\mid G=1}(y\mid x)
	<\infty.
\end{align*}
\end{as}

\begin{as}[Data]\label{as:iid}
$B_i$ is i.i.d. with a bounded fourth moment.
\end{as}

Assumption~\ref{as:smoothness} imposes regularity conditions on the potential-outcome quantiles uniformly over $\Upsilon$. The lower bound on the density permits uniform quantile linearization, while the Lipschitz continuity and boundedness of the conditional densities control the associated remainder terms. Assumption~\ref{as:iid} provides the sampling and moment conditions required for uniform laws of large numbers, central limit theory, and Gaussian approximation for the LTQTE estimator.

\section{Identification with Efficient Influence Functions}\label{sec:identification}

In this section, we establish identification of the LTQTE defined in \eqref{eq:QTE}. 
Whenever no confusion arises, we suppress the individual subscript $i$ throughout this section.

We introduce the following notation for conditional probabilities and their integrated counterparts:
\begin{align*}
m_w(s, x ; q) &= \Pr(Y \leq q \mid W=w, S=s, X=x, G=1)
\qquad\text{and}\\
\overline{m}_w(x;q) &= \mathbb{E}\left[m_w(S, X;q) \mid W=w, X=x, G=0\right].
\end{align*}
We also introduce the following notation for the various propensity scores:
\begin{align*}
\rho_w(s, x) &= \Pr(W=w \mid S(w)=s, X=x, G=1)
~~\text{and}~~
\varrho(x) = \Pr(W=1 \mid X=x, G=0).
\end{align*}
Similarly, denote the conditional and unconditional probabilities of selection into the observational sample by
\begin{align*}
\gamma(x) = \Pr(G=1 \mid X=x)
~~\text{and}~~
\pi = \Pr(G=1),~~\text{respectively}.
\end{align*}
Although $\rho_w(s,x)$ conditions on the potential short-term outcome $S(w)$, it can be identified from observed data under Assumptions \ref{as:unconfoundedness} and \ref{as:external_validity}. Specifically, following \citet{chen2023semiparametric}, Bayes' rule yield
\begin{align}
	&\rho_w(s,x) =
	\nonumber\\
	&\frac{\Pr(G=1\mid S=s,W=w,X=x)}{\Pr(G=0\mid S=s,W=w,X=x)}\frac{\Pr(G=0\mid W=w,X=x)}{\Pr(G=1\mid W=w,X=x)}\Pr(W=w\mid X=x,G=1).
	\label{eq:rho-identification}
\end{align}
Under the overlap condition in Assumption \ref{as:overlap}, $\rho_w(s,x)$ is identified through conditional probabilities involving only observed variables from the experimental and observational samples.

Collect the outcome regression functions as
\begin{align*}
	\eta_{\rm out}(B;q_0,q_1)=\left(\overline{m}_w(X;q_w),\, m_w(S,X;q_w)\right)_{w=0,1},
\end{align*}
and the propensity-score-type functions as
\begin{align*}
	\eta_{\rm prop}(B)=\left(\rho_0(S,X),\,\rho_1(S,X),\,\varrho(X),\,\gamma(X),\,\pi\right).
\end{align*}
Consequently, the full collection of nuisance functions is given by
\begin{align*}
	\eta=\left(\eta_{\rm out},\eta_{\rm prop}\right),~~\eta_{\rm out}=\eta_{\rm out}(B;q_0,q_1),~~\text{and}~~
	\eta_{\rm prop}=\eta_{\rm prop}(B).
\end{align*}

Under the stated assumptions, we can identify the quantiles of interest, $q_{1,\tau}$ and $q_{0,\tau}$, through the following inverse-probability-weighted (IPW) moment restrictions:
\begin{align}
	&\mathbb{E}[\phi_{1}(B_i, q_{1,\tau}, \zeta_{1})]:=\mathbb{E}\left[\frac{G_i}{\pi}\left(\frac{W_i \mathbf{1}\{Y_i\leq q_{1,\tau}\}}{\rho_1(S_i, X_i)}\right) - \tau\right] = 0 \qquad\text{and} \label{eq:ipw0}\\
	&\mathbb{E}[\phi_{0}(B_i, q_{0,\tau}, \zeta_{0})]:=\mathbb{E}\left[
	\frac{G_i}{\pi}\left(\frac{(1-W_i)\mathbf{1}\{Y_i\leq q_{0,\tau}\}}{\rho_0(S_i, X_i)}\right) - \tau
	\right] = 0, 	\label{eq:ipw1}
\end{align}
respectively, where $\zeta_w = (\pi, \rho_w)$ for $w\in\{0,1\}$. 
These moment conditions not only identify the target quantiles but also provide a basis for deriving their efficient influence functions (EIFs). 
By viewing $q_{1,\tau}$ and $q_{0,\tau}$ as solutions to mean-zero moment conditions, we can characterize their pathwise derivatives and thereby derive their EIFs, as formally stated in the following theorem.

\begin{thm}\label{thm:EIF1}
Suppose that Assumptions \ref{as:unconfoundedness}--\ref{as:smoothness} hold. The EIFs for $q_{1,\tau}$ and $q_{0,\tau}$ are given by
\begin{align}
	\psi_1(B, q_{1,\tau}, \eta) 
	&= -\frac{G}{\pi } \Bigg[
	\frac{W \,( \mathbf{1}\{Y \le q_{1,\tau}\} - m_1(S,X;q_{1,\tau}) )}{\rho_1(S,X) f_{Y(1) \vert G=1}(q_{1,\tau})}+\frac{\overline{m}_1(X;q_{1,\tau})-\tau}{f_{Y(1) \vert G=1}(q_{1,\tau})}
	\Bigg] \notag\\
	&\label{eq:eif1}-\frac{1-G}{\pi } \frac{\gamma(X)}{1-\gamma(X)} \Bigg[
	\frac{W \,( m_1(S,X;q_{1,\tau}) - \overline{m}_1(X;q_{1,\tau}) )}{\varrho(X)f_{Y(1) \vert G=1}(q_{1,\tau})}\Bigg]\qquad\text{and}\\
	\psi_0(B,  q_{0,\tau}, \eta) 
	&= -\frac{G}{\pi } \Bigg[\frac{(1-W)\,( \mathbf{1}\{Y \le q_{0,\tau}\} - m_0(S,X;q_{0,\tau}) )}{\rho_0(S,X)f_{Y(0) \vert G=1}(q_{0,\tau})}+ \frac{\overline{m}_0(X;q_{0,\tau})-\tau}{f_{Y(0) \vert G=1}(q_{0,\tau})}
	\Bigg] \notag\\
	&\label{eq:eif2}-\frac{1-G}{\pi } \frac{\gamma(X)}{1-\gamma(X)} \Bigg[
	\frac{(1-W)\, ( m_0(S,X;q_{0,\tau}) - \overline{m}_0(X;q_{0,\tau}) )}{(1-\varrho(X))f_{Y(0) \vert G=1}(q_{0,\tau})}
	\Bigg],
\end{align}
respectively. 
Consequently, the EIF for the LTQTE, $\Delta_\tau$, is given by
$$\psi(B, q_{1,\tau}, q_{0,\tau}, \eta)=\psi_1(B, q_{1,\tau}, \eta)-\psi_0(B, q_{0,\tau}, \eta).$$
\end{thm}

While the original moment restrictions \eqref{eq:ipw0}--\eqref{eq:ipw1} only serve to identify $q_{1,\tau}$ and $q_{0,\tau}$, Theorem \ref{thm:EIF1} further establishes their efficient counterparts through the associated EIFs in \eqref{eq:eif1}--\eqref{eq:eif2}.
In particular, $q_{1,\tau}$ and $q_{0,\tau}$ are characterized as solutions to
\begin{align}\label{kk1}
\mathbb{E}\big[\psi_w(B, q_{w,\tau}, \eta_0)\big] = 0,
\end{align}
for $w = 1$ and $w = 0$, respectively, where $\eta_0$ denotes the true nuisance function.

These EIFs can be shown to be Neyman orthogonal. That is, for any perturbation $h$,
\begin{align*}
\left.\frac{\partial}{\partial r}
\mathbb{E}\big[\psi_w(B, q_{w,\tau}, \eta_0 + r h)\big]
\right|_{r=0} = 0.
\end{align*}
In other words, these moment conditions are locally insensitive to first-order perturbations of the nuisance functions, a key property underlying valid inference with flexible first-step estimators of $\eta$; see \citet{chernozhukov2022locally, ichimura2022influence}.
See also \citet*{chen2026equivalence} for a general characterization of the connection between EIFs and Neyman orthogonality.

Furthermore, they are doubly robust in the sense that
$\mathbb{E}\big[\psi_w(B, q_{w,\tau}, \eta)\big] = 0$
holds if either
\begin{enumerate}
\item[(a)] the propensity-score-type functions $(\pi, \rho_w, \gamma, \varrho)$ are correctly specified; or
\item[(b)] the outcome regression functions $(m_w, \overline{m}_w)$ are correctly specified.
\end{enumerate}
We formally state these properties in the following theorem.
		
\begin{thm}[Local and Double Robustness]\label{thm:dr}
Suppose that Assumptions \ref{as:unconfoundedness}--\ref{as:smoothness} hold. The EIFs $\psi_w(B,q_{w,\tau},\eta)$ defined in \eqref{eq:eif1}--\eqref{eq:eif2} are Neyman orthogonal and doubly robust.
\end{thm}

In summary, the EIFs in \eqref{eq:eif1}--\eqref{eq:eif2} provide the basis for efficient estimation while also delivering Neyman orthogonality and double robustness.
Motivated by these results, the next section develops a semiparametric procedure of efficient estimation and inference on the LTQTE.

\section{Integrated M-Estimation}\label{sec:estimation}

The identification results in Section \ref{sec:identification} characterize $q_{1,\tau}$ and $q_{0,\tau}$ as zeros of Neyman-orthogonal and doubly robust moment functions, suggesting estimation via flexible first-step plug-in estimators. However, the EIFs depend on the density evaluated at $q_{w,\tau}$, requiring nonparametric density estimation and additional tuning parameters.

While the EIF-based approach accommodates modern machine learning methods \citep*[e.g., double machie learning;][]{chernozhukov2018double}, for the LTQTEs, these standard methods would require estimating conditional densities given high-dimensional covariates, which is challenging in practice rather than in theory. We instead adapt the two-step gradient-based approach of \citet*{jiangetal2023}, which reformulates the EIF-based estimating equation as an optimization problem and evaluates nuisance components at a consistent preliminary estimator of the target quantile, thereby circumventing density estimation. Moreover, because nuisance functions are evaluated at a consistent estimator, the score retains both Neyman orthogonality and convexity, yielding closed-form estimators in the original and bootstrap samples. With this approach, we establish asymptotic normality and show that the estimator attains the semiparametric efficiency bound.

\subsection{Population Integrated Objective Functions}
\label{subsec:integratedM}

Recall that the quantiles $q_{w,\tau}$ for $w=0$,$1$ show up in the EIFs \eqref{eq:eif1}--\eqref{eq:eif2} via the indicator $\mathbf 1\{Y\le q_{w,\tau}\}$, the nuisance functions $m_w$ and $\overline m_w$, and the density function $f_{Y(w)\vert G=1}(q_{w,\tau})$.
At the true quantiles and nuisance functions, the EIFs satisfy
\begin{align*}
	\psi_1(B,q_{1,\tau},\eta_0)
	&= -\frac{\Phi_1(B;q_{1,\tau},q_{1,\tau},\eta_0)}{f_{Y(1)\vert G=1}(q_{1,\tau})}
	\qquad\text{and}\\
	\psi_0(B,q_{0,\tau},\eta_0)
	&= -\frac{\Phi_0(B;q_{0,\tau},q_{0,\tau},\eta_0)}{f_{Y(0)\vert G=1}(q_{0,\tau})},
\end{align*}
where
\begin{align}
	\Phi_1(B;q,a,\eta)
	&=\frac{G}{\pi}\left[
	\frac{W\{\mathbf 1(Y\le q)-m_1(S,X;a)\}}{\rho_1(S,X)}
	+\overline m_1(X;a)-\tau
	\right]\notag\\
	&\quad+\frac{1-G}{\pi}\frac{\gamma(X)}{1-\gamma(X)}
	\left[
	\frac{W\{m_1(S,X;a)-\overline m_1(X;a)\}}{\varrho(X)}
	\right]
	\label{eq:Phi1}
	\qquad\text{and}\\
	\Phi_0(B;q,a,\eta)
	&=\frac{G}{\pi}\left[
	\frac{(1-W)\{\mathbf 1(Y\le q)-m_0(S,X;a)\}}{\rho_0(S,X)}
	+\overline m_0(X;a)-\tau
	\right]\notag\\
	&\quad+\frac{1-G}{\pi}\frac{\gamma(X)}{1-\gamma(X)}
	\left[
	\frac{(1-W)\{m_0(S,X;a)-\overline m_0(X;a)\}}{1-\varrho(X)}
	\right]
	\label{eq:Phi0}
\end{align}
with $\eta=(m_0,m_1,\overline m_0,\overline m_1,\rho_0,\rho_1,\varrho,\gamma,\pi)$, where $\eta$ collects the nuisance functions defined in Section \ref{sec:identification}, and $\eta_0$ denotes their true values.
When $a=q_{w,\tau}$ and $\eta=\eta_0$, the derivative of $\mathbb E[\Phi_w(B;q,a,\eta)]$ with respect to $q$, evaluated at $q=q_{w,\tau}$, is the density $f_{Y(w)\vert G=1}(q_{w,\tau})$. Consequently, by integrating the moment function $\Phi_w(B;q,a,\eta)$ with respect to $q$, while keeping $a$ and $\eta$ fixed, we can construct M-estimation objective functions: 
\begin{align}
	\mathcal L_{1,\tau}(q;a,\eta)
	&=\mathbb E\Bigg[
	\frac{G}{\pi}\frac{W}{\rho_1(S,X)}
	\left\{\rho_\tau(Y-q)+\big(\tau-m_1(S,X;a)\big)q\right\}
	+\frac{G}{\pi}\left(\overline m_1(X;a)-\tau\right)q\notag\\
	&\qquad+\frac{1-G}{\pi}\frac{\gamma(X)}{1-\gamma(X)}
	\frac{W}{\varrho(X)}
	\big\{m_1(S,X;a)-\overline m_1(X;a)\big\}q
	\Bigg]
	\label{eq:poploss1}
	\qquad\text{and}\\
	\mathcal L_{0,\tau}(q;a,\eta)
	&=\mathbb E\Bigg[
	\frac{G}{\pi}\frac{1-W}{\rho_0(S,X)}
	\left\{\rho_\tau(Y-q)+\big(\tau-m_0(S,X;a)\big)q\right\}
	+\frac{G}{\pi}\left(\overline m_0(X;a)-\tau\right)q\notag\\
	&\qquad+\frac{1-G}{\pi}\frac{\gamma(X)}{1-\gamma(X)}
	\frac{1-W}{1-\varrho(X)}
	\big\{m_0(S,X;a)-\overline m_0(X;a)\big\}q
	\Bigg].
	\label{eq:poploss0}
\end{align}
Then, the corresponding population score, given by the first derivative of $\mathcal L_{w,\tau}(q;a,\eta)$ with respect to $q$, is $\mathbb E[\Phi_w(B;q,a,\eta)]$, while its Hessian, evaluated at $q=a=q_{w,\tau}$ and $\eta=\eta_0$, is precisely the density $f_{Y(w)\vert G=1}(q_{w,\tau})$.
Consequently, the influence function associated with the M-estimator based on the objective function $\mathcal L_{w,\tau}(q;q_{w,\tau},\eta_0)$ coincides with the EIF. The density serves as the Hessian of the M-estimation objective, and becomes implicit in the optimization problem, thereby circumventing the need for explicit density estimation and its associated tuning parameters.

Another advantage of the integrated M-estimation approach is that the objective function $\mathcal L_{w,\tau}(q;a,\eta)$ is convex in $q$ for any fixed $a$ and $\eta$, since the check loss $\rho_\tau(Y-q)$ is convex in $q$ and the remaining terms are linear in $q$. The sample analogue of the optimization problem can therefore be characterized through subgradient conditions and solved in closed form by sorting the $Y_i$'s, as discussed in more detail in the next subsection.

\subsection{Cross-fitting Estimation}

Let $\mathcal I_1,...,\mathcal I_K$ be a partition of $[n]:=\{1,...,n\}$. For each fold $k$, estimate the nuisance functions and an initial quantile estimator using observations outside $\mathcal I_k$. 

We adopt the DML2 implementation of \citet*{chernozhukov2018double}, which solves a single estimating equation after averaging the cross-fitted orthogonal scores. Compared with DML1, DML2 is first-order equivalent but yields better high-order asymptotics; see also \citet{velez2024asymptotic}.

Write these estimates as
\begin{align*}
	\widehat\eta^{(-k)}=\left(\widehat m_0^{(-k)},\widehat m_1^{(-k)},\widehat{\overline m}_0^{(-k)},\widehat{\overline m}_1^{(-k)},\widehat\rho_0^{(-k)},\widehat\rho_1^{(-k)},\widehat\varrho^{(-k)},\widehat\gamma^{(-k)},\widehat\pi^{(-k)}\right),
\end{align*}
and let $\widetilde q_{1,\tau}^{(-k)}$ and $\widetilde q_{0,\tau}^{(-k)}$ denote initial estimators of quantiles, $q_{1,\tau}$ and $q_{0,\tau}$, respectively, constructed using observations outside $\mathcal I_k$. In particular, the propensity-score function $\rho_w(s,x)$ is estimated using the observable representation in \eqref{eq:rho-identification}. For each training sample $\mathcal I_k^c$, we separately estimate the conditional probabilities appearing on the right-hand side of \eqref{eq:rho-identification} and combine them to construct $\widehat\rho_w^{(-k)}(s,x)$.

For $w=1$, define the cross-fitted sample objective function as 
\begin{align}
	\widehat{\mathcal L}_{1,\tau}(q)&=\frac{1}{n}{\sum_{k=1}^K \sum_{i\in\mathcal I_k}}\Bigg[\frac{G_i}{\widehat\pi^{(-k)}}\frac{W_i}{\widehat\rho_1^{(-k)}(S_i,X_i)}\left\{\rho_\tau(Y_i-q)+\big(\tau-\widehat m_1^{(-k)}(S_i,X_i;\widetilde q^{(-k)}_{1,\tau})\big)q \right\} \notag\\
	&\qquad+\frac{G_i}{\widehat\pi^{(-k)}} \left( \widehat{\overline m}_1^{(-k)}(X_i;\widetilde q^{(-k)}_{1,\tau}) - \tau \right) q \notag\\
	&\qquad+\frac{1-G_i}{\widehat\pi^{(-k)}}\frac{\widehat\gamma^{(-k)}(X_i)}{1-\widehat\gamma^{(-k)}(X_i)}\frac{W_i}{\widehat\varrho^{(-k)}(X_i)}\big\{\widehat m_1^{(-k)}(S_i,X_i;\widetilde q^{(-k)}_{1,\tau})-\widehat{\overline m}_1^{(-k)}(X_i;\widetilde q^{(-k)}_{1,\tau})\big\}q\Bigg].
	\label{eq:sampleloss1}
\end{align}
Then, we define the quantile estimate $\widehat q_{1,\tau}$ given by
\begin{align}
	\widehat q_{1,\tau}=\arg\min_{q}\widehat{\mathcal L}_{1,\tau}(q).
	\label{eq:qhat1-new}
\end{align}
Similarly, for $w=0$, we can obtain 
\begin{align}
	\widehat{\mathcal L}_{0,\tau}(q)&=\frac{1}{n}{\sum_{k=1}^K \sum_{i\in\mathcal I_k}} \Bigg[\frac{G_i}{\widehat\pi^{(-k)}}\frac{1-W_i}{\widehat\rho_0^{(-k)}(S_i,X_i)} \left\{\rho_\tau(Y_i-q)+\big(\tau-\widehat m_0^{(-k)}(S_i,X_i;\widetilde q^{(-k)}_{0,\tau})\big)q \right\} \notag\\
	&\qquad+\frac{G_i}{\widehat\pi^{(-k)}} \left( \widehat{\overline m}_0^{(-k)}(X_i;\widetilde q^{(-k)}_{0,\tau}) - \tau \right) q \notag\\
	&\qquad+\frac{1-G_i}{\widehat\pi^{(-k)}}\frac{\widehat\gamma^{(-k)}(X_i)}{1-\widehat\gamma^{(-k)}(X_i)}\frac{1-W_i}{1-\widehat\varrho^{(-k)}(X_i)}\big\{\widehat m_0^{(-k)}(S_i,X_i;\widetilde q^{(-k)}_{0,\tau})-\widehat{\overline m}_0^{(-k)}(X_i;\widetilde q^{(-k)}_{0,\tau}) \big\}q\Bigg].
	\label{eq:sampleloss0}
\end{align}
We define the quantile estimate $\widehat q_{0,\tau}$ given by
\begin{align}
	\widehat q_{0,\tau} = \arg\min_{q} \widehat{\mathcal L}_{0,\tau}(q),
	\label{eq:qhat0-new}
\end{align}
and $\widehat\Delta_\tau=\widehat q_{1,\tau}-\widehat q_{0,\tau}$.

The details of the cross-fitted estimation procedure are provided in Appendix \ref{sec:implementation1}. The corresponding objective functions in \eqref{eq:sampleloss1} and \eqref{eq:sampleloss0} are convex in $q$, and the estimators can be computed using the subgradient characterization described in Appendix~\ref{sec:implementation2}.

\subsection{Asymptotic Theory}

This section presents the asymptotic distribution of the integrated M-estimator.

As the Neyman orthogonality condition ensures that the first-order impact of nuisance function estimation is asymptotically negligible, establishing uniform asymptotic linearity of the integrated estimator only requires the nuisance functions to be estimated at sufficiently fast rates so that their second-order estimation error remains asymptotically negligible. Specifically, we can divide the nuisance functions into propensity score-type functions 
\begin{align*}
\eta_{\rm prop}(B) = (\pi,\rho_0(S,X),\rho_1(S,X),\gamma(X), \varrho(X))	
\end{align*}
and the outcome regression-type functions 
\begin{align*}
\eta_{\rm out}(B;q_0,q_1) = (m_w(S,X;q_w),\overline m_w(X;q_w))_{w=0,1},	
\end{align*}
and require the following conditions on their cross-fit estimators $(\widehat \eta_{\rm prop}^{(-k)}(B), \widehat \eta_{\rm out}^{(-k)}(B;\widetilde q_{0,\tau}^{(-k)},\widetilde q_{1,\tau}^{(-k)}))$ to hold, where  $(\widetilde q_{0,\tau}^{(-k)},\widetilde q_{1,\tau}^{(-k)})$ are the initial quantile estimators for $(q_{0,\tau},q_{1,\tau})$. 

\begin{as}[First-Step Rate Condition]\label{as:first-stage-rate}
For $w = 0,1$, denote $\mathcal Q_w$ as the the compact $\epsilon$-enlargement of $\{q_{w,\tau}:\tau\in\Upsilon\}$ for some $\delta>\epsilon>0$. Suppose there exists some deterministic sequence $r_n\to0$ such that the following rates hold:
	\begin{align}
		&\max_{1\le k\le K}\|\widehat\eta_{\rm prop}^{(-k)}(B)-\eta_{\rm prop}(B)\|_{\mathbb P,2}=O(r_n),
		\label{eq:dml-first-stage-rate-propensity}
\\
		&\max_{1\le k\le K} \sup_{\tau\in\Upsilon} \left| \widetilde q_{w,\tau}^{(-k)}-q_{w,\tau} \right|=O(r_n), \qquad w=0,1,
		\label{eq:initial-q-rate-simple}
\\
			&\max_{1\le k\le K} \sup_{(q_0,q_1) \in \mathcal Q_0 \times \mathcal Q_1}\|\widehat\eta_{\rm out}^{(-k)}(B;q_0,q_1)-\eta_{\rm out}(B;q_0,q_1)\|_{\mathbb P,2}=O(r_n), 	
			\label{eq:dml-first-stage-rate-outcome}
\qquad\text{and}
\\
		&r_n\sqrt{\log(1/r_n)}=o(1)~~\text{and}~~\sqrt n r_n^2=o(1).
		\label{eq:dml-rate-condition-simple}
	\end{align}
Moreover, there exists a constant $c_*>0$ such that
\begin{align*}
	\Pr\Bigg(\min_{k\in[K]}\inf_{w\in\{0,1\}}\inf_{s,x}\widehat\rho_w^{(-k)}(s,x)\ge c_*,~~\min_{k\in[K]}\inf_x
	\Big\{\widehat\varrho^{(-k)}(x),1-\widehat\varrho^{(-k)}(x),1-\widehat\gamma^{(-k)}(x)\Big\}\ge c_*\Bigg)\rightarrow1.
\end{align*}

\end{as}

\begin{as}[Uniform $L_2$-Continuity in Quantile Index]
	\label{as:outcome-lipschitz-q}
	There exists a constant $C<\infty$ such that, for all $(q_0,q_1),(q_0',q_1')\in\mathcal Q_0\times\mathcal Q_1$,
	\begin{align*}
		\left\Vert\eta_{\rm out}(B;q_0,q_1)-\eta_{\rm out}(B;q_0',q_1')\right\Vert_{\mathbb P,2}
		\leq C\max_{w=0,1}|q_w-q_w'|.
	\end{align*}
\end{as}

\begin{as}[Uniform Complexity]
	\label{as:uniform_score_complexity}
	Let $\mathcal D_{-k}=\sigma(B_i:i\notin I_k)$ be the training-sample sigma-field for fold $k$. For $w\in\{0,1\}$, $\tau\in\Upsilon$ and $v\in\mathbb R$, define
	\begin{align*}
		h_{n,k,w,\tau,v}(B)=\Phi_w\left(B;q_{w,\tau}+\frac{v}{\sqrt n},\widetilde q_{w,\tau}^{(-k)},\widehat\eta^{(-k)}
		\right)-\Phi_w\left(B;q_{w,\tau}+\frac{v}{\sqrt n},q_{w,\tau},\eta_0\right).
	\end{align*}
	For each fixed $M\in(0,\infty)$, let
	\begin{align*}
	\mathcal H_{n,k,w}(M):=\left\{h_{n,k,w,\tau,v}:\tau\in\Upsilon,\ |v|\le M\right\}.
	\end{align*}
	Suppose that the following conditions hold for every fixed $M$.
	
	\begin{enumerate}
	
		\item[(i)] \emph{Conditional uniform envelope and entropy condition.}
		There exist a deterministic constant value $F_M\in[1,\infty)$ independent of $n$ and deterministic sequences $A_n\ge e$ and $v_n\ge1$, which may diverge with $n$, such that, with probability approaching one,
		\begin{align}
			&\max_{k\in[K]}\max_{w\in\{0,1\}}\sup_{h\in\mathcal H_{n,k,w}(M)}\|h\|_{\mathbb P,\infty}\le F_M,
			\label{eq:uc_envelope}
			\\
			&\max_{k\in[K]}\max_{w\in\{0,1\}}\sup_Q\mathcal N(\varepsilon F_M,\mathcal H_{n,k,w}(M),L_2(Q))\le\left(\frac{A_n}{\varepsilon}\right)^{v_n}
			\label{eq:uc_entropy}
		\end{align}
		simultaneously for every $0<\varepsilon\le1$.
		
		\item[(ii)] 
		\emph{Complexity-adjusted rate restrictions.}
		With $J_n:=v_n\log(A_n/r_n)$, we assume that
		\begin{equation}
			\label{eq:uc_growth}
			r_n\sqrt{J_n}+\frac{J_n\log(n\vee e)}{\sqrt n}+\sqrt n\,r_n^2\rightarrow0.
		\end{equation}
	\end{enumerate}
\end{as}

Assumption \ref{as:first-stage-rate} imposes boundedness and rate conditions on the first-step estimators. While these conditions are stated at a high level, Section \ref{sec:np-first-stage-uniform} provides concrete examples of first-step estimators together with primitive conditions under which these rate requirements are satisfied.
Assumption \ref{as:outcome-lipschitz-q} requires the outcome-regression nuisance functions to be Lipschitz continuous over quantiles. A sufficient condition is that the conditional density of $Y(w)$ given $(S(w),X,G=1)$ is uniformly Lipschitz continuous over quantiles. This property carries over to the outcome-regression nuisance functions by integration. 

Assumption~\ref{as:uniform_score_complexity} imposes the uniform envelope and entropy conditions on the cross-fitted score differences. The rate restrictions ensure that the empirical and multiplier process remainders are asymptotically negligible. For the indicator-response Lasso estimators, {Theorem \ref{thm:np-uniform-first-stage}} verifies both conditions, following \citet{belloni2017program}.

\begin{thm}[Asymptotic Properties of the Integrated Estimator]
	\label{thm:integratedM}
	Suppose that Assumptions \ref{as:unconfoundedness}--\ref{as:iid}, \ref{as:first-stage-rate}, \ref{as:outcome-lipschitz-q}--\ref{as:uniform_score_complexity} hold. Then, we obtain
	\begin{align}
\sup_{\tau \in \Upsilon}		\left\vert \sqrt n(\widehat\Delta_\tau-\Delta_\tau)- \frac{1}{\sqrt n}\sum_{i=1}^n\psi_\Delta^{M}(B_i;\tau) \right\vert  = o_p(1),
		\label{eq:Delta-bahadur-new}
	\end{align}
 	where
	\begin{align}
		\psi_\Delta^{M}(B;\tau)=-\frac{\Phi_1(B;q_{1,\tau},q_{1,\tau},\eta_0)}{f_{Y(1)\mid G=1}(q_{1,\tau})}+\frac{\Phi_0(B;q_{0,\tau},q_{0,\tau},\eta_0)}{f_{Y(0)\mid G=1}(q_{0,\tau})}.
		\label{eq:M-IF-Delta}
	\end{align}

	Moreover, the following functional law holds:
	\begin{align*}
		\sqrt n(\widehat\Delta_\tau-\Delta_\tau) \rightsquigarrow \mathcal B(\tau)~~ \text{in }L_\infty(\Upsilon),
	\end{align*}
	where $\mathcal B$ is a tight mean-zero Gaussian process with the covariance kernel
	\begin{align}
		\Sigma_\Delta(\tau,\tau')=\mathbb E[\psi_\Delta^{M}(B;\tau)\psi_\Delta^{M}(B;\tau')].
		\label{eq:M-Sigma}
	\end{align}
\end{thm}

The oracle influence function in \eqref{eq:M-IF-Delta} governs the first-order behavior of estimator uniformly over $\Upsilon$, with the first-stage estimation entering through the $o_p(1)$ remainder in \eqref{eq:Delta-bahadur-new}. The continuous mapping theorem gives $\|\sqrt{n}(\widehat\Delta-\Delta)\|_{\mathbb P,\infty}\rightsquigarrow \|\mathcal B\|_{\mathbb P,\infty}$, yielding the distribution for simultaneous confidence bands.

\section{Bootstrap Inference}\label{sec:bootstrap_uniform_inference}

Building on the asymptotic theory developed above, this section presents the multiplier bootstrap procedure for pointwise and uniform inference on the LTQTE process.

\subsection{Multiplier Bootstrap Construction and Validity}

We implement the multiplier bootstrap inference by reweighting the integrated objectives. Let $\{\xi_i\}_{i=1}^n$ be nonnegative i.i.d. multiplier weights, independent of the data, satisfying
\begin{align}\label{eq:multiplier}
\mathbb E[\xi_i]=1~~\text{and}~~\mathbb E[(\xi_i-1)^2]=1,
\end{align}
and having a sub-exponential upper tail. The nonnegativity of the multiplier weights preserves the convexity of the weighted objective functions. In practice, we recommend drawing the multiplier weights from an $\operatorname{Exp}(1)$ distribution. Importantly, the high-dimensional nuisance functions need not be re-estimated in each bootstrap sample.

For $w=1$, define the bootstrap weighted objective by
\begin{align*}
	\widehat{\mathcal L}_{1,\tau}^*(q) &=\frac{1}{n}{\sum_{k=1}^K \sum_{i\in\mathcal I_k}}
	\xi_i \Bigg[\frac{G_i}{\widehat\pi^{(-k)}}\frac{W_i}{\widehat\rho_1^{(-k)}(S_i,X_i)}
	\left\{\rho_\tau(Y_i-q)+\big(\tau-\widehat m_1^{(-k)}(S_i,X_i;\widetilde q_{1,\tau}^{(-k)})\big)q\right\} \\
	&\qquad +\frac{G_i}{\widehat\pi^{(-k)}}\left( \widehat{\overline m}_1^{(-k)}(X_i;\widetilde q_{1,\tau}^{(-k)}) - \tau \right)q \\
	&\qquad +\frac{1-G_i}{\widehat\pi^{(-k)}}\frac{\widehat\gamma^{(-k)}(X_i)}{1-\widehat\gamma^{(-k)}(X_i)} \frac{W_i}{\widehat\varrho^{(-k)}(X_i)}\big\{\widehat m_1^{(-k)}(S_i,X_i;\widetilde q_{1,\tau}^{(-k)})-\widehat{\overline m}_1^{(-k)}(X_i;\widetilde q_{1,\tau}^{(-k)})\big\}q \Bigg],
\end{align*}
and define
\begin{align*}
\widehat q_{1,\tau}^* = \arg\min_{q\in\mathcal Q} \widehat{\mathcal L}_{1,\tau}^*(q).
\end{align*}
Similarly, for $w=0$, define
\begin{align*}
	\widehat{\mathcal L}_{0,\tau}^*(q)&=\frac{1}{n}{\sum_{k=1}^K \sum_{i\in\mathcal I_k}}
	\xi_i \Bigg[\frac{G_i}{\widehat\pi^{(-k)}}\frac{1-W_i}{\widehat\rho_0^{(-k)}(S_i,X_i)} \left\{\rho_\tau(Y_i-q)+\big(\tau-\widehat m_0^{(-k)}(S_i,X_i;\widetilde q_{0,\tau}^{(-k)})\big)q\right\} \\
	&\qquad +\frac{G_i}{\widehat\pi^{(-k)}}\left( \widehat{\overline m}_0^{(-k)}(X_i;\widetilde q_{0,\tau}^{(-k)}) - \tau \right)q \\
	&\qquad +\frac{1-G_i}{\widehat\pi^{(-k)}}\frac{\widehat\gamma^{(-k)}(X_i)} {1-\widehat\gamma^{(-k)}(X_i)}\frac{1-W_i}{1-\widehat\varrho^{(-k)}(X_i)} \big\{\widehat m_0^{(-k)}(S_i,X_i;\widetilde q_{0,\tau}^{(-k)})-\widehat{\overline m}_0^{(-k)}(X_i;\widetilde q_{0,\tau}^{(-k)})\big\}q \Bigg],
\end{align*}
and
\begin{align*}
\widehat q_{0,\tau}^* = \arg\min_{q\in\mathcal Q} \widehat{\mathcal L}_{0,\tau}^*(q).
\end{align*}
The bootstrap estimator of the LTQTE is
\begin{align*}
\widehat\Delta_\tau^*=\widehat q_{1,\tau}^*-\widehat q_{0,\tau}^*.
\end{align*}
Implementation details of the multiplier bootstrap procedure, including the subgradient characterization of the bootstrap estimators, are provided in Appendix \ref{sec:implementation3}.

The next theorem states that the multiplier bootstrap consistently approximates the sampling law of the LTQTE process.

\begin{thm}[Multiplier Bootstrap Validity]\label{thm:multiplier-bootstrap-integrated}
Suppose that Assumptions \ref{as:unconfoundedness}--\ref{as:iid} and \ref{as:first-stage-rate}--\ref{as:uniform_score_complexity} hold. Let $\{\xi_i\}_{i\in[n]}$ be nonnegative multiplier weights that are independent of the data, satisfy \eqref{eq:multiplier}, and satisfy a sub-exponential tail condition. Then, conditionally on the data
\begin{align}\label{eq:boot_converge}
	\sqrt n(\widehat\Delta_\tau^*-\widehat\Delta_\tau) \underset{\xi}{\overset{\mathbb{P}}{\rightsquigarrow}} \mathcal B(\tau)\quad\text{in }L_\infty(\Upsilon)
\end{align}
in probability, where $\mathcal B$ is the same Gaussian process as in Theorem \ref{thm:integratedM}, with covariance kernel defined in \eqref{eq:M-Sigma}.\footnote{\label{remark1}We view $\sqrt n(\widehat\Delta_\tau^*-\widehat\Delta_\tau)$ and $\mathcal{B}(\tau)$ as two processes indexed by $\tau \in \Upsilon$ and denote them as $G_n$ and $G$, respectively. Then, following \citet[Chapter 2.9]{VW96}, we say $G_n \underset{\xi}{\overset{\mathbb{P}}{\rightsquigarrow}} G$ uniformly over $\tau \in \Upsilon$ if
	$
		\sup_{h \in \text{BL}_1}|\mathbb{E}_\xi h(G_n) - \mathbb{E}h(G)| \convP 0,
	$
	where $\text{BL}_1$ is the set of all functions $h:L_\infty(\Upsilon) \mapsto [0,1]$ such that $|h(z_1)-h(z_2)| \leq |z_1-z_2|$ for every $z_1,z_2 \in L_\infty(\Upsilon)$, and $\mathbb{E}_\xi$ denotes expectation with respect to the bootstrap weights $\{\xi\}_{i \in [n]}$.}
\end{thm}

Theorem \ref{thm:multiplier-bootstrap-integrated} establishes the conditional weak convergence of the bootstrap process in \eqref{eq:boot_converge}. Together with Theorem \ref{thm:integratedM}, which gives the weak convergence of the original LTQTE estimation process, this implies that the multiplier bootstrap consistently approximates the sampling law of the LTQTE process. Pointwise and uniform inference are obtained by applying the bootstrap approximation to the studentized functionals described below.

\subsection{Pointwise Bootstrap Inference} \label{subsec:pointwise_boot}

Fix $\tau\in\Upsilon$. For $\nu\in(0,1)$, let $\widehat{\mathcal C}_\tau(\nu)$ denote the conditional $\nu$-th quantile of $\widehat\Delta_\tau^*$, defined by
\begin{align}
	\widehat{\mathcal C}_\tau(\nu)
	:=\inf\left\{z:\Pr\left(\widehat\Delta_\tau^*\leq z\mid\mathcal D_n\right)\geq\nu\right\}.
	\label{eq:conditional-quantile}
\end{align}
Here, $\mathcal D_n$ denotes the observed data, including the cross-fitting partition. 
	Let $\widehat\Delta_\tau^{*,1},\ldots,\widehat\Delta_\tau^{*,R}$ be the bootstrap LTQTE estimates from $R$ multiplier draws. In practice, $\widehat{\mathcal C}_\tau(\nu)$ can be approximated by the empirical $\nu$-th quantile of $\{\widehat\Delta_\tau^{*,r}\}_{r=1}^R$ for sufficiently large $R$.
Let $z_\nu$ denote the $\nu$-th quantile of the standard normal distribution.

We estimate the pointwise standard error by the normal calibrated interquantile range
\begin{align}
	\widehat\sigma_\tau=\frac{\widehat{\mathcal C}_\tau(0.975)-\widehat{\mathcal C}_\tau(0.025)}
	{z_{0.975}-z_{0.025}}.
	\label{eq:bootstrap_scale_pointwise}
\end{align}
If the bootstrap distribution is locally Gaussian, then $\widehat\sigma_\tau$ is asymptotically equivalent to the bootstrap standard deviation. The interquantile range estimator is less sensitive to extreme bootstrap draws than the conventional bootstrap standard deviation.

For testing the null hypothesis
\begin{align*}
	\mathcal H_{0,\tau}:\Delta_\tau=\Delta_\tau^0,
\end{align*}
define the pointwise studentized statistic
\begin{align}
	T_n(\tau)=\frac{\widehat\Delta_\tau-\Delta_\tau^0}{\widehat\sigma_\tau}.
	\label{eq:pointwise_t}
\end{align}
The null hypothesis is rejected at level $\alpha$ if
\begin{align*}
	|T_n(\tau)|>z_{1-\alpha/2}.
\end{align*}
Then, the pointwise $(1-\alpha)$ confidence interval is
\begin{align}
	CI_\tau(1-\alpha)=[\widehat\Delta_\tau+z_{\alpha/2}\widehat\sigma_\tau,~~
	\widehat\Delta_\tau+z_{1-\alpha/2}\widehat\sigma_\tau].
	\label{eq:pointwise_ci}
\end{align}

\begin{thm}[Pointwise size and coverage]\label{thm:pointwise_ci}
	Suppose that the conditions of Theorem \ref{thm:multiplier-bootstrap-integrated}
	hold and define $\widehat{\mathcal C}_\tau(\nu)$ and $\widehat\sigma_\tau$ by \eqref{eq:conditional-quantile} and \eqref{eq:bootstrap_scale_pointwise}. Impose the uniform nondegeneracy condition
	\begin{align}
		0<\inf_{\tau\in\Upsilon}\Sigma_\Delta(\tau,\tau)\leq \sup_{\tau\in\Upsilon}\Sigma_\Delta(\tau,\tau)<\infty.
		\label{eq:nondegeneracy}
	\end{align} 
	Then, it follows that
	
	\begin{align}
		\sup_{\tau\in\Upsilon}\left|\sqrt n\,\widehat\sigma_\tau\Sigma_\Delta(\tau,\tau)^{-1/2}-1\right|\rightarrow_{p}0.
		\label{eq:uniform-scale-consistency}
	\end{align}
		
	Consequently, for every fixed $\tau\in\Upsilon$, under $\mathcal H_{0,\tau}:\Delta_\tau=\Delta_\tau^0$,
	\begin{align}
		T_n(\tau)
		=\frac{\widehat\Delta_\tau-\Delta_\tau^0}
		{\widehat\sigma_\tau}
		\rightsquigarrow \mathcal N(0,1).
		\label{eq:pointwise-normality}
	\end{align}
	It follows that, for every $\alpha\in(0,1)$,
	\begin{align}
		\Pr\left\{|T_n(\tau)|>z_{1-\alpha/2}\right\}
		&\rightarrow \alpha,
		\label{eq:pointwise-size}\\
		\Pr\left\{\Delta_\tau\in
		[\widehat\Delta_\tau+z_{\alpha/2}\widehat\sigma_\tau,
		\widehat\Delta_\tau+z_{1-\alpha/2}\widehat\sigma_\tau]
		\right\}
		&\rightarrow 1-\alpha.
		\label{eq:pointwise-coverage}
	\end{align}
	Hence, the pointwise test has asymptotic size $\alpha$, and the pointwise confidence interval has asymptotic coverage $1-\alpha$.
\end{thm}

\subsection{Uniform Bootstrap Inference} \label{subsec:uniform_boot}

We next consider inference on the entire LTQTE curve over $\Upsilon$. The null hypothesis is
\begin{align}
	H_0: \Delta_\tau=\Delta_\tau^0\quad \text{for all } \tau\in\Upsilon.
\end{align}
For each $\tau\in \Upsilon$, define $\widehat\sigma_\tau$ as in \eqref{eq:bootstrap_scale_pointwise}. Let $\widetilde\Delta_\tau^*=\widehat{\mathcal C}_\tau(0.5)$ be the conditional median of $\widehat\Delta_\tau^*$ given $\mathcal D_n$. This centering is first-order equivalent to centering at $\widehat\Delta_\tau$, but may improve finite sample performance. This is the centering recommended by \citet[Appendix B.2]{jiangetal2023}.

Define the standardized bootstrap sup-statistic
\begin{align*}
	T^*=\sup_{\tau\in\Upsilon}\left|\frac{\widehat\Delta_\tau^*-\widetilde\Delta_\tau^*}{\widehat\sigma_\tau}\right|.
\end{align*}
Let $\widehat c_{1-\alpha}$ be the conditional $(1-\alpha)$-quantile of $T^*$ given $\mathcal D_n$, defined by
\begin{align}
	\widehat c_{1-\alpha}:=\inf\left\{z:\Pr\left(T^*\le z\mid\mathcal D_n\right)\ge 1-\alpha\right\}.
	\label{eq:uniform_critical}
\end{align}
In practical implementation, we approximate the continuum $\Upsilon$ by a fine grid
	\begin{align*}
		\mathcal G=\{\tau_1,...,\tau_G\}\subset\Upsilon.
	\end{align*}
	The conditional median $\widetilde\Delta_\tau^*$ and standard error $\widehat\sigma_\tau$ can be approximated using the empirical quantiles from $R$ multiplier draws. For each bootstrap replication $r$, we approximate $T^*$ by
\begin{align*}
	T_r^*=\sup_{\tau\in\mathcal G}\left|\frac{\widehat\Delta_\tau^{*,r}-\widetilde\Delta_\tau^*}{\widehat\sigma_\tau}\right|,
\end{align*}
where the conditional median and standard error are replaced by their empirical counterparts. The critical value $\widehat c_{1-\alpha}$ can then be approximated by the empirical $(1-\alpha)$-th quantile of $\{T_r^*\}_{r=1}^R$ for sufficiently large $R$.

The uniform statistic for testing $\mathcal H_0$ is
\begin{align}
	T_n=\sup_{\tau\in\Upsilon}\left|\frac{\widehat\Delta_\tau-\Delta_\tau^0} {\widehat\sigma_\tau}\right|.
	\label{eq:uniform_t}
\end{align}
The null hypothesis is rejected at level $\alpha$ if
\begin{align*}
	T_n>\widehat c_{1-\alpha}.
\end{align*}
The preceding bootstrap validity result implies that this test has the asymptotic size $\alpha$ uniformly over the quantile region $\Upsilon$. The associated $(1-\alpha)$ uniform confidence band is
\begin{align}
	\mathrm{CB}(1-\alpha)=\left\{\left[\widehat\Delta_\tau-\widehat c_{1-\alpha}\widehat\sigma_\tau,~~ \widehat\Delta_\tau+\widehat c_{1-\alpha}\widehat\sigma_\tau\right]: \tau\in\Upsilon\right\}.
	\label{eq:uniform_confidence_band}
\end{align}
Equivalently, the null curve $\Delta^0_\tau$ is rejected if it is not contained in the confidence band in \eqref{eq:uniform_confidence_band}.

\begin{thm}[Uniform size and coverage]\label{thm:uniform}
	Suppose that the conditions of Theorem \ref{thm:multiplier-bootstrap-integrated} hold and define $\widehat\sigma_\tau$ by \eqref{eq:bootstrap_scale_pointwise}. Impose the uniform nondegeneracy condition in \eqref{eq:nondegeneracy}. For any fixed $\alpha\in(0,1)$, under the uniform null hypothesis $\mathcal H_0:\Delta_\tau=\Delta_\tau^0$ for all $\tau\in\Upsilon$, the test statistic $T_n$ and the bootstrap critical value $\widehat c_{1-\alpha}$ defined in \eqref{eq:uniform_t} and \eqref{eq:uniform_critical} satisfy
	\begin{align}
		\Pr(T_n>\widehat c_{1-\alpha})\rightarrow\alpha.
		\label{eq:uniform-size}
	\end{align}
	In addition, the uniform confidence band
	\begin{align*}
		\mathrm{CB}(1-\alpha)=\left\{[\widehat\Delta_\tau-\widehat c_{1-\alpha}\cdot\widehat\sigma_\tau,~~\widehat\Delta_\tau+\widehat c_{1-\alpha}\cdot\widehat\sigma_\tau]:\tau\in\Upsilon\right\}
	\end{align*}
	satisfies
	\begin{align}
		\Pr\left\{ \{\Delta_\tau: \tau \in \Upsilon\} \in \mathrm{CB}(1-\alpha) \right\} 	= \Pr\left\{\sup_{\tau \in \Upsilon}\left|\frac{\widehat\Delta_\tau-\Delta_\tau}{\widehat\sigma_\tau}\right|\leq\widehat c_{1-\alpha}\right\}\rightarrow1-\alpha.
		\label{eq:uniform-coverage}
	\end{align}
\end{thm}

\section{Auxiliary Regressions} \label{sec:np-first-stage-uniform}

Our estimator introduced in Section \ref{sec:estimation} accommodates a general class of first-step estimators, subject to the high-level conditions stated in Assumption~\ref{as:first-stage-rate}.
This section provides the high-dimensional first-step estimators and primitive conditions under which Assumption~\ref{as:first-stage-rate} holds, with particular attention to convergence rates that are uniform over $q\in\mathcal Q_w$. Following \citet[Assumptions~6.1--6.2 and Theorems~6.1--6.2]{belloni2017program}, we estimate the conditional-mean functions by least-squares Lasso and the propensity-score-type functions by logistic Lasso.
 
Let
\begin{align*}
	H_{p_n}(z)=(h_{1n}(z),...,h_{p_n n}(z))^\prime
\end{align*}
be a $p_n$-dimensional dictionary of regressors, where $p_n$ may grow with $n$ and may exceed $n$. The dictionary may include original covariates, powers, interactions and other technical regressors. The argument $z$ equals $(s,x)$ for the functions depending on $(S,X)$ and equals $x$ for functions depending only on $X$. For notational simplicity, we use the same dimension $p_n$ for all nuisance functions; allowing nuisance-function-specific dictionaries is straightforward.

For $w=0,1$ and $q\in\mathcal Q_w$, recall that
\begin{align*}
	m_w(s,x;q)=&\Pr(Y\leq q\vert~ W=w,S=s,X=x,G=1)
	\\
	=&\mathbb E[\mathbf 1\{Y\leq q\}\vert~ W=w,S=s,X=x,G=1].
\end{align*}
We estimate this $m_w$ by linear Lasso. Specifically, define the preliminary linear estimator:
\begin{align*}
	\widetilde m_w^{L,(-k)}(s,x;q)=H_{p_n}(s,x)^\prime\widehat\beta_w^{m,(-k)}(q),
\end{align*}
where
\begin{align*}
	\widehat\beta_w^{m,(-k)}(q)=\arg\min_{\beta\in\mathbb R^{p_n}}
	\frac{1}{n_{1w}^{(-k)}}\sum_{i\in\mathcal I_k^c:G_i=1,W_i=w}\left[\mathbf 1\left\{Y_i\leq q\right\}-H_{p_n}(S_i,X_i)^\prime\beta
	\right]^2+\frac{\ell_{n,w}^{m,(-k)}}{n_{1w}^{(-k)}}\left\Vert\widehat\Omega_w^{m,(-k)}\beta\right\Vert_1,
\end{align*}
and
\begin{align*}
	n_{1w}^{(-k)}=\sum_{i\in\mathcal I_k^c}\mathbf 1\left\{G_i=1,W_i=w\right\}.
\end{align*}

Since $m_w$ is a conditional expectation of a binary variable and is bounded between zero and one, we use the truncated estimator
\begin{align*}
	\widehat m_w^{(-k)}(s,x;q)=\Pi_{[0,1]}[\widetilde m_w^{L,(-k)}(s,x;q)],
\end{align*}
where
\begin{align*}
	\Pi_{[0,1]}(u)=\min\{\max\{u,0\},1\}.
\end{align*}
The truncation operator is nonexpansive and therefore does not worsen the $L_2$ convergence rate.

Next, recall that
\begin{align*}
	\overline m_w(x;q)=\mathbb E\left[m_w(S,X;q)\vert W=w,X=x,G=0\right].
\end{align*}
We estimate this $\overline m_w$ by a second-step linear Lasso regression. Specifically, define
\begin{align*}
	\widetilde{\overline m}_w^{L,(-k)}(x;q)=H_{p_n}(x)^\prime\widehat\beta_w^{\overline m,(-k)}(q),
\end{align*}
where
\begin{align*}
	\widehat\beta_w^{\overline m,(-k)}(q)=\arg\min_{\beta\in\mathbb R^{p_n}}\frac{1}{n_{0w}^{(-k)}}
	\sum_{i\in\mathcal I_k^c:G_i=0,W_i=w}(\widehat m_w^{(-k)}(S_i,X_i;q)-H_{p_n}(X_i)^\prime\beta)^2 +\frac{\ell_{n,w}^{\overline m,(-k)}}{n_{0w}^{(-k)}}\Vert\widehat\Omega_w^{\overline m,(-k)}\beta\Vert_1
\end{align*}
and
\begin{align*}
	n_{0w}^{(-k)}=\sum_{i\in\mathcal I_k^c}\mathbf 1\{G_i=0,W_i=w\}.
\end{align*}
Again, because $\overline m_w$ is bounded between zero and one, we define
\begin{align*}
	\widehat{\overline m}_w^{(-k)}(x;q)=\Pi_{[0,1]}[\widetilde{\overline m}_w^{L,(-k)}(x;q)].
\end{align*}

The propensity-score-type functions are estimated by the logistic Lasso. Let
$\lambda(u)=\exp(u)/\{1+\exp(u)\}$ denote the logistic CDF. For example,
the initial logistic-Lasso estimate of $\gamma$ is
\begin{align*}
	\widetilde\gamma^{(-k)}(x)
	&=\lambda\big(H_{p_n}(x)'\widehat\theta^{\gamma,(-k)}\big),
\end{align*}
where
\begin{align*}
	\widehat\theta^{\gamma,(-k)}
	=\arg\min_{\theta\in\mathbb R^{p_n}}
	\Bigg\{&-\frac1{n^{(-k)}}\sum_{i\in\mathcal I_k^c}
	\Big[G_i\log\lambda\big(H_{p_n}(X_i)'\theta\big)
	+(1-G_i)\log\big\{1-\lambda(H_{p_n}(X_i)'\theta)\big\}\Big]\\
	&+\frac{\ell_n^{\gamma,(-k)}}{n^{(-k)}}
	\big\|\widehat\Omega^{\gamma,(-k)}\theta\big\|_1
	\Bigg\}.
\end{align*}
Similarly, $\varrho(x)=\Pr(W=1\vert X=x,G=0)$ is initially estimated by the logistic Lasso of $W_i$ on $H_{p_n}(X_i)$ using data in $\{i\in\mathcal I_k^c:G_i=0\}$; denote the resulting fitted
probability by $\widetilde\varrho^{(-k)}(x)$.

For $w\in\{0,1\}$, the observed-data representation
\eqref{eq:rho-identification} involves
\begin{align*}
	e_w(s,x)&=\Pr(G=1\mid S=s,W=w,X=x),&
	h_w(x)&=\Pr(G=1\mid W=w,X=x),\\
	t(x)&=\Pr(W=1\mid X=x,G=1),&
	t_w(x)&=wt(x)+(1-w)\{1-t(x)\}.
\end{align*}
On each training sample $\mathcal I_k^c$, we estimate $e_w$, $h_w$ and $t$ by the logistic Lasso on their respective conditioning subsamples; write $\widetilde e_w^{(-k)}$, $\widetilde h_w^{(-k)}$ and
$\widetilde t^{(-k)}$ for the corresponding fitted probabilities.

To ensure that all the estimated inverse-probability weights are uniformly bounded, we use truncations. After reducing the constants if necessary, let $c\in(0,1/2)$ be a common lower bound compatible with Assumptions~\ref{as:overlap} and \ref{as:np-sieve-logit}(v), and fix
$c_*\in(0,c]$. For any real $z$, set
\begin{align*}
	\Pi_{[c_*,1-c_*]}(z)=\min\{1-c_*,\max\{c_*,z\}\}.
\end{align*}
We first truncate the probabilities entering the observed-data
representation:
\begin{align*}
	\widehat e_w^{(-k)}(s,x)
	&=\Pi_{[c_*,1-c_*]}\big[\widetilde e_w^{(-k)}(s,x)\big],&
	\widehat h_w^{(-k)}(x)
	&=\Pi_{[c_*,1-c_*]}\big[\widetilde h_w^{(-k)}(x)\big],\\
	\widehat t^{(-k)}(x)
	&=\Pi_{[c_*,1-c_*]}\big[\widetilde t^{(-k)}(x)\big],&
	\widehat t_w^{(-k)}(x)
	&=w\widehat t^{(-k)}(x)+(1-w)\{1-\widehat t^{(-k)}(x)\}.
\end{align*}
The propensity-score estimates used are then defined by
\begin{align*}
	\widehat\gamma^{(-k)}(x)
	&=\Pi_{[c_*,1-c_*]}\big[\widetilde\gamma^{(-k)}(x)\big],\\
	\widehat\varrho^{(-k)}(x)
	&=\Pi_{[c_*,1-c_*]}\big[\widetilde\varrho^{(-k)}(x)\big],\\
	\widehat\rho_w^{(-k)}(s,x)
	&=\Pi_{[c_*,1-c_*]}\left[
	\frac{\widehat e_w^{(-k)}(s,x)}{1-\widehat e_w^{(-k)}(s,x)}
	\frac{1-\widehat h_w^{(-k)}(x)}{\widehat h_w^{(-k)}(x)}
	\widehat t_w^{(-k)}(x)
	\right],~~ w\in\{0,1\}.
\end{align*}

The truncations preserve the first-stage convergence rates. Indeed, Assumption~\ref{as:overlap} directly bounds $e_w$ and, by iterated expectations, also bounds $h_w$ and $t$ away from zero and one; Assumption~\ref{as:np-sieve-logit}(v) imposes the corresponding bounds on $\gamma$, $\varrho$, and $\rho_w$. Thus, all these true probabilities belong to $[c_*,1-c_*]$. Hence, for any such true probability $p$ and its untruncated estimate $\widetilde p^{(-k)}$,
\begin{align*}
	\big|\Pi_{[c_*,1-c_*]}(\widetilde p^{(-k)})-p\big|
	&\le|\widetilde p^{(-k)}-p|.
\end{align*}
Moreover, for each $w\in\{0,1\}$ the mapping
$(e,h,t)\mapsto e(1-h)\{wt+(1-w)(1-t)\}/\{(1-e)h\}$ is Lipschitz on
$[c_*,1-c_*]^3$, so \eqref{eq:rho-identification} implies that
\begin{align*}
	|\widehat\rho_w^{(-k)}(s,x)-\rho_w(s,x)|\le C\big\{|\widehat e_w^{(-k)}(s,x)-e_w(s,x)|+|\widehat h_w^{(-k)}(x)-h_w(x)|+|\widehat t^{(-k)}(x)-t(x)|\big\},
\end{align*}
where $C<\infty$ relies only on $c_*$. Consequently, any common population $L_2$ rate established for these three logistic-Lasso components also holds for the resulting estimator of $\rho_w$.

By construction, uniformly over all arguments and $k\in[K]$,
\begin{align*}
	\widehat\rho_w^{(-k)}\ge c_*,~~c_*\le\widehat\varrho^{(-k)}\le1-c_*,~~1-\widehat\gamma^{(-k)}\ge c_*.
\end{align*}
Consequently, we obtain that
\begin{align*}
	\left|\frac1{\widehat\rho_w^{(-k)}}-\frac1{\rho_w}\right|
	&\le c_*^{-2}|\widehat\rho_w^{(-k)}-\rho_w|,\\
	\left|\frac1{\widehat\varrho^{(-k)}}-\frac1\varrho\right|
	+\left|\frac1{1-\widehat\varrho^{(-k)}}-\frac1{1-\varrho}\right|
	&\le2c_*^{-2}|\widehat\varrho^{(-k)}-\varrho|,\\
	\left|\frac{\widehat\gamma^{(-k)}}{1-\widehat\gamma^{(-k)}}
	-\frac\gamma{1-\gamma}\right|
	&\le c_*^{-2}|\widehat\gamma^{(-k)}-\gamma|.
\end{align*}

The following assumption adapts primitive Lasso conditions to our LTQTE setting. 

\begin{as}[Primitive Conditions for Lasso First-Step Estimation]
	\label{as:np-sieve-logit}
	Let $s_n$ denote the maximal sparsity among sparse approximations of nuisance functions. Let $L_n=\log(p_n\vee n)$ and define $r_n=\sqrt{s_nL_n/n}=\sqrt{s_n\log(p_n\vee n)/n}$.
	
	\bigskip\noindent
	\textnormal{(i)} For every relevant conditioning event $\mathcal C$ used in the first-step regressions, there exists $c_{\mathcal C}>0$ such that
	\begin{align*}
		\inf_{k\in[K]}
		\frac{n_{\mathcal C}^{(-k)}}{n}
		\geq
		c_{\mathcal C}+o_p(1),
	\end{align*}
	where $n_{\mathcal C}^{(-k)}=\sum_{i\in\mathcal I_k^c}\mathbf 1\{\mathcal C_i=1\}$.

	\bigskip\noindent
	\textnormal{(ii)} The dictionary satisfies
	\begin{align*}
		\sup_i\Vert H_{p_n}(Z_i)\Vert_{\infty}\leq\zeta_n
	\end{align*}
	with probability approaching one, and for some $\bar\nu\geq 3$ and the $j$th entry,
	\begin{align*}
		\sup_{j\in[p_n]}\mathbb E\left[\left|H_{p_n,j}(Z_i)\right|^{\bar\nu}
		\vert~ \mathcal C\right]<C<\infty,
	\end{align*}
	for every relevant conditioning event $\mathcal C$.
	
	\bigskip\noindent
	\textnormal{(iii)} The following approximation-error, residual and empirical regularity conditions hold.
	
	\begin{enumerate}[(a)]
		\item \textit{Uniform approximation errors.}
		The population approximation errors satisfy
		\begin{align*}
			&\max_{w\in\{0,1\}}\sup_{q\in\mathcal Q_w}
			\mathbb E\left[
			r_w^m(S,X;q)^2
			\mid G=1,W=w
			\right]+
			\max_{w\in\{0,1\}}\sup_{q\in\mathcal Q_w}
			\mathbb E\left[
			r_w^{\overline m}(X;q)^2
			\mid G=0,W=w
			\right]
			=O(r_n^2).
		\end{align*}
		Moreover, for some constant $C<\infty$, with probability
		approaching one,
		\begin{align*}
			&\max_{k\in[K],\,w\in\{0,1\}}\sup_{q\in\mathcal Q_w}\mathbb E_{n_{1w}^{(-k)}}\left[r_w^m(S,X;q)^2\right]+\max_{k\in[K],\,w\in\{0,1\}}\sup_{q\in\mathcal Q_w}\mathbb E_{n_{0w}^{(-k)}}\left[r_w^{\overline m}(X;q)^2\right]\leq Cr_n^2,
		\end{align*}
		where
		\begin{align*}
			n_{gw}^{(-k)}:=\sum_{i\in\mathcal I_k^c}\mathbf 1\{G_i=g,W_i=w\}~~\text{and}~~\mathbb E_{n_{gw}^{(-k)}}[f(B)]:=\frac{1}{n_{gw}^{(-k)}}\sum_{i\in\mathcal I_k^c}\mathbf 1\{G_i=g,W_i=w\}f(B_i),
		\end{align*}
		for $g,w\in\{0,1\}$ and whenever $n_{gw}^{(-k)}>0$.
		
		\item \textit{First-stage residual nondegeneracy.}
		For some constant $c_m\in(0,1/2)$,
		\begin{align*}
			c_m\leq m_w(S,X;q)\leq 1-c_m
		\end{align*}
		uniformly over $w\in\{0,1\}$ and $q\in\mathcal Q_w$, almost surely given $G=1,W=w$.
		
	\item \textit{First-stage empirical regularity.} For any $w\in\{0,1\}$, $\mathcal Q_w$ is a bounded subset of $\mathbb R$ with 
	\begin{align*}
		\log\mathcal N(\epsilon,\mathcal Q_w,|\cdot|)
		\lesssim\log(e/\epsilon),
		\qquad \epsilon\in(0,1].
	\end{align*}
	There exist deterministic sequences $\delta_n,\Delta_n\downarrow0$, decreasing no faster than polynomial rates, such that, with probability at least $1-\Delta_n$, the following conditions hold simultaneously for all $k\in[K]$ and $w\in\{0,1\}$:
	\begin{align*}
		&\sup_{q\in\mathcal Q_w}\max_{j\in[p_n]}
		\Big|
		\mathbb E_{n_{1w}^{(-k)}}\left[
		H_{p_n,j}(S,X)^2
		\{\mathbf 1\{Y\le q\}-m_w(S,X;q)\}^2
		\right]\\
		&-
		\mathbb E\left[
		H_{p_n,j}(S,X)^2
		\{\mathbf 1\{Y\le q\}-m_w(S,X;q)\}^2
		\mid G=1,W=w
		\right]
		\Big|\leq\delta_n,\\
		&\sup_{q\in\mathcal Q_w}\max_{j\in[p_n]}
		\Big|
		\mathbb E_{n_{1w}^{(-k)}}\left[
		H_{p_n,j}(S,X)^2\mathbf 1\{Y\le q\}
		\right]-
		\mathbb E\left[
		H_{p_n,j}(S,X)^2\mathbf 1\{Y\le q\}
		\mid G=1,W=w
		\right]
		\Big|\leq\delta_n,
	\end{align*}
	and
	\begin{align*}
		&\sup_{\substack{
				q,q'\in\mathcal Q_w\\
				|q-q'|\leq 1/n_{1w}^{(-k)}
		}}
		\max_{j\in[p_n]}
		\Bigg(
		\mathbb E_{n_{1w}^{(-k)}}\Big[
		H_{p_n,j}(S,X)^2\times
		\{\mathbf 1\{Y\le q\}-\mathbf 1\{Y\le q'\}
		-m_w(S,X;q)+m_w(S,X;q')\}^2
		\Big]
		\Bigg)^{1/2}\\
		&\leq\frac{\delta_n}{\sqrt{L_n}},\\
		&\sup_{\substack{
				q,q'\in\mathcal Q_w\\
				|q-q'|\leq 1/n_{1w}^{(-k)}
		}}
		\Bigg\Vert
		\mathbb E_{n_{1w}^{(-k)}}\Big[
		H_{p_n}(S,X)\times
		\{\mathbf 1\{Y\le q\}-\mathbf 1\{Y\le q'\}
		-m_w(S,X;q)+m_w(S,X;q')\}
		\Big]
		\Bigg\Vert_\infty
		\leq\frac{\delta_n}{\sqrt{n_{1w}^{(-k)}}}.
	\end{align*}
		
	\end{enumerate}

	\textnormal{(iv)} The propensity-score-type functions allow for sparse logistic approximations: There exist sparse vectors $\theta^\gamma$, $\theta^\varrho$ and $\theta_w^\rho$, $w=0,1$, such that
	\begin{align*}
		&\gamma(x)=\lambda\left(H_{p_n}(x)^\prime\theta^\gamma\right)+r^\gamma(x),~~\varrho(x)=\lambda\left(H_{p_n}(x)^\prime\theta^\varrho\right)+r^\varrho(x)
		~~\text{and}\\
		&\rho_w(s,x)=\lambda\left(H_{p_n}(s,x)^\prime\theta_w^\rho\right)+r_w^\rho(s,x),~~w=0,1,
	\end{align*}
	with
	\begin{align*}
		\|\theta^\gamma\|_0\leq s_n,~~\|\theta^\varrho\|_0\leq s_n~~\text{and}~~\max_{w\in\{0,1\}}\|\theta_w^\rho\|_0\leq s_n.
	\end{align*}
	The corresponding logistic approximation errors satisfy the $L_2$ bounds
	\begin{align*}
		&\mathbb E[(r^\gamma(X))^2]+\mathbb E[(r^\varrho(X))^2\mid G=0]+\max_{w\in\{0,1\}}\mathbb E[(r_w^\rho(S,X))^2\mid G=1]=O(r_n^2).
	\end{align*}
	Moreover, the approximation errors for the directly estimated logistic regressions satisfy
	\begin{align*}
		\|r^\gamma\|_{\mathbb P,\infty}+\|r^\varrho\|_{\mathbb P,\infty}\leq\delta_n.
	\end{align*}

	\bigskip\noindent
	\textnormal{(v)} The propensity-score-type functions are uniformly bounded away from zero and one. That is, for some constant $c\in(0,1/2)$,
	\begin{align*}
		&c\leq\inf_x\gamma(x)\leq\sup_x\gamma(x)\leq1-c,~~c\leq\inf_x\varrho(x)\leq\sup_x\varrho(x)\leq1-c,\\
		&c\leq\inf_{w\in\left\{0,1\right\}}\inf_{s,x}\rho_w(s,x)
		\leq\sup_{w\in\left\{0,1\right\}}\sup_{s,x}\rho_w(s,x)\leq1-c.
	\end{align*}
	The same bounds, after reducing $c$ if necessary, apply to $e_w,h_w,t$ in (ix). All fitted probabilities entering the weights are clipped to $[c_*,1-c_*]$ for a fixed $c_*\in(0,c]$.
	
	\bigskip\noindent
	\textnormal{(vi)} Sparse restricted eigenvalue conditions. Define $\ell_n=\log(n)$. There exist constants $0<\kappa_1<\kappa_2<\infty$ such that, with probability approaching one, for every relevant conditioning event $\mathcal C$ and every fold $k$ and any $Z_i$,
	\begin{align*}
		\kappa_1\leq&
		\inf_{\|v\|_0\leq s_n\ell_n}\frac{v^\prime\left[\frac{1}{n_{\mathcal C}^{(-k)}}\sum_{i\in\mathcal I_k^c:\mathcal C_i=1}H_{p_n}(Z_i)H_{p_n}(Z_i)^\prime\right]v}{\left\|v\right\|_2^2
}
\\
\leq&\sup_{\left\|v\right\|_0\leq s_n\ell_n}\frac{v^\prime\left[\frac{1}{n_{\mathcal C}^{(-k)}}\sum_{i\in\mathcal I_k^c:\mathcal C_i=1}
			H_{p_n}(Z_i)H_{p_n}(Z_i)^\prime\right]v}{\|v\|_2^2}
			\leq\kappa_2.
	\end{align*}
	The corresponding population restricted eigenvalue condition also holds:
	\begin{align*}
		\kappa_1\leq\inf_{\left\|v\right\|_0\leq s_n\ell_n}\frac{v^\prime\mathbb E[H_{p_n}(Z_i)H_{p_n}(Z_i)^\prime\vert~\mathcal C]v}{\|v\|_2^2}\leq
		\sup_{\left\|v\right\|_0\leq s_n\ell_n}\frac{v^\prime\mathbb E[H_{p_n}(Z_i)H_{p_n}(Z_i)^\prime\vert \mathcal C]v}{\|v\|_2^2}\leq\kappa_2.
	\end{align*}
	
	\bigskip\noindent
	\textnormal{(vii)} For each regression $\upsilon$ and fold $k\in[K]$, write $N=n_{\mathcal C}^{(-k)}$.
	
	\begin{enumerate}[(a)]
		
		\item For logistic Lasso and first-stage indicator-response Lasso, the penalty level is
		\begin{align*}
			\ell_{n,\upsilon}^{(-k)}=a_\upsilon c_\ell\sqrt N\,\Phi^{-1}\left(1-\frac{1}{2p_nN^{1+d_\upsilon}}\right),
		\end{align*}
		where $c_\ell>1$, $\Phi$ is the standard normal distribution function and $(a_\upsilon,d_\upsilon)=(1,0)$ for logistic Lasso and $(2,1)$ for first-stage indicator-response Lasso. 
		
		\item For second-stage predicted-outcome Lasso, the penalty level and loadings are
		\begin{align*}
			\ell_{n,\upsilon}^{(-k)}=4\sqrt N\left\{4+\sqrt{\frac{\log(p_nN)}{2}}\right\},~~\widehat\Omega_{\upsilon,jj}^{(-k)}=\left\{\mathbb E_N[H_{p_n,j}(X)^2]\right\}^{1/2},~~\text{with}~~ j\in[p_n].
		\end{align*}
		
		\item For some constants $0<c_\Omega<C_\Omega<\infty$,
		with probability approaching one,
		\begin{align*}
			c_\Omega\leq\min_{j\in[p_n]}\widehat\Omega_{\upsilon,jj}^{(-k)}\leq\max_{j\in[p_n]}\widehat\Omega_{\upsilon,jj}^{(-k)}\leq C_\Omega
		\end{align*}
		uniformly over regressions, folds and $q\in\mathcal Q_w$ whenever applicable.
		
		\item For each stage, the estimation procedure, including the penalty tuning and solution selection, is deterministic: the same design matrix and response vector always yield the same fitted function. The first-stage estimation uses only observations with $G_i=1$ and $W_i=w$ and depends on $q$ only through the response vector $(\mathbf 1\{Y_i\leq q\})_i$. The second-stage estimation uses observations with $G_i=0$ and $W_i=w$ and depends on $q$ only through the generated response vector $(\widehat m_w^{(-k)}(S_i,X_i;q))_i$.
		
	\end{enumerate}

	\bigskip\noindent
	\textnormal{(viii)} The following growth conditions hold:
	\begin{align*}
		\frac{L_n}{n^{1/3}}\rightarrow0,~~\frac{\zeta_n^2s_n^2L_n^2}{n}\rightarrow0~~\text{and}~~\frac{s_n^2L_n^2\log^2 n}{n}\rightarrow0.
	\end{align*}

	\bigskip\noindent
	\textnormal{(ix)} Let $\mathcal V_{\mathrm{logit}}=\{\gamma,\varrho,e_0,e_1,h_0,h_1,t\}$. For each $\upsilon\in\mathcal V_{\mathrm{logit}}$, suppose
	\begin{align*}
		p_\upsilon(Z)=\lambda(H_{p_n}(Z)^\prime\theta_\upsilon)+r_\upsilon(Z),
		~~\|\theta_\upsilon\|_0\leq s_n,~~\varepsilon_\upsilon=R_\upsilon-p_\upsilon(Z),
	\end{align*}
	where $R_\upsilon$, $Z$ and $\mathcal C_\upsilon$ denote the corresponding
	binary response, regressors and conditioning event, respectively.
	The approximation errors satisfy
	\begin{align*}
		\max_{\upsilon\in\mathcal V_{\mathrm{logit}}}\mathbb E[r_\upsilon(Z)^2\mid\mathcal C_\upsilon]=O(r_n^2),~~\max_{\upsilon\in\mathcal V_{\mathrm{logit}}}\|r_\upsilon\|_{\mathbb P,\infty}\leq\delta_n.
	\end{align*}
	Moreover, writing $N=n_{\mathcal C_\upsilon}^{(-k)}$ and letting
	$\mathbb E_N$ denote the corresponding subsample average, with probability
	at least $1-\Delta_n$, uniformly over $k\in[K]$ and
	$\upsilon\in\mathcal V_{\mathrm{logit}}$,
	\begin{align*}
		\mathbb E_N[r_\upsilon(Z)^2]\leq Cr_n^2,~~\max_{j\in[p_n]}\Big|\mathbb E_N[H_{p_n,j}(Z)^2\varepsilon_\upsilon^2]-\mathbb E[H_{p_n,j}(Z)^2\varepsilon_\upsilon^2\mid\mathcal C_\upsilon]\Big|\leq\delta_n,
	\end{align*}
	where $C<\infty$ is fixed and $\delta_n,\Delta_n$ are as in Condition (iii).
	Conditions (ii) and (vi)--(viii) apply to all these regressions.
\end{as}

Assumption~\ref{as:np-sieve-logit} provides primitive conditions under which first-step lasso estimators satisfy the required high-level rates. Its structure follows the functional-response Lasso theory of \citet[Assumptions~6.1--6.2 and Theorems~6.1--6.2]{belloni2017program}.  The conditional-mean components are controlled by sparse linear approximation and least-squares Lasso rates, while the propensity-score-type components are controlled by sparse logistic approximation and logistic-Lasso rates. The sparse restricted eigenvalue condition replaces the full Gram-matrix condition used in low-dimensional estimation. The growth restrictions ensure that the stochastic error, approximation error and model-selection error are negligible regarding the Neyman orthogonal moment.

The following theorem verifies Assumptions~\ref{as:first-stage-rate} and  \ref{as:uniform_score_complexity} under primitive conditions. 

\begin{thm}[Uniform Rates for Lasso First-Step Estimators]
	\label{thm:np-uniform-first-stage}
	Suppose that Assumptions~\ref{as:unconfoundedness}--\ref{as:iid}, \ref{as:outcome-lipschitz-q}
	and \ref{as:np-sieve-logit} hold. The Lasso estimators of the conditional-mean components and the logistic-Lasso estimators of the propensity-score-type components satisfy
	\begin{align*}
		\max_{1\leq k\leq K}\Vert\widehat\eta_{\rm prop}^{(-k)}-\eta_{\rm prop}\Vert_{\mathbb P,2}=
		O_p\left(\sqrt{\frac{s_n\log p_n}{n}}\right),
	\end{align*}
	and
\begin{align*}
	\max_{1\leq k\leq K}\max_{w\in\{0,1\}}\sup_{q\in\mathcal Q_w}
	[\Vert\widehat m_w^{(-k)}(\cdot,\cdot;q)-m_w(\cdot,\cdot;q)\Vert_{\mathbb P,2}+\Vert
	\widehat{\overline m}_w^{(-k)}(\cdot;q)-\overline m_w(\cdot;q)\Vert_{\mathbb P,2}]
	=O_p\left(
	\sqrt{\frac{s_n\log p_n}{n}}\right).
\end{align*}
	Consequently, Equations~\eqref{eq:dml-first-stage-rate-propensity} and \eqref{eq:dml-first-stage-rate-outcome} of Assumption~\ref{as:first-stage-rate} hold with
	\begin{align*}
		r_n
		=
		\sqrt{\frac{s_n\log p_n}{n}}.
	\end{align*}

In addition, denote the joint fitted path as 
	\begin{align}
		\mathcal P_{k,w}=\left\{\bigl(\widehat m_w^{(-k)}(\cdot;b),\widehat{\overline m}_w^{(-k)}(\cdot;b)\bigr): b\in Q_w\right\}
	\end{align}
Then, we have
	\begin{equation}
		\label{eq:lasso_path_second}
		\#\mathcal P_{k,w}\le n_{1w}^{(-k)}+1\le n+1.
	\end{equation}
	For each fixed $M>0$, write $\mathcal H_{k,w}=\mathcal H_{n,k,w}(M)$ for the class in Assumption~\ref{as:uniform_score_complexity}. There exist deterministic constants $F_M\ge 1$ and $C<\infty$ such that, with probability tending to one,
	\begin{align}
		\label{eq:lasso_conditional_entropy}
		&\max_{k,w}\sup_{h\in\mathcal H_{k,w}}\|h\|_{\mathbb P,\infty}\le F_M,\notag\\
		&\max_{k,w}\sup_Q
		\mathcal N(\varepsilon F_M,\mathcal H_{k,w},L_2(Q))
		\le C(n+1)\varepsilon^{-5},~~ 0<\varepsilon\le 1.
	\end{align}

\end{thm}

Theorem~\ref{thm:np-uniform-first-stage} adapts the uniform rates of \citet[Theorems~6.1--6.2]{belloni2017program} to our first-step estimators. In their scheme, the index corresponds to $q\in\mathcal Q_w$, the dictionary corresponds to $H_{p_n}$ and sparsity index corresponds to $s_n$. Assumption~\ref{as:np-sieve-logit} imposes their corresponding regularity conditions uniformly over the relevant conditioning subsamples and cross-fitting folds: \citet[Theorem~6.1]{belloni2017program} yields the uniform rates over $q$, while \citet[Theorem~6.2]{belloni2017program} yields the rates for logistic-Lasso regressions. Combining these results with the error-propagation bound for sieve approximation, the proposed Lasso procedure satisfies Assumption~\ref{as:first-stage-rate} with
\begin{align*}
	r_n=\sqrt{\frac{s_n\log(p_n\vee n)}{n}}.
\end{align*}

Theorem~\ref{thm:np-uniform-first-stage} further shows Assumption \ref{as:uniform_score_complexity} holds with $v_n \lesssim 1$ and $A_n = n$, which, combined with the rate $r_n$ derived above, implies that \eqref{eq:uc_growth} holds with $J_n=v_n\log(A_n/r_n) \lesssim \log (n)$ and 
\begin{align*}
\frac{s_n^2\log(p_n\vee n)^2}{n} \to 0.
\end{align*}



We further construct a preliminary quantile estimator directly from the non-Neyman-orthogonal IPW moments of \eqref{eq:ipw0} and \eqref{eq:ipw1}. This estimator is used only to obtain an initial estimator $\widetilde{q}_{w,\tau}^{(-k)}$. 

For $w\in\{0,1\}$, let $D_w=\mathbf 1\{W=w\}$. Given $\widehat{\zeta}_w =(\widehat{\pi},\widehat{\rho}_w)$, define the feasible IPW distribution estimator
\begin{align*}
	\widehat F_w(q)=\mathbb E_n
	\left[\frac{G_i D_w \mathbf 1\{Y_i\le q\}}{\widehat{\pi}\widehat{\rho}_w(S_i,X_i)}\right].
\end{align*}
The preliminary IPW quantile estimator is then defined by
\begin{align}
	\label{eq:initial-ipw-qw}
	\widetilde q_{w,\tau}\in\arg\min_{q\in\mathcal Q_w}\vert\widehat F_w(q)-\tau\vert.
\end{align}

\begin{thm}[Initial Rate of the IPW Quantile Estimator]
	\label{thm:initial-ipw-quantile-rate}
	Suppose that Assumptions~\ref{as:unconfoundedness}--\ref{as:iid}, \ref{as:outcome-lipschitz-q} and \ref{as:np-sieve-logit} hold. Then,
	\begin{align}
		\max_{w \in \{0,1\}}\sup_{\tau\in\Upsilon}\left|
		\widetilde{q}_{w,\tau}-q_{w,\tau}\right|=O_p(n^{-1/2}+r_n).
	\end{align}
	In particular, if $r_n\gtrsim n^{-1/2}$, then \eqref{eq:initial-q-rate-simple} of Assumption \ref{as:first-stage-rate} holds as
	\begin{align*}
		\max_{w \in \{0,1\}}\sup_{\tau\in\Upsilon}
		\left|\widetilde{q}_{w,\tau}-q_{w,\tau}\right|=O_p(r_n).
	\end{align*}
\end{thm}

The estimator $\widetilde q_{w,\tau}^{(-k)}$ is used only as a preliminary input. Although the IPW moment in \eqref{eq:initial-ipw-qw} is not Neyman-orthogonal regarding $\zeta_w=(\pi,\rho_w)$, Neyman orthogonality is unnecessary at this step. Under imposed conditions, $\widetilde q_{w,\tau}$ satisfies the rate $O_p(n^{-1/2}+r_n)$, and if $r_n\gtrsim n^{-1/2}$, this becomes $O_p(r_n)$, as required by Assumption~\ref{as:first-stage-rate}.

\section{Simulation Studies}\label{sec:simulation}

In this section, we examine the finite-sample performance of the proposed estimation and inference methods through simulation studies. Our design follows the data-generating framework of \citet{chen2023semiparametric}, which was calibrated to the randomized poverty-alleviation evaluation studied by \citet{banerjee2015multifaceted}. Our main departure is that we focus on LTQTEs rather than long-term average treatment effects and evaluate both pointwise and uniform inference for the LTQTE curve.

\subsection{Data Generating Process}

Following \citet{chen2023semiparametric}, we generate an i.i.d. sample, $B_i=(G_iY_i,W_i,G_i,S_i,X_i), i=1,\dots,n$, in each Monte Carlo replication. We use 20 pre-treatment measurements as covariates $X_i$, two-year post-treatment measurements as short-term outcomes $S_i$, and a three-year post-treatment measurement of total asset value as the long-term outcome $Y_i$. We also adopt the same basic experimental-versus-observational structure: treatment $W_i$ is randomized in the experimental sample ($G_i=0$) and potentially confounded in the observational sample ($G_i=1$). 

Data generation is based on the latent-unconfounded-treatment WGAN design of \citet{chen2023semiparametric}. We use a pre-generated synthetic population containing the baseline covariates and both potential short- and long-term outcomes, 	$\left\{X_i,S_i(0),S_i(1),Y_i(0),Y_i(1)\right\}$.
The WGAN construction approximates the corresponding population distributions according to
	\begin{align*}
		X_i&\sim\widehat F_X,\\
		S_i(w)&\sim\widehat F_{S\mid X,W}\left(\cdot\mid X_i,w\right),\qquad w\in\{0,1\}, \qquad\text{and}\\
		Y_i(w)&\sim\widehat F_{Y\mid S,X,W}\left(\cdot\mid S_i(w),X_i,w\right),\qquad w\in\{0,1\}.
	\end{align*}
The calibration uses 20 baseline covariates, two-year post-treatment as the short-term outcomes, and three-year follow-up as the long-term outcome.

For each simulated unit, we first draw the data-type indicator
\begin{align*}
	G_i \sim \mathrm{Bernoulli}(1/2),
\end{align*}
so that the experimental and observational samples have equal expected size. This implies that the unconditional sampling probability satisfies $\pi=\Pr(G=1)=1/2$.

Treatment assignment differs across the two samples. In the experimental sample, treatment is randomly assigned:
\begin{align*}
	W_i \mid G_i=0 \sim \mathrm{Bernoulli}(1/2).
\end{align*}
Accordingly, under this design the experimental-sample propensity scores satisfy
\begin{align*}
	\varrho(x) = \Pr(W=1 \mid X=x, G=0)=1/2.
\end{align*}

In the observational sample, by contrast, treatment assignment is allowed to be confounded. The observational treatment assignment depends on heterogeneous short-term treatment effects. Let
\begin{align*}
	D_{ic}&=S_{ic}(1)-S_{ic}(0) \qquad\text{and}\\
	D_{ia}&=S_{ia}(1)-S_{ia}(0),
\end{align*}
where $c$ denotes two-year total consumption and $a$ denotes the two-year asset index. Let $\widetilde D_{ic}$ and $\widetilde D_{ia}$ denote the corresponding standardized treatment effects. The observational propensity score is then generated as
\begin{align*}
	p_i^{O}=\Lambda\left\{\phi\left(\widetilde D_{ic}+\widetilde D_{ia}\right)\right\},
\end{align*}
where $\Lambda(t)=\frac{1}{1+e^{-t}}$ denotes the logistic function. We then draw
\begin{align*}
	W_i\mid G_i=1,X_i,S_i(0),S_i(1)\sim \mathrm{Bernoulli}(p_i^{O}).
\end{align*}
We set $\phi=0.5$ in the baseline design. Thus, unlike the experimental propensity score, the observational treatment probability varies with latent individual-level treatment heterogeneity.

Finally, the observed short-term and long-term outcomes are constructed as
\begin{align*}
	S_i&=W_iS_i(1)+(1-W_i)S_i(0) \qquad\text{and}\\
	Y_i&=W_iY_i(1)+(1-W_i)Y_i(0).
\end{align*}
For each quantile index $\tau$, the true LTQTE is
\begin{align*}
	\Delta_\tau=q_{1,\tau}-q_{0,\tau},
\end{align*}
and the population values of $q_{1,\tau}$, $q_{0,\tau}$, and $\Delta_\tau$ are computed directly from the large WGAN-generated population.

\subsection{Estimators}\label{sec:sim_est}

Our primary object of interest is the LTQTE:
\begin{align*}
	\Delta_\tau := q_{1,\tau}-q_{0,\tau}.
\end{align*}
We estimate $q_{1,\tau}$ and $q_{0,\tau}$ using the integrated M-estimation procedure introduced in Section \ref{sec:estimation}. This estimator combines the experimental and observational samples through the orthogonal moments derived under latent unconfoundedness. For nuisance function estimation, we use five-fold cross-validated Lasso. Estimated propensity scores are trimmed to $[0.05,0.95]$.

We consider cross-fitted estimators with $K=2$ and $K=3$ folds. In each case, the folds are constructed separately within treatment groups to maintain treatment balance. For a given held-out fold, both the initial quantile estimates and the nuisance functions are constructed using observations outside that fold, so that the final moment is evaluated with out-of-fold nuisance function estimates. This design follows the DML principle of separating nuisance function estimation from the final estimating equation. For comparison, we also report a full-sample implementation without sample splitting.

As a benchmark, we also report conventional QTE estimates based solely on the observational sample, using $X$ as the set of control variables,
\begin{align*}
	\Delta_{G=1}(\tau):=q_{1,\tau}^{(G=1)}-q_{0,\tau}^{(G=1)}.
\end{align*}
This estimator uses only observations with \(G=1\), and therefore does not exploit the identifying information contained in the short-term experimental outcomes. Moreover, because treatment assignment in the observational sample depends on latent heterogeneous short-term treatment effects, conditioning on baseline covariates alone does not generally eliminate the resulting selection. The comparison between $\Delta_\tau$ and $\Delta_{G=1}(\tau)$, therefore, illustrates the contribution of combining experimental and observational data.

\subsection{Pointwise Inference Output}

For each replication, we generate a sample of size $n=500$. We evaluate pointwise inference at the quantile indices
\begin{align*}
	\tau\in\{0.1,0.3,0.5,0.7,0.9\}.
\end{align*}
The nominal coverage probability is set to $95\%$ throughout, and the Monte Carlo experiment is repeated $1,000$ times, with $1,000$ multiplier draws per sample. For each replication, estimator, nuisance function learner, and quantile index, we compute the studentized pointwise statistic using the bootstrap scale estimator defined in \eqref{eq:pointwise_t}. The pointwise coverage probability is computed as the fraction of Monte Carlo replications for which the absolute value of the statistic, $T_n(\tau)$, is no larger than the 95\% standard normal critical value $1.96$.


\begin{table}[htbp]
	\centering
	\caption{Pointwise Coverage Probabilities}
	\label{tab:size}
	\begin{tabular}{lcccccc}
		\hline
		& \multicolumn{3}{c}{Combined Data} & \multicolumn{3}{c}{Observational Only} \\
		\cline{2-4} \cline{5-7}
		$\tau$ & Full & CF \(K=2\) & CF \(K=3\) & Full & CF \(K=2\) & CF \(K=3\) \\
		\hline
	0.1 & 0.953 & 0.956 & 0.955 & 0.852 & 0.854 & 0.854 \\
	0.3 & 0.934 & 0.944 & 0.950 & 0.888 & 0.898 & 0.903 \\
	0.5 & 0.926 & 0.943 & 0.949 & 0.887 & 0.897 & 0.892 \\
	0.7 & 0.938 & 0.951 & 0.957 & 0.873 & 0.876 & 0.869 \\
	0.9 & 0.952 & 0.937 & 0.941 & 0.804 & 0.806 & 0.793 \\
		\hline
	\end{tabular}
	\begin{minipage}{0.95\textwidth}
		\footnotesize 	\vspace{0.5em}
		\textit{Notes:}
		The table reports pointwise coverage probabilities of nominal \(95\%\) confidence intervals using Lasso for nuisance function estimation. The columns labeled ``Full,'' ``CF \(K=2\),'' and ``CF \(K=3\)'' correspond, respectively, to the full-sample implementation and the localized cross-fitting implementations with two and three folds. ``Observational Only'' denotes the benchmark QTE estimator based solely on the observational sample, while ``Combined Data'' denotes the proposed LTQTE estimator that combines the experimental and observational samples. Results are reported for the quantile indices \(\tau\in\{0.1,0.3,0.5,0.7,0.9\}\). The number of Monte Carlo replications is \(1,000\).
	\end{minipage}
\end{table}

Table \ref{tab:size} reports the pointwise coverage probabilities of nominal \(95\%\) confidence intervals using Lasso nuisance function estimation. The results show two main patterns. First, the benchmark estimator based only on the observational sample exhibits substantial undercoverage across all quantile indices and implementations. This coverage distortion is consistent with the simulation design, where treatment assignment in the observational sample is affected by latent selection and adjustment based only on observed covariates is insufficient.

Second, our proposed LTQTE estimator based on the combined experimental and observational data delivers much better coverage. For $\tau\in\{0.1,0.3,0.5,0.7\}$, the coverage probabilities across the full-sample and cross-fitted implementations are close to the nominal \(95\%\) level. The results indicate that incorporating the experimental sample through the proposed LTQTE estimator substantially improves finite-sample coverage relative to relying on the observational sample alone.

\subsection{Uniform Inference Output}

We next evaluate the finite-sample performance of the uniform inference procedure. The uniform inference grid is
\begin{align*}
	\mathcal G=\{0.10,0.11,\ldots,0.90\},
\end{align*}
which contains 81 quantile indices. We set the sample size to $n=500$ and conduct $1,000$ Monte Carlo replications, with $1,000$ bootstrap replications for each Monte Carlo sample. We use Lasso for nuisance function estimation and report results for the full-sample implementation and the cross-fitted implementations with $K=2$ and $K=3$ folds.

For each Monte Carlo replication, we construct the proposed uniform confidence band using the multiplier-bootstrap critical value $\widehat c_{0.95}$ defined in \eqref{eq:uniform_critical}. Uniform coverage is defined as the event that the confidence band contains the true population LTQTE $\Delta_\tau$ simultaneously for every $\tau\in\mathcal G$. For comparison, we also report simultaneous coverage obtained by using the normal critical value $1.96$ at every quantile. This comparison illustrates the importance of using a uniform critical value when inference is conducted over the entire quantile process.

\begin{table}[htbp]
	\centering
	\caption{Uniform Coverage Probabilities}
	\label{tab:uniform_size}
	\begin{tabular}{lcccc}
		\hline
		& \multicolumn{2}{c}{Combined Data} & \multicolumn{2}{c}{Observational Data Only} \\
		\cline{2-3} \cline{4-5}
		Method & $1.96$ cv & Bootstrap cv & $1.96$ cv & Bootstrap cv \\
		\hline
		Full sample & 0.490 & 0.957 & 0.100 & 0.494 \\
		Cross-fitting $K=2$ & 0.516 & 0.963 & 0.102 & 0.528 \\
		Cross-fitting $K=3$ & 0.523 & 0.976 & 0.100 & 0.509 \\
		\hline
	\end{tabular}
	\begin{minipage}{0.95\textwidth}
		\footnotesize 	\vspace{0.5em}
		\textit{Notes:}
		The table reports simultaneous coverage probabilities over the uniform grid $\mathcal G=\{0.10,0.11,\ldots,0.90\}$. The columns labeled ``$1.96$ cv'' report the probability that the confidence intervals constructed using the pointwise normal critical value $1.96$ cover the true LTQTE simultaneously at all 81 quantile indices. The columns labeled ``Bootstrap cv'' report the simultaneous coverage probability of the confidence band constructed using the multiplier-bootstrap critical value. ``Combined Data'' denotes the proposed LTQTE estimator that combines the experimental and observational samples, while ``Observational Data Only'' denotes the benchmark estimator based solely on the observational sample. All results use Lasso for nuisance function estimation. The number of Monte Carlo replications is 1,000, with 1,000 bootstrap replications for each Monte Carlo sample.
	\end{minipage}
\end{table}

Table \ref{tab:uniform_size} shows that the bootstrap uniform confidence band performs well for the combined-data estimator. The simultaneous coverage probabilities based on the bootstrap critical value are slightly above the nominal \(95\%\) level for the full-sample, $K=2$ cross-fitted, and $K=3$ cross-fitted implementations, indicating that the bootstrap uniform bands are somewhat conservative in this design. By contrast, using the pointwise critical value $1.96$ leads to substantial undercoverage when simultaneous coverage over the full quantile grid is required. This comparison highlights the distinction between pointwise and uniform inference. Although $1.96$ is appropriate for a single fixed quantile under a standard normal approximation, it does not account for requiring coverage at all 81 quantile indices. The multiplier-bootstrap critical value corrects for this multiplicity by approximating the distribution of the supremum statistic. 

The results for the observational-only estimator are considerably worse. This pattern is consistent with the observational treatment-assignment mechanism in the DGP: the observational-only estimator does not account for latent selection through the short-term potential outcomes, and the resulting estimation bias cannot be corrected simply by using a larger uniform critical value.

\begin{figure}[tbp]
	\centering
	\includegraphics[width=0.9\textwidth]{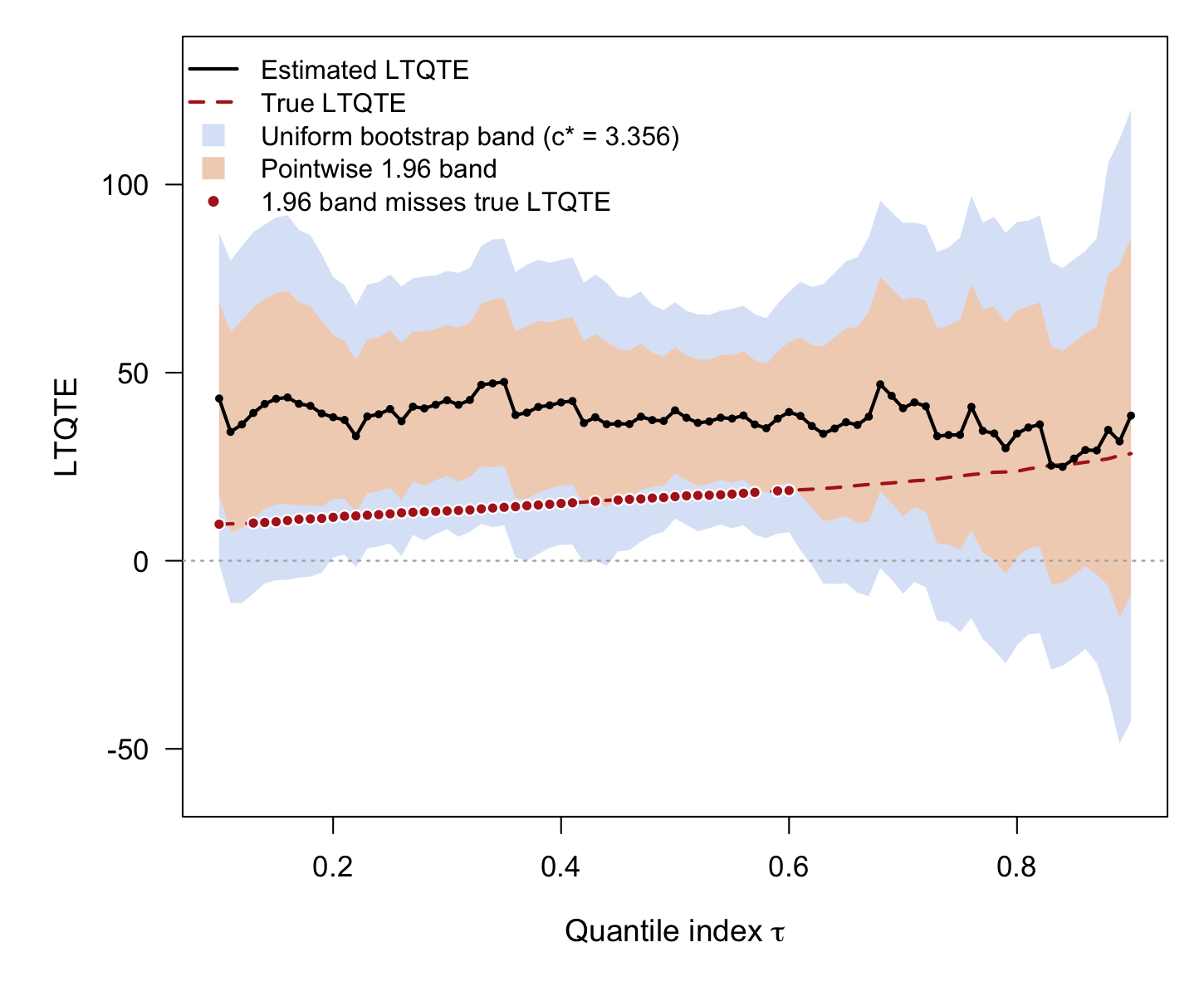}
	\caption{Estimated LTQTE Curve and Confidence Bands}
	\label{fig:ltqte_cb_k2}
	\begin{minipage}{0.95\textwidth}
		\footnotesize 	\vspace{0.5em}
		\textit{Notes:}
		The figure reports one representative replication for the cross-fitted LTQTE estimator with $K=2$ folds. The solid black curve represents the estimated LTQTE curve, while the dashed red line represents the true LTQTE curve computed from the generated population. The orange shaded region represents the pointwise $1.96$ confidence band, and the blue shaded region represents the uniform bootstrap confidence band. The uniform confidence band is constructed over the grid $\mathcal G=\{0.10,0.11,\ldots,0.90\}$. All nuisance functions are estimated using five-fold cross-validated Lasso. The number of bootstrap replications is 1,000.
	\end{minipage}
\end{figure}

We also report graphical evidence from a representative replication. Figure \ref{fig:ltqte_cb_k2} plots the estimated LTQTE curve, the true LTQTE curve, the pointwise $1.96$ confidence band, and the uniform bootstrap confidence band for the cross-fitted estimator with $K=2$ folds. In this typical replication, the pointwise $1.96$ band is narrower and fails to cover the true curve at several quantile indices, as indicated by the red points. By contrast, the uniform bootstrap band is wider and contains the true curve over the full quantile grid. The pointwise band is narrower because it controls coverage separately at each quantile index. The uniform bootstrap band is wider, reflecting simultaneous coverage over the full grid $\mathcal G$.

\section{Conclusion}\label{sec:conclusion}

This paper develops methods for identification, estimation, and inference on long-term quantile treatment effects (LTQTEs) when short-term experimental data are combined with long-term observational data. The setting is motivated by empirical environments in which random assignment is available only for short-term outcomes, while the long-term outcomes of interest are observed in a separate observational sample. We study this problem under the latent unconfoundedness framework, in which short-term latent outcomes signal treatment selection in the observational sample and its relationship with long-term potential outcomes.

We establish identification of the LTQTE curve and derive the associated efficient influence functions. Building on these influence functions, we develop a semiparametrically efficient estimator that accommodates flexible machine learning methods for nuisance function estimation while preserving first-order validity through Neyman orthogonality and cross-fitting. We further develop a multiplier bootstrap procedure for pointwise and uniform inference on the LTQTE curve. Simulation evidence demonstrates the importance of combining experimental and observational data for addressing confounding in the observational sample and shows that the multiplier-bootstrap critical value provides reliable uniform inference over the quantile curve.

	\renewcommand\bibname{\LARGE \textbf {Bibliography}}
	\bibliographystyle{chicago}
	\bibliography{IF}	

\newpage

\appendix
\section*{Appendices}

Throughout the following proofs we use the same notation as in the paper. The technical lemmas listed in the next section play central roles in the proofs of the main theorems.

\section{Proofs}\label{sec:proofs}

\subsection{Proofs for Section \ref{sec:identification}}

\noindent \textbf{Proof of Equations \eqref{eq:ipw0} and \eqref{eq:ipw1}}: We follow the idea of \citet*{athey2020combining} and \citet{athey2025surrogate_index_restud} and give the full derivation for the case of $w=1$; the case of $w=0$ is identical once $(1-W)$ and $\rho_0$ replace both $W$ and $\rho_1$. For simplicity, write $q^{\ast}_{w,\tau}$ as $q_{w,\tau}$. Recall that $\pi=\Pr(G=1)$ and
\begin{align*}
	\rho_1(s,x)=\Pr(W=1\vert S(1)=s,X=x,G=1).
\end{align*}
Under Assumptions \ref{as:unconfoundedness}--\ref{as:LUTM}, Bayes' rule also gives the representation
\begin{align*}
	\rho_w(s,x)=\frac{\Pr(G=1\vert S=s,X=x,W=w)}{\Pr(G=0\vert S=s,X=x,W=w)}\frac{1-\gamma(x)}{\gamma(x)}(\varrho(x))^{w}(1-\varrho(x))^{1-w}.
\end{align*}
Thus, $\rho_w$ is identified from the observed data and the overlap condition (Assumption \ref{as:overlap}) ensures that the denominator below is bounded away from zero. Since
$\pi=\Pr(G=1)$, Assumptions \ref{as:unconfoundedness}--\ref{as:LUTM} and the fact that the summand disappears unless
$W=1$ give
\begin{align*}
	&\mathbb{E}\left[
	\frac{G}{\pi}
	\frac{W\mathbf{1}(Y\leq q_{1,\tau})}{\rho_1(S,X)}
	\right] \\
	&=
	\mathbb{E}\left[
	\left.
	\frac{W\mathbf{1}(Y\leq q_{1,\tau})}{\rho_1(S,X)}
	\right\rvert G=1
	\right] \\
	&=
	\mathbb{E}\left[
	\left.
	\frac{W\mathbf{1}(Y(1)\leq q_{1,\tau})}
	{\rho_1(S(1),X)}
	\right\rvert G=1
	\right] \\
	&=
	\mathbb{E}\left[
	\left.
	\frac{1}{\rho_1(S(1),X)}
	\mathbb{E}\left[
	W\mathbf{1}(Y(1)\leq q_{1,\tau})
	\vert S(1),X,G=1
	\right]
	\right\rvert G=1
	\right],
\end{align*}
where the first equality follows from the definition of conditional expectation, the second follows from consistency and the last follows from the law of iterated expectations because $1/\rho_1(S(1),X)$ is measurable with respect to $(S(1),X,G)$. Assumptions~\ref{as:unconfoundedness} and \ref{as:external_validity} yield the observed-data representation of $\rho_1$ established above, while Assumption~\ref{as:overlap} ensures that $\rho_1(S(1),X)$ is bounded away from zero, so the inverse weight is well-defined and integrable.

By latent unconfoundedness (Assumption \ref{as:LUTM}), $W\perp Y(1)\vert S(1),X,G=1$. Hence,
\begin{align*}
	&\mathbb{E}\left[
	W\mathbf{1}(Y(1)\leq q_{1,\tau})
	\vert S(1),X,G=1
	\right] \\
	&=
	\Pr\left(W=1\vert S(1),X,G=1\right)
	\Pr\left(Y(1)\leq q_{1,\tau}\vert S(1),X,G=1\right) \\
	&=
	\rho_1(S(1),X)
	\Pr\left(Y(1)\leq q_{1,\tau}\vert S(1),X,G=1\right).
\end{align*}
Substitution into the above expression and the law of iterated expectations
therefore yield
\begin{align*}
	\mathbb{E}\left[
	\frac{G}{\pi}
	\frac{W\mathbf{1}(Y\leq q_{1,\tau})}{\rho_1(S,X)}
	\right]
	&=
	\mathbb{E}\left[
	\Pr(Y(1)\leq q_{1,\tau}\vert S(1),X,G=1)
	\vert G=1
	\right] \\
	&=\Pr\left(Y(1)\leq q_{1,\tau}\vert G=1\right)=F_{Y(1)\vert G=1}(q_{1,\tau}).
\end{align*}
The smoothness condition (Assumption \ref{as:smoothness}) implies that the distribution function of
$Y(1)\vert G=1$ is continuous at $q_{1,\tau}$. By the definition of $q_{1,\tau}$, it follows that
\begin{align*}
	F_{Y(1)\mid G=1}(q_{1,\tau})=\tau,
\end{align*}
which proves Equation~\eqref{eq:ipw0}.

For $w=0$, recall that
\begin{align*}
	\rho_0(s,x)=\Pr\left(W=0\vert S(0)=s,X=x,G=1\right).
\end{align*}
The same argument, using consistency and $W\perp Y(0)\vert S(0),X,G=1$ (Assumption \ref{as:LUTM}), gives 
\begin{align*}
	&\mathbb{E}\left[\frac{G}{\pi}\frac{(1-W)\mathbf{1}(Y\leq q_{0,\tau})}{\rho_0(S,X)}\right] \\
	&=\mathbb{E}\left[\left.\frac{(1-W)\mathbf{1}(Y(0)\leq q_{0,\tau})}{\rho_0(S(0),X)}\right\rvert G=1\right] \\
	&=\mathbb{E}\left[\Pr(Y(0)\leq q_{0,\tau}\vert S(0),X,G=1)\vert G=1\right] \\
	&=F_{Y(0)\vert G=1}(q_{0,\tau})=\tau,
\end{align*}
in which the final equality follows from smoothness (Assumption \ref{as:smoothness}) and the definition of $q_{0,\tau}$. This proves Equation~\eqref{eq:ipw1}. Then, the proof is complete. $\blacksquare$

\bigskip

\noindent\textbf{Proof of Theorem \ref{thm:EIF1}}: This proof is an extension to the unconditional quantile case of the proofs by \citet{chen2023semiparametric}.
We similarly separate our proofs into four steps:  First, we factorize the density of the observed data. Second, for any smooth parametric submodel, we give the tangent space of the distribution of the observed data. Third, we show the functional of interest is pathwise differentiable and propose a functional form of the tangent space. Fourth, we verify that the proposed efficient influence function satisfies these conditions.

Let the subscript ``$\star$'' denote objects associated with the distribution of complete data.
	Fix a measurable space $(\Omega,\mathcal F)$ and a $\sigma$-finite measure $\lambda$ defined on the appropriate product space. All probability measures introduced below are assumed to be dominated by $\lambda$. Define the complete data vector
	\begin{align*}
	B_\star := (Y^*, S^*, W, G, X),
	\end{align*}
	and let $P_\star$ denote the probability measure induced by $B_\star$ on $(\Omega,\mathcal F)$, with Radon--Nikodym derivative
	\begin{align*}
	p_\star := \frac{dP_\star}{d\lambda}.
	\end{align*}
	The observed data vector is
	\begin{align*}
	B_1 := (W G Y^*,~W S^*,~W,~G,~X),
	\end{align*}
	where the observed outcomes are generated via the measurable mappings
	\begin{align*}
	Y := W G Y^*,~~~S := W S^* .
	\end{align*}
	Let $\mathcal A \subseteq \mathcal F$ denote the $\sigma$-algebra generated by $B_1$. 
	The distribution of the observed data is defined as the pushforward of $P_\star$ under the mapping
	$B_\star \mapsto B_1$; denote this distribution by $P$, with density
	\begin{align*}
	p := \frac{dP}{d\lambda}.
	\end{align*}
	By construction, $P$ coincides with $P_\star$ on the observable $\sigma$-algebra $\mathcal A$; that is,
	\begin{align*}
	P(E)=P_\star(E), \qquad \forall E\in\mathcal A .
	\end{align*}
	Moreover, since the complete and observed data agree almost surely on the events $\{W=1,G=1\}$ and $\{W=1\}$,
	their conditional laws coincide on these events. 
	In particular, for all $y\in\mathbb R$, $s\in\mathbb R^d$, and all $E\in\mathcal A$,
	\begin{align*}
	p_\star(y \mid W=1,G=1,E)=p(y \mid W=1,G=1,E),~~~p_\star(s \mid W=1,E)=p(s \mid W=1,E).
	\end{align*}
	Consequently, the observed-data density $p$ is fully characterized by the complete-data density $p_\star$
	restricted to the observable $\sigma$-algebra. Further,
	\begin{align*}
		s(b_1) & =w g \cdot s_1(y \vert x, s)+w \cdot s_2(s \vert x)+s_3(g, x) \\
		& -g(1-w)\left(\frac{\mathbb{E}_{P_{\star}}[\rho(S^*, X) s_2(S^* \vert X) \vert G=0, W=1, X=x]}{1-\mathbb{E}_{P_{\star}}[\rho(S^*, X) \vert G=0, W=1, X=x]}\right) \\
		& +g\left(w \cdot \frac{s_4(s, x)}{\rho(s, x)}-(1-w) \frac{\mathbb{E}_{P_{\star}}[s_4\left(S^*, X\right) \vert G=0, W=1, X=x]}{1-\mathbb{E}_{P_{\star}}[\rho(S^*, X) \vert G=0, W=1, X=x]}\right) \\
		& +(1-g)\left(s_5(x) \frac{w-\varrho(x)}{\varrho(x)(1-\varrho(x))}\right),
	\end{align*}
in which 
\begin{align*}
	&\mathbb{E}_P[s_1(Y \vert S, X) \vert W=1, G=1, S, X]=\mathbb{E}_{P_{\star}}[s_1(Y^* \vert S^*, X) \vert S^*=s, X=x]=0, \\ 
	&\mathbb{E}_P[s_2(S \vert X) \vert W=1, G=0, X]=\mathbb{E}_{P_{\star}}[s_2(S^* \vert X) \vert X=x] =0~~\text{and}~~\mathbb{E}_P[s_3(G, X)] =0.
\end{align*}

Next establish pathwise differentiability and derive the gradient of the parameter of interest. The argument adapts \citet[Theorem~3.1, Part~I]{chen2023semiparametric} to an unconditional quantile. Since efficiency is determined by the observed-data tangent space, we verify the required Riesz representation directly in terms of observed data.

The unconditional $\tau$-quantile of $Y(1)$ given $G=1$ is
\begin{equation*}
	q_{1,\tau}=Q_{Y(1)\vert G=1}(\tau)=\inf\{q:F_1(q)\ge\tau\},~~F_1(q):=\Pr\{Y(1)\le q\vert G=1\}.
\end{equation*}
Let $\{P_\varepsilon:\varepsilon\in(-\delta,\delta)\}$ be an arbitrary regular parametric submodel of the observed-data model satisfying the imposed identifying restrictions, with $P_0=P$. Write
\begin{equation*}
	\ell'(B):=\left.\frac{\partial}{\partial\varepsilon}\log p_\varepsilon(B)\right|_{\varepsilon=0}\in\mathcal T,
\end{equation*}
for its score, where $B=(GY,S,W,G,X)$ and $\mathcal T\subset L_2^0(P)$ denotes the observed-data tangent space. Let $F_{1,\varepsilon}$ and $q_{1,\tau,\varepsilon}$ denote the related distribution function and $\tau$-quantile under $P_\varepsilon$.

Under consistency, latent unconfoundedness, experimental unconfoundedness and
external validity (Assumptions \ref{as:unconfoundedness}--\ref{as:overlap}) along the submodel is identified
by
\begin{equation}
	F_{1,\varepsilon}(q)=\mathbb E_{P_\varepsilon}\{\overline m_{1,\varepsilon}(X;q)\vert G=1\},
	\label{eq:identified-F1}
\end{equation}
where
\begin{align*}
	m_{1,\varepsilon}(s,x;q)
	&:=\Pr_{P_\varepsilon}(Y\le q\vert W=1,S=s,X=x,G=1),\\
	\overline m_{1,\varepsilon}(x;q)
	&:=\mathbb E_{P_\varepsilon}
	\{m_{1,\varepsilon}(S,X;q)\vert W=1,X=x,G=0\}.
\end{align*}
At $\varepsilon=0$, suppress the subscript $0$ on the nuisance functions and
define
\begin{align}
	\widetilde\psi_{1,1}(B;q)
	:={}&\frac{G}{\pi}
	\left[
	\frac{W}{\rho_1(S,X)}
	\{\mathbf 1(Y\le q)-m_1(S,X;q)\}
	+\overline m_1(X;q)-F_1(q)
	\right] \notag\\
	&+\frac{1-G}{\pi}
	\frac{\gamma(X)}{1-\gamma(X)}
	\frac{W}{\varrho(X)}
	\{m_1(S,X;q)-\overline m_1(X;q)\}.
	\label{eq:F1-gradient-candidate}
\end{align}

We first verify that \eqref{eq:F1-gradient-candidate} is a mean-zero,
square-integrable function. By definition,
\begin{align*}
	\mathbb E[\mathbf 1(Y\le q)-m_1(S,X;q)\vert W=1,S,X,G=1]=0~~\text{and}~~
	\mathbb E[m_1(S,X;q)-\overline m_1(X;q)\vert W=1,X,G=0]=0.
\end{align*}
Moreover, \eqref{eq:identified-F1} implies
\begin{equation*}
	\mathbb E[\overline m_1(X;q)-F_1(q)\vert G=1]=0,
\end{equation*}
which further gives
\begin{equation*}
	\mathbb E[\widetilde\psi_{1,1}(B;q)]=0.
\end{equation*}
The overlap conditions (Assumption \ref{as:overlap}) bound $\pi^{-1}$, $\rho_1^{-1}$,
$\varrho^{-1}$ and $\gamma/(1-\gamma)$. Since the remaining quantities in
\eqref{eq:F1-gradient-candidate} are bounded, it follows that
\begin{equation}
	\widetilde\psi_{1,1}(\cdot;q)\in L_2(P).
	\label{eq:F1-L2}
\end{equation}

We now verify the pathwise-derivative identity. For any regular conditional
distribution, the conditional-score identity gives
\begin{equation}
	\left.\frac{\partial}{\partial\varepsilon}\mathbb E_{P_\varepsilon}[h(Z)\vert V]
	\right|_{\varepsilon=0}=\mathbb E\bigl[(h(Z)-\mathbb E(h(Z)\vert V))\ell'(B)\vert V\bigr].
	\label{eq:conditional-score-identity}
\end{equation}
Furthermore, consistency and latent unconfoundedness (Assumptions \ref{as:unconfoundedness} and \ref{as:LUTM}) imply
\begin{equation}
	\mathbb E[\frac{G}{\pi}\frac{W}{\rho_1(S,X)}h(Y,S,X)]
	=\mathbb E[h(Y(1),S(1),X)\vert G=1].
	\label{eq:observational-change-measure}
\end{equation}
Experimental unconfoundedness (Assumption \ref{as:unconfoundedness}) and external validity (Assumption \ref{as:external_validity}) imply
\begin{equation}
	\mathbb E\left[
	\frac{1-G}{\pi}
	\frac{\gamma(X)}{1-\gamma(X)}
	\frac{W}{\varrho(X)}h(S,X)
	\right]
	=\mathbb E[h(S(1),X)\vert G=1].
	\label{eq:experimental-change-measure}
\end{equation}
The factor $W/\varrho(X)$ recovers the conditional distribution of $S(1)$ given $X$ in the experimental population, whereas $\gamma(X)/[1-\gamma(X)]$ transports the distribution of $X$ from $G=0$ to $G=1$.

Following \citet{chen2023semiparametric}, differentiating the three terms in \eqref{eq:identified-F1} and applying
\eqref{eq:conditional-score-identity}--\eqref{eq:experimental-change-measure}, yields
\begin{equation}
	\left.\frac{\partial}{\partial\varepsilon}F_{1,\varepsilon}(q)\right|_{\varepsilon=0}=A_{Y,1}(q)+A_{S,1}(q)+A_{X,1}(q),
	\label{eq:F1-path-decomposition}
\end{equation}
where
\begin{align}
	A_{Y,1}(q)
	=&\mathbb E\left[\frac{G}{\pi}\frac{W}{\rho_1(S,X)}\{\mathbf 1(Y\le q)-m_1(S,X;q)\}\ell'(B)\right],
	\label{eq:F1-outcome-derivative}\\
	A_{S,1}(q)
	=&\mathbb E\left[\frac{1-G}{\pi}\frac{\gamma(X)}{1-\gamma(X)}\frac{W}{\varrho(X)}(m_1(S,X;q)-\overline m_1(X;q))\ell'(B)\right],
	\label{eq:F1-surrogate-derivative}\\
	A_{X,1}(q)=&\mathbb E\left[\frac{G}{\pi}(\overline m_1(X;q)-F_1(q))\ell'(B)\right].
	\label{eq:F1-covariate-derivative}
\end{align}
The first term is produced by perturbations of the conditional distribution of $Y$ given $W=1,S,X,G=1$; the second is generated by perturbations of the distribution of $S(1)$ given $X$; and the third is generated by perturbations of target distribution of $X\vert G=1$.

Combining \eqref{eq:F1-path-decomposition}--
\eqref{eq:F1-covariate-derivative} gives the Riesz representation
\begin{equation}
	\left.\frac{\partial}{\partial\varepsilon}
	F_{1,\varepsilon}(q)\right|_{\varepsilon=0}
	=\mathbb E\{\widetilde\psi_{1,1}(B;q)\ell'(B)\}
	\qquad\text{for every }\ell'\in\mathcal T.
	\label{eq:F1-Riesz}
\end{equation}

It remains to verify that the gradient lies in the tangent space. The three terms in \eqref{eq:F1-gradient-candidate} have the respective forms
\begin{align*}
	&GW a(Y,S,X),
	&&\mathbb E[a(Y,S,X)\vert W=1,S,X,G=1]=0,\\
	&(1-G)W b(S,X),
	&&\mathbb E[b(S,X)\vert W=1,X,G=0]=0,\\
	&G c(X),
	&&\mathbb E[c(X)\vert G=1]=0.
\end{align*}
They then belong to the observed-data tangent space, so that
\begin{equation}
	\widetilde\psi_{1,1}(\cdot;q)\in\overline{\mathcal T}.
	\label{eq:F1-tangent-membership}
\end{equation}
Equations~\eqref{eq:F1-L2}, \eqref{eq:F1-Riesz} and \eqref{eq:F1-tangent-membership} prove that
$\widetilde\psi_{1,1}(B;q)$ is the gradient and hence the efficient influence function of the distribution-function parameter $F_1(q)$.

It remains to apply the quantile functional chain rule. By definition,
\begin{equation*}
	F_{1,\varepsilon}(q_{1,\tau,\varepsilon})=\tau
\end{equation*}
for all $\varepsilon$ in a neighborhood of zero. Under Assumption \ref{as:smoothness}, the marginal distribution function is continuously differentiable and uniformly strictly increasing on a fixed neighborhood of quantiles, with its derivative uniformly bounded from zero. So, the quantile functional is locally Hadamard differentiable. Applying the functional chain rule along regular parametric submodels yields

\begin{align*}
	0=\left.\frac{d}{d\varepsilon}
	F_{1,\varepsilon}(q_{1,\tau,\varepsilon})
	\right|_{\varepsilon=0}=\mathbb E[\widetilde\psi_{1,1}(B;q_{1,\tau})\ell'(B)]
	+f_{Y(1)\mid G=1}(q_{1,\tau})
	\left.\frac{d}{d\varepsilon}q_{1,\tau,\varepsilon}
	\right|_{\varepsilon=0}.
\end{align*}
Consequently, it follows that
\begin{align}
	\left.\frac{d}{d\varepsilon}q_{1,\tau,\varepsilon}
	\right|_{\varepsilon=0}
	&=\mathbb E\{\psi_1(B,q_{1,\tau},\eta_0)\ell'(B)\},
	\notag\\
	\psi_1(B,q_{1,\tau},\eta_0)
	&=-\frac{\widetilde\psi_{1,1}(B;q_{1,\tau})}
	{f_{Y(1)\mid G=1}(q_{1,\tau})}.
	\label{eq:q1-gradient}
\end{align}
Because $F_1(q_{1,\tau})=\tau$, the EIF in \eqref{eq:q1-gradient} can be written as
\begin{align}
	\psi_1(B,q_{1,\tau},\eta_0)
	={}&-\frac{G}{\pi}
	\left[
	\frac{W\{\mathbf 1(Y\le q_{1,\tau})-m_1(S,X;q_{1,\tau})\}}
	{\rho_1(S,X)f_{Y(1)\mid G=1}(q_{1,\tau})}
	+\frac{\overline m_1(X;q_{1,\tau})-\tau}
	{f_{Y(1)\mid G=1}(q_{1,\tau})}
	\right] \notag\\
	&-\frac{1-G}{\pi}\frac{\gamma(X)}{1-\gamma(X)}
	\frac{W\{m_1(S,X;q_{1,\tau})-\overline m_1(X;q_{1,\tau})\}}
	{\varrho(X)f_{Y(1)\mid G=1}(q_{1,\tau})}.
	\label{eq:q1-gradient-expanded}
\end{align}

The same argument applies to $q_{0,\tau}$. In particular, letting $f_0=f_{Y(0)\mid G=1}(q_{0,\tau})$, we obtain
\begin{align}
	\psi_0(B,q_{0,\tau},\eta_0)
	=&-\frac{G}{\pi}
	\left[
	\frac{(1-W)\{\mathbf 1(Y\le q_{0,\tau})-m_0(S,X;q_{0,\tau})\}}
	{\rho_0(S,X)f_0}
	+\frac{\overline m_0(X;q_{0,\tau})-\tau}{f_0}
	\right] \notag\\
	&-\frac{1-G}{\pi}\frac{\gamma(X)}{1-\gamma(X)}
	\frac{(1-W)(m_0(S,X;q_{0,\tau})-\overline m_0(X;q_{0,\tau}))}
	{\{1-\varrho(X)\}f_0}.
	\label{eq:q0-gradient-expanded}
\end{align}
Thus, by linearity, the EIF of $\Delta_\tau=q_{1,\tau}-q_{0,\tau}$ is given by
\begin{equation*}
	\psi(B,q_{1,\tau},q_{0,\tau},\eta_0)=\psi_1(B,q_{1,\tau},\eta_0)-\psi_0(B,q_{0,\tau},\eta_0).
\end{equation*}
The proof is then complete. $\blacksquare$

\bigskip

\noindent\textbf{Proof of Theorem \ref{thm:dr}}: We prove the result for $\psi_1$; the argument for $\psi_0$ follows from the same calculations. Throughout the proof, let
\begin{align*}
f_1:=f_{Y(1)\vert G=1}(q_{1,\tau}),
\end{align*}
which is the marginal density of $Y(1)\vert G=1$ evaluated at $q_{1,\tau}$. By smoothness (Assumption \ref{as:smoothness}), $f_1$ is a strictly positive and finite constant. Double robustness means that this score continues to have mean zero at $q_{1,\tau}$ when either the outcome regression functions or the propensity-score-type functions are correctly specified.

Using the definition of $\psi_1$, we have
\begin{align}
	\mathbb E[\psi_1(B,q_{1,\tau},\eta)]=&-\frac{1}{f_1}\mathbb E\left[
	\frac{G}{\pi}\frac{W}{\rho_1(S,X)}[\mathbf 1(Y\le q_{1,\tau})-m_1(S,X;q_{1,\tau})]\right] \notag\\
	&-\frac{1}{f_1}\mathbb E\left[\frac{G}{\pi}[\overline m_1(X;q_{1,\tau})-\tau]\right] \notag\\
	&-\frac{1}{f_1}\mathbb E\left[\frac{1-G}{\pi}\frac{\gamma(X)}{1-\gamma(X)}\frac{W}{\varrho(X)}(m_1(S,X;q_{1,\tau})-\overline m_1(X;q_{1,\tau}))\right].
	\label{eq:mean-psi1}
\end{align}

\medskip
\noindent
\textit{Case (i): the outcome regression functions are correctly specified.}
In this case,
\begin{align*}
	m_1(S,X;q_{1,\tau})
	&=\mathbb E[\mathbf 1(Y\le q_{1,\tau})\vert W=1,S,X,G=1],\\
	\overline m_1(X;q_{1,\tau})
	&=\mathbb E[m_1(S,X;q_{1,\tau})\vert W=1,X,G=0].
\end{align*}
Consequently, it follows that
\begin{equation*}
	\mathbb E[\mathbf 1(Y\le q_{1,\tau})-m_1(S,X;q_{1,\tau})\vert W=1,S,X,G=1]=0.
\end{equation*}
It follows by iterated expectations that, for any admissible
$\rho_1(S,X)$,
\begin{align}
	\mathbb E\left[\frac{G}{\pi}\frac{W}{\rho_1(S,X)}[\mathbf 1(Y\le q_{1,\tau})-m_1(S,X;q_{1,\tau})]
	\right]=0.
	\label{eq:outcome-first-zero}
\end{align}
Similarly,
\begin{equation*}
	\mathbb E[m_1(S,X;q_{1,\tau})-\overline m_1(X;q_{1,\tau})\vert W=1,X,G=0]=0,
\end{equation*}
and hence, for any admissible $\varrho(X)$ and $\gamma(X)$,
\begin{align}
	\mathbb E\left[\frac{1-G}{\pi}\frac{\gamma(X)}{1-\gamma(X)}\frac{W}{\varrho(X)}(m_1(S,X;q_{1,\tau})-\overline m_1(X;q_{1,\tau}))\right]=0.
	\label{eq:outcome-third-zero}
\end{align}
Finally, the identification formula
\begin{equation*}
	\mathbb E[\overline m_1(X;q_{1,\tau})\vert G=1]=\Pr(Y(1)\le q_{1,\tau}\vert G=1)=\tau
\end{equation*}
implies that
\begin{align}
	\mathbb E\left[\frac{G}{\pi}(\overline m_1(X;q_{1,\tau})-\tau)\right]=0.
	\label{eq:outcome-second-zero}
\end{align}
Combining \eqref{eq:mean-psi1}--\eqref{eq:outcome-second-zero} yields that
\begin{equation}
	\mathbb E[\psi_1(B,q_{1,\tau},\eta)]=0,
	\label{eq:outcome-robustness}
\end{equation}
regardless of whether the propensity-score-type functions are correctly specified.

\medskip
\noindent
\textit{Case (ii): the propensity-score-type functions are correctly
	specified.}
The first weighted component in \eqref{eq:mean-psi1} satisfies
\begin{align}
	&\mathbb E\left[\frac{G}{\pi}\frac{W}{\rho_1(S,X)}[\mathbf 1(Y\le q_{1,\tau})-m_1(S,X;q_{1,\tau})]
	\right] \notag\\
	&=\mathbb E\left[\Pr(Y(1)\le q_{1,\tau}\vert S(1),X,G=1)-m_1(S(1),X;q_{1,\tau})
	\,\middle|\,G=1\right],
	\label{eq:propensity-first}
\end{align}
which follows from consistency and latent unconfoundedness (Assumption \ref{as:unconfoundedness}),
because
\begin{equation*}
	\mathbb E[W\vert Y(1),S(1),X,G=1]=\mathbb E[W\vert S(1),X,G=1]=\rho_1(S(1),X).
\end{equation*}

For the experimental sample, experimental unconfoundedness (Assumption \ref{as:unconfoundedness}) gives
\begin{equation*}
	\mathbb E[W\vert S(1),X,G=0]=\varrho(X),
\end{equation*}
and external validity (Assumption \ref{as:external_validity}) identifies the conditional distribution of $S(1)$ given $X$ across the two samples. The odd factor $\gamma(X)/(1-\gamma(X))$ transports the distribution of $X$ from $G=0$ to $G=1$. Therefore,
\begin{align}
	&\mathbb E\left[\frac{1-G}{\pi}\frac{\gamma(X)}{1-\gamma(X)}\frac{W}{\varrho(X)}(m_1(S,X;q_{1,\tau})-\overline m_1(X;q_{1,\tau}))\right] =\mathbb E\left[m_1(S(1),X;q_{1,\tau})-\overline m_1(X;q_{1,\tau})\,\middle|\,G=1\right].
	\label{eq:propensity-third}
\end{align}
Moreover, we obtain that
\begin{align}
	\mathbb E\left[\frac{G}{\pi}(\overline m_1(X;q_{1,\tau})-\tau)\right]=\mathbb E[\overline m_1(X;q_{1,\tau})-\tau\vert G=1].
	\label{eq:propensity-second}
\end{align}
Substituting \eqref{eq:propensity-first}--\eqref{eq:propensity-second} into \eqref{eq:mean-psi1}, the terms involving $m_1$ and $\overline m_1$ cancel, giving
\begin{align*}
	\mathbb E[\psi_1(B,q_{1,\tau},\eta)]
	&=-\frac{1}{f_1}\left[\mathbb E\left[\Pr(Y(1)\le q_{1,\tau}\vert S(1),X,G=1)\vert G=1\right]-\tau\right]\\
	&=-\frac{1}{f_1}(\Pr(Y(1)\le q_{1,\tau}\vert G=1)-\tau)
	=0.
\end{align*}
The score has mean zero when the propensity-score-type functions are
correctly specified, regardless of whether $m_1$ and $\overline m_1$ are
correctly specified. With \eqref{eq:outcome-robustness}, this proves
double robustness.

It remains to establish Neyman orthogonality. The double robustness shows that the expected score is identically zero when either group of nuisance functions is fixed at its true value. Then, under the stated differentiability conditions, the Gateaux derivative regarding each group vanishes at $\eta_0$. By linearity of the Gateaux differential, for every admissible perturbation $h$,
\begin{equation*}
	\left.
	\frac{\partial}{\partial r}
	\mathbb E\bigl[\psi_1(B,q_{1,\tau},\eta_0+r h)\bigr]
	\right|_{r=0}
	=0.
\end{equation*}

For $\psi_0$, the same proof applies after replacing $W$ by $1-W$, $\rho_1(S,X)$ by $\rho_0(S,X)$, $\varrho(X)$ by $1-\varrho(X)$, $m_1$ by $m_0$, $\overline m_1$ by $\overline m_0$ and $Y(1)$ by $Y(0)$. Then, $\psi_0$ is also doubly robust and Neyman orthogonal. The proof is complete. $\blacksquare$

\bigskip
\subsection{Proofs for Section \ref{sec:estimation}}

To prove Theorem \ref{thm:integratedM}, we first establish Lemmas \ref{lem:generated-q-outcome-rate} and \ref{lem:dml-rate-oracle-equivalence}.

\begin{lm}[Rate upper bound]
	\label{lem:generated-q-outcome-rate}
	Suppose Assumptions \ref{as:first-stage-rate} and \ref{as:outcome-lipschitz-q} hold. We obtain that
	\begin{align*}
		\max_{1\le k\le K}\sup_{\tau\in\Upsilon}
		\left\Vert\widehat\eta_{\rm out}^{(-k)}
		\left(B;\widetilde q_{0,\tau}^{(-k)},\widetilde q_{1,\tau}^{(-k)}\right)-\eta_{\rm out}
		\left(B;q_{0,\tau},q_{1,\tau}\right)\right\Vert_{\mathbb P,2}=O_p(r_n).
	\end{align*}   
\end{lm} 

\noindent\textbf{Proof of Lemma \ref{lem:generated-q-outcome-rate}}: Because $r_n\to0$, Assumption \ref{as:first-stage-rate} implies
\begin{align*}
	\Pr\Bigg(&
	\left(\widetilde q_{0,\tau}^{(-k)},\widetilde q_{1,\tau}^{(-k)}\right)\in\mathcal Q_0\times\mathcal Q_1
	\text{ for every }1\le k\le K\text{ and }\tau\in\Upsilon
	\Bigg)\longrightarrow1,
\end{align*}
since $\mathcal Q_w$ is a fixed $\epsilon$-enlargement of $\{q_{w,\tau}:\tau\in\Upsilon\}$ for $w=0,1$. We work on this event below.

By the triangle inequality, it follows that
\begin{align*}
	&\max_{1\le k\le K}\sup_{\tau\in\Upsilon}\left\Vert\widehat\eta_{\rm out}^{(-k)}\left(B;\widetilde q_{0,\tau}^{(-k)},\widetilde q_{1,\tau}^{(-k)}\right)-\eta_{\rm out}\left(B;q_{0,\tau},q_{1,\tau}\right)\right\Vert_{\mathbb P,2}\\
	&\le
	\max_{1\le k\le K}\sup_{\tau\in\Upsilon}\left\Vert\widehat\eta_{\rm out}^{(-k)}\left(B;\widetilde q_{0,\tau}^{(-k)},\widetilde q_{1,\tau}^{(-k)}\right)-\eta_{\rm out}\left(B;\widetilde q_{0,\tau}^{(-k)},\widetilde q_{1,\tau}^{(-k)}\right)\right\Vert_{\mathbb P,2}\\
	&+\max_{1\le k\le K}\sup_{\tau\in\Upsilon}\left\Vert\eta_{\rm out}\left(B;\widetilde q_{0,\tau}^{(-k)},\widetilde q_{1,\tau}^{(-k)}\right)-\eta_{\rm out}\left(B;q_{0,\tau},q_{1,\tau}\right)\right\Vert_{\mathbb P,2}\\
	&=:I_{1n}+I_{2n}. 
\end{align*}

For $I_{1n}$, the generated quantiles lie in $\mathcal Q_0\times\mathcal Q_1$ on the event under consideration. Therefore,
\begin{align*}
	I_{1n}
	&\le
	\max_{1\le k\le K}
	\sup_{(q_0,q_1)\in\mathcal Q_0\times\mathcal Q_1}
	\left\Vert
	\widehat\eta_{\rm out}^{(-k)}(B;q_0,q_1)
	-\eta_{\rm out}(B;q_0,q_1)
	\right\Vert_{\mathbb P,2}
	=O_p(r_n)
\end{align*}
by Assumption \ref{as:first-stage-rate}. For $I_{2n}$, Assumption \ref{as:outcome-lipschitz-q} gives
\begin{align*}
	I_{2n}
	&\le
	C\max_{1\le k\le K}\sup_{\tau\in\Upsilon}
	\max_{w=0,1}
	\left|\widetilde q_{w,\tau}^{(-k)}-q_{w,\tau}\right|
	=O_p(r_n),
\end{align*}
where the last equality follows from Assumption \ref{as:first-stage-rate}. Combining these two bounds and using the fact that the event under consideration has probability approaching one yields
\begin{align*}
	&\max_{1\le k\le K}\sup_{\tau\in\Upsilon}
	\left\Vert
	\widehat\eta_{\rm out}^{(-k)}
	\left(B;\widetilde q_{0,\tau}^{(-k)},
	\widetilde q_{1,\tau}^{(-k)}\right)
	-\eta_{\rm out}\left(B;q_{0,\tau},q_{1,\tau}\right)
	\right\Vert_{\mathbb P,2}
	=O_p(r_n).
\end{align*}
This completes the proof. $\blacksquare$

\bigskip

Define $\mathcal L_{w,\tau,n}^{0}(q;q_{w,\tau},\eta_0)$ as the oracle sample objective obtained from \eqref{eq:sampleloss1}--\eqref{eq:sampleloss0} by replacing all estimated nuisance functions by their true values and $\widetilde q_{w,\tau}^{(-k)}$ by $q_{w,\tau}$. That is, $\mathcal L_{w,\tau,n}^{0}$ is the sample analogue of $\mathcal L_{w,\tau}(\cdot;q_{w,\tau},\eta_0)$. For notational convenience, let $\widehat{\mathcal L}_{w,\tau}(\cdot)$ denote the feasible cross-fitted sample objective defined in \eqref{eq:sampleloss1} and \eqref{eq:sampleloss0}.


\begin{lm}[Local oracle equivalence]
\label{lem:dml-rate-oracle-equivalence} Suppose Assumptions \ref{as:unconfoundedness}--\ref{as:iid}, \ref{as:first-stage-rate} and \ref{as:outcome-lipschitz-q}--\ref{as:uniform_score_complexity} hold. Then, for each fixed $u\in\mathbb R$ and for $w=0,1$, the local oracle-equivalence condition holds:
	\begin{align}
		&\sup_{\tau\in\Upsilon} \left|n\left[\widehat{\mathcal L}_{w,\tau}\left(q_{w,\tau}+\frac{u}{\sqrt n}\right)-\widehat{\mathcal L}_{w,\tau}(q_{w,\tau})\right]\right.\notag\\
		&\quad
		\left.-n\left[\mathcal L_{w,\tau,n}^{0}\left(q_{w,\tau}+\frac{u}{\sqrt n};q_{w,\tau},\eta_0\right)-\mathcal L_{w,\tau,n}^{0} \left(q_{w,\tau};q_{w,\tau},\eta_0\right)\right]\right|=o_p(1)
		\label{eq:loss-oracle-equiv-new}
	\end{align}
\end{lm}

\noindent\textbf{Proof of Lemma \ref{lem:dml-rate-oracle-equivalence}}: The key point is that the objective difference can be represented exactly as an integral of the score. Indeed, for each fixed observation, the map
\begin{align}
	q\mapsto \rho_\tau(Y_i-q)
	\notag
\end{align}
is convex, Lipschitz and absolutely continuous in $q$, with the derivative
\begin{align*}
	\frac{\partial}{\partial q}\rho_\tau(Y_i-q)=\mathbf 1\{Y_i\leq q\}-\tau
\end{align*}
for Lebesgue-almost every $q$. Since the remaining terms in \eqref{eq:poploss1}--\eqref{eq:poploss0} are linear in $q$, then the fundamental theorem of calculus for absolutely continuous functions implies that, for
\begin{align*}
	q_{w,\tau,v}=q_{w,\tau}+\frac{v}{\sqrt n},
\end{align*}
we have
\begin{align}
	&n\left[\widehat{\mathcal L}_{w,\tau}\left(q_{w,\tau}+\frac{u}{\sqrt n}\right)-\widehat{\mathcal L}_{w,\tau}(q_{w,\tau})\right]=\sqrt n\int_0^u\mathbb E_n\left[\Phi_w\left(
	B_i;q_{w,\tau,v},\widetilde q_{w,\tau}^{(-k)},\widehat\eta^{(-k)}\right)\right]dv.
	\label{eq:feasible-loss-integral}
\end{align}
Similarly, we obtain that
\begin{align}
	n\left[\mathcal L_{w,\tau,n}^{0}\left(q_{w,\tau}+\frac{u}{\sqrt n};q_{w,\tau},\eta_0\right)-\mathcal L_{w,\tau,n}^{0}\left(q_{w,\tau};q_{w,\tau},\eta_0\right)
	\right]=\sqrt n\int_0^u\mathbb E_n\left[\Phi_w\left(
	B_i;q_{w,\tau,v},q_{w,\tau},\eta_0\right)\right]dv.
	\label{eq:oracle-loss-integral}
\end{align}
Therefore, by \eqref{eq:feasible-loss-integral} and \eqref{eq:oracle-loss-integral}, it suffices to show that
\begin{align}
	\sup_{\tau\in\Upsilon}\left|\sqrt n\int_0^u\mathbb E_{n}
	\left[\Phi_w\left(B;q_{w,\tau}+\frac{v}{\sqrt n},\widetilde q_{w,\tau}^{(-k)},\widehat\eta^{(-k)}\right)
	-\Phi_w\left(B;q_{w,\tau}+\frac{v}{\sqrt n},q_{w,\tau},\eta_0\right)\right]dv\right|=o_p(1).
	\label{eq:dml-proof-main-target}
\end{align}
In particular, because $u$ is fixed, \eqref{eq:dml-proof-main-target} follows from the stronger bound obtained by taking the supremum of the absolute value of its integrand over $\tau \in \Upsilon$ and $|v| \leq|u|$.

To prove \eqref{eq:dml-proof-main-target}, we perform the following decomposition:
\begin{align}
	&\sqrt n
	\mathbb E_{n}\left[\Phi_w\left(B;q_{w,\tau}+\frac{v}{\sqrt n},\widetilde{q}_{w,\tau}^{(-k)},
	\widehat\eta^{(-k)}\right)-\Phi_w\left(B;q_{w,\tau}+\frac{v}{\sqrt n},
	q_{w,\tau},\eta_0\right)\right]\notag\\
	& =\sqrt n(\mathbb E_{n}-\mathbb E)\left[\Phi_w\left(B;q_{w,\tau}+\frac{v}{\sqrt n},
	\widetilde q_{w,\tau}^{(-k)},\widehat\eta^{(-k)}\right)-\Phi_w\left(B;
	q_{w,\tau}+\frac{v}{\sqrt n},q_{w,\tau},\eta_0\right)\right]\notag\\
	&\quad +\sqrt n
	\mathbb E\left[\Phi_w\left(B;q_{w,\tau}+\frac{v}{\sqrt n},\widetilde q_{w,\tau}^{(-k)},
	\widehat\eta^{(-k)}\right)-\Phi_w\left(B;q_{w,\tau}+\frac{v}{\sqrt n},q_{w,\tau},\eta_0\right)\right].
	\label{eq:dml-proof-decomposition}
\end{align}

First, control the centered empirical-process term in
\eqref{eq:dml-proof-decomposition}.  For each fold $k$, condition on the
training sample $\{B_j:j\notin\mathcal I_k\}$.  Conditional on this sample,
$\widetilde q_{w,\tau}^{(-k)}$ and $\widehat\eta^{(-k)}$ are fixed, while the
validation observations $\{B_i:i\in\mathcal I_k\}$ are i.i.d. and independent
of the fitted quantities.

By the uniform positivity condition in Assumption~\ref{as:first-stage-rate}, with $\pi>0$ and the consistency of $\widehat\pi^{(-k)}$ under Assumptions~\ref{as:iid} and \ref{as:first-stage-rate}, the estimated denominators are uniformly bounded away from zero over all arguments and folds, w.p.a.1.
Therefore, on this event, Assumption~\ref{as:overlap} yields, uniformly over $k\in[K]$
and $w\in\{0,1\}$,
\begin{align*}
	&\left|
	\frac{1}{\widehat\pi^{(-k)}\widehat\rho_w^{(-k)}}
	-\frac{1}{\pi\rho_w}
	\right|
	\le C\left(
	|\widehat\pi^{(-k)}-\pi|
	+|\widehat\rho_w^{(-k)}-\rho_w|
	\right),\\
	&\left|
	\frac{\widehat\gamma^{(-k)}}
	{\widehat\pi^{(-k)}(1-\widehat\gamma^{(-k)})
		(\widehat\varrho^{(-k)})^w
		(1-\widehat\varrho^{(-k)})^{1-w}}
	-
	\frac{\gamma}
	{\pi(1-\gamma)\varrho^w(1-\varrho)^{1-w}}
	\right|\lesssim \left(
	|\widehat\pi^{(-k)}-\pi|
	+|\widehat\gamma^{(-k)}-\gamma|
	+|\widehat\varrho^{(-k)}-\varrho|
	\right),
\end{align*}
where $C<\infty$ is a constant value. Since the true outcome-regression functions take values in $[0,1]$, these inequalities, with \eqref{eq:dml-first-stage-rate-propensity}, \eqref{eq:dml-first-stage-rate-outcome}, \eqref{eq:initial-q-rate-simple} and Lemma~\ref{lem:generated-q-outcome-rate}, imply
\begin{align}
	&\max_{1\leq k\leq K}\sup_{\tau\in\Upsilon}\sup_{|v|\leq |u|}\left\|\Phi_w\left(B;q_{w,\tau}+\frac{v}{\sqrt n},\widetilde q_{w,\tau}^{(-k)},\widehat\eta^{(-k)}\right)-\Phi_w\left(B;q_{w,\tau}+\frac{v}{\sqrt n},q_{w,\tau},\eta_0\right)\right\|_{\mathbb P,2}=O_p(r_n).
	\label{eq:dml-proof-l2-size}
\end{align}

Fix $M>0$ with $|u|\le M$ and let $J_n=v_n\log(A_n/r_n)$. Recall that the score-difference class
$\mathcal H_{n,k,w}(M)$ and its elements $h_{n,k,w,\tau,v}$ from Assumption~\ref{as:uniform_score_complexity}. Denote $\mathbb G_{n_k,k}h:=n_k^{-1/2}\sum_{i\in I_k}[h(B_i)-\mathbb Eh]$. The centered process in \eqref{eq:dml-proof-decomposition} is
\begin{align*}
	\sqrt n(\mathbb E_n-\mathbb E)[h_{n,k,w,\tau,v}]=\sum_{k=1}^K\sqrt{\frac{n_k}{n}}\,
	\mathbb G_{n_k,k}h_{n,k,w,\tau,v}.
\end{align*}
Conditional on the training sigma-field $\mathcal D_{-k}$, the class $\mathcal H_{n,k,w}(M)$ is fixed and the validation observations are i.i.d. By Assumption~\ref{as:uniform_score_complexity}(i), there exists a constant value $F_M<\infty$, independent of $n$, such that, on the same event with probability approaching one,
	\begin{align*}
		&\max_{k,w}\sup_{h\in\mathcal H_{n,k,w}(M)}
		\Vert h\Vert_{\mathbb P,\infty}\le F_M,~~\max_{k,w}\sup_Q\mathcal N(\varepsilon F_M,\mathcal H_{n,k,w}(M),L_2(Q))\le(A_n/\varepsilon)^{v_n}~~\text{with}~~ 0<\varepsilon\le1.
	\end{align*}
	With the $L_2(\mathbb P)$ bound in \eqref{eq:dml-proof-l2-size}, \citet[Corollary~5.1]{chernozhukovetal2014} implies, conditionally on $\mathcal D_{-k}$,
	\begin{align*}
		\mathbb E\!\left[\sup_{h\in\mathcal H_{n,k,w}(M)}|\mathbb G_{n_k,k}h|\;\middle|\;\mathcal D_{-k}\right]=O_p\!\left(r_n\sqrt{J_n}+\frac{J_n}{\sqrt n}\right),
	\end{align*}
	where $n_k\asymp n$. Conditional Markov's inequality and the fixed $K$ then give
	\begin{align}
		&\sup_{\tau\in\Upsilon}\sup_{|v|\le |u|}\left|\sqrt n(\mathbb E_n-\mathbb E)[h_{n,k,w,\tau,v}]\right|=O_p\!\left(r_n\sqrt{J_n}+\frac{J_n}{\sqrt n}\right).
		\label{eq:dml-proof-maximal}
	\end{align}
	By \eqref{eq:uc_growth}, $r_n\sqrt{J_n}+J_n/\sqrt n\to0$; therefore,
	\begin{align}
		&\sup_{\tau\in\Upsilon}\sup_{|v|\le |u|}
		\left|\sqrt n(\mathbb E_n-\mathbb E)
		[h_{n,k,w,\tau,v}]\right|=o_p(1).
		\label{eq:dml-proof-centered-op}
	\end{align}
	This establishes the required uniform bound for the centered empirical-process term in \eqref{eq:dml-proof-decomposition}.


Next control the bias directly, without differentiating the
quantile-indexed nuisance functions. Fix $k$, $w$ and $\tau$, and write
$a=q_{w,\tau}$ and $\widehat a=\widetilde q_{w,\tau}^{(-k)}$.
In the next displays, suppress the superscript $(-k)$ on nuisance
estimates. All expectations integrate over an independent observation
$B$, treating the training sample as fixed. Define
\begin{align*}
 D_w&=\mathbf 1\{W=w\}, &
 p_w&=\varrho^w(1-\varrho)^{1-w}, &
 \widehat p_w&=\widehat\varrho^w(1-\widehat\varrho)^{1-w},\\
 A_w&=\frac{GD_w}{\pi\rho_w}, &
 U&=\frac G\pi, &
 C_w&=\frac{(1-G)\gamma D_w}{\pi(1-\gamma)p_w},\\
 \widehat A_w&=\frac{GD_w}{\widehat\pi\widehat\rho_w}, &
 \widehat U&=\frac G{\widehat\pi}, &
 \widehat C_w&=
 \frac{(1-G)\widehat\gamma D_w}
 {\widehat\pi(1-\widehat\gamma)\widehat p_w}.
\end{align*}
Also let
\begin{align*}
 \delta m&=\widehat m_w(S,X;\widehat a)-m_w(S,X;a),\\
 \delta\overline m&=
 \widehat{\overline m}_w(X;\widehat a)-\overline m_w(X;a),\\
 R_{k,w,\tau}(t)&=
 \mathbb E\left[
 \Phi_w(B;t,\widehat a,\widehat\eta)
 -\Phi_w(B;t,a,\eta_0)\right].
\end{align*}
The conditional mean-zero identities for the true outcome regressions,
together with the change-of-measure identities used in the proof of
Theorem~\ref{thm:dr}, give
\begin{align*}
 \mathbb E[(C_w-A_w)\delta m]&=0, &
 \mathbb E[(U-C_w)\delta\overline m]&=0.
\end{align*}
Consequently, expanding the score at $t=a$ yields the exact identity
\begin{align*}
 R_{k,w,\tau}(a)
 =\mathbb E\Big[&
 \{(\widehat C_w-C_w)-(\widehat A_w-A_w)\}\delta m\\
 &+\{(\widehat U-U)-(\widehat C_w-C_w)\}
 \delta\overline m\Big].
\end{align*}
The positivity bounds, Assumption~\ref{as:first-stage-rate} and
Lemma~\ref{lem:generated-q-outcome-rate} imply that the $L_2(\mathbb P)$
norms of both outcome errors and all three weight errors in this
display are $O_p(r_n)$, uniformly over folds and quantile indices.
Cauchy--Schwarz therefore gives
\begin{align*}
 \max_{k,w}\sup_{\tau\in\Upsilon}
 |R_{k,w,\tau}(q_{w,\tau})|=O_p(r_n^2).
\end{align*}
For $t=a+v/\sqrt n$, the first argument of $\Phi_w$ enters only through
$\mathbf 1\{Y\le t\}$. Conditioning on $(S,X,G=1,W=w)$ thus gives
\begin{align*}
 R_{k,w,\tau}(t)-R_{k,w,\tau}(a)
 =\mathbb E\left[
 (\widehat A_w-A_w)
 \{m_w(S,X;t)-m_w(S,X;a)\}\right].
\end{align*}
For each fixed $u$, both $a$ and $t$ belong to $\mathcal Q_w$ for all
$|v|\le |u|$ and all sufficiently large $n$.
Assumption~\ref{as:outcome-lipschitz-q} and Cauchy--Schwarz imply,
uniformly over these indices,
\begin{align*}
 |R_{k,w,\tau}(t)-R_{k,w,\tau}(a)|
 &\le \|\widehat A_w-A_w\|_{\mathbb P,2}
       \|m_w(\cdot;t)-m_w(\cdot;a)\|_{\mathbb P,2}\\
 &\le C\|\widehat A_w-A_w\|_{\mathbb P,2}|t-a|.
\end{align*}
Combining the preceding bounds yields
\begin{align}
 &\max_{1\le k\le K}\sup_{\tau\in\Upsilon}\sup_{|v|\le |u|}
 \left|\mathbb E\left[
 \Phi_w\left(B;q_{w,\tau}+\frac v{\sqrt n},
 \widetilde q_{w,\tau}^{(-k)},\widehat\eta^{(-k)}\right)
 -\Phi_w\left(B;q_{w,\tau}+\frac v{\sqrt n},
 q_{w,\tau},\eta_0\right)\right]\right|\notag\\
 &\qquad=O_p\left(r_n^2+\frac{|u|r_n}{\sqrt n}\right).
 \label{eq:dml-proof-pop-bias-rate}
\end{align}
Therefore, we have
\begin{align}
	&\max_{1\le k\le K}\sup_{\tau\in\Upsilon}\sup_{|v|\le |u|}
	\left|\sqrt n \mathbb E\left[\Phi_w\left(B;q_{w,\tau}+\frac{v}{\sqrt n},
	\widetilde q_{w,\tau}^{(-k)},\widehat\eta^{(-k)}\right)-\Phi_w\left(B;q_{w,\tau}+\frac{v}{\sqrt n},
	q_{w,\tau},\eta_0\right)\right]\right|\notag\\
	&=O_p\left(|u|r_n+\sqrt n r_n^2\right)=o_p(1),
	\label{eq:dml-proof-bias-op}
\end{align}
by \eqref{eq:dml-rate-condition-simple}. Combining \eqref{eq:dml-proof-centered-op} and
\eqref{eq:dml-proof-bias-op} gives
\begin{align}
	&\sup_{\tau\in\Upsilon}\sup_{|v|\le |u|}\bigg|\sqrt n\mathbb E_{n}\left[\Phi_w\left(B;q_{w,\tau}+\frac{v}{\sqrt n},\widetilde q_{w,\tau}^{(-k)},\widehat\eta^{(-k)}\right)-\Phi_w\left(B;q_{w,\tau}+\frac{v}{\sqrt n},q_{w,\tau},\eta_0\right)\right]\bigg|=o_p(1).
	\label{eq:dml-proof-total-op}
\end{align}
Consequently, we have
\begin{align*}
	&\sup_{\tau\in\Upsilon}
	\left|\sqrt n\int_0^u\mathbb E_{n}\left[\Phi_w\left(B;q_{w,\tau}+\frac{v}{\sqrt n},\widetilde q_{w,\tau}^{(-k)},\widehat\eta^{(-k)}\right)-\Phi_w\left(B;q_{w,\tau}+\frac{v}{\sqrt n},
	q_{w,\tau},\eta_0\right)\right]\,dv\right|\\
	&\le|u|\sup_{\tau\in\Upsilon}
	\sup_{|v|\le |u|}\bigg|\sqrt n\mathbb E_{n}\left[\Phi_w
	\left(B;q_{w,\tau}+\frac{v}{\sqrt n},\widetilde q_{w,\tau}^{(-k)},\widehat\eta^{(-k)}\right)-
	\Phi_w\left(B;q_{w,\tau}+\frac{v}{\sqrt n},q_{w,\tau},\eta_0\right)\right]\bigg|\\
	&=o_p(1).
\end{align*}
This completes the proof. $\blacksquare$

\bigskip

\noindent\textbf{Proof of Theorem \ref{thm:integratedM}}: We prove the case for $w=1$. The proof for $w=0$ holds by replacing $W$, $\rho_1(S,X)$, $\varrho(X)$, $m_1$, $\overline m_1$ and $q_{1,\tau}$ with $1-W$, $\rho_0(S,X)$, $1-\varrho(X)$, $m_0$, $\overline m_0$ and $q_{0,\tau}$, respectively. This proof follows the steps of \citet[Theorem 3.1]{jiangetal2023}.
	
Let the true sample loss be
	\begin{align*}
		\mathcal L_{1,\tau,n}^{0}(q;q_{1,\tau},\eta_0)=\frac1n\sum_{i=1}^n\ell_{1,\tau}^{0}(B_i;q,q_{1,\tau},\eta_0),
	\end{align*}
where
	\begin{align*}
		\ell_{1,\tau}^{0}(B_i;q,q_{1,\tau},\eta)&=\frac{G_i}{\pi}\frac{W_i}{\rho_1(S_i,X_i)}\left[\rho_\tau(Y_i-q)+(\tau-m_1(S_i,X_i;q_{1,\tau}))q\right]+\frac{G_i}{\pi}(\overline m_1(X_i;q_{1,\tau})-\tau)q  \\
		&+\frac{1-G_i}{\pi}\frac{\gamma(X_i)}{1-\gamma(X_i)}\frac{W_i}{\varrho(X_i)}[m_1(S_i,X_i;q_{1,\tau})-\overline m_1(X_i;q_{1,\tau})]q.
	\end{align*}
Define the empirical local objective
	\begin{align*}
		L_n(u,\tau)=n\left[\widehat{\mathcal L}_{1,\tau}\left(q_{1,\tau}+\frac{u}{\sqrt n}\right)-\widehat{\mathcal L}_{1,\tau}(q_{1,\tau})\right],
	\end{align*}
and its oracle counterpart
	\begin{align*}
		L_n^0(u,\tau)=n\left[\mathcal L_{1,\tau,n}^{0}\left(q_{1,\tau}+\frac{u}{\sqrt n};q_{1,\tau},\eta_0\right)-\mathcal L_{1,\tau,n}^{0}(q_{1,\tau};q_{1,\tau},\eta_0)
		\right].
	\end{align*}
By Lemma \ref{lem:dml-rate-oracle-equivalence}, we obtain that
	\begin{align*}
		\sup_{\tau\in\Upsilon}\left\vert L_n(u,\tau)-L_n^0(u,\tau)\right\vert=o_p(1)
	\end{align*}
	for each fixed $u$ and further establish that
	\begin{align*}
		\sup_{\tau\in\Upsilon}\left|\sqrt n(\widehat q_{1,\tau}-q_{1,\tau})-\arg\min_{u\in\mathbb R}L_n^0(u,\tau)\right|=o_p(1).
	\end{align*}
	We divide the proof into three steps.
	
	\paragraph{Step (1): Uniform quadratic approximation}
	
	Here we prove the result for $w=1$; the case $w=0$ follows a similar derivation. 
	
	Under Assumptions \ref{as:unconfoundedness}--\ref{as:overlap}, we establish the following identity for every
	$y\in\mathbb R$:
	\begin{align*}
	\mathbb E\!\left[\frac{GW}{\pi\rho_1(S,X)}\mathbf 1\{Y\le y\}\right]=F_{Y(1)\mid G=1}(y).
	\end{align*}
	Indeed, by the consistency and $\pi=\Pr(G=1)$ as defined in Section 3, and iterated expectations,
	\begin{align}\label{kk0}
		&\mathbb E\left[\frac{GW}{\pi\rho_1(S,X)}\mathbf 1\{Y\le y\}\right]\notag\\
		&=\mathbb E\!\left[\left.
		\frac{W}{\rho_1(S(1),X)}\mathbf 1\{Y(1)\le y\}\right|G=1\right]\notag\\
		&=\mathbb E\left[\left.
		\frac{\mathbb E[W\mathbf 1\{Y(1)\le y\}\mid S(1),X,G=1]}
		{\rho_1(S(1),X)}\right|G=1\right]\notag\\
		&=\mathbb E\left[\Pr(Y(1)\le y\mid S(1),X,G=1)\mid G=1\right]\notag\\
		&=F_{Y(1)\mid G=1}(y).
	\end{align}
	where the third equality uses Assumption \ref{as:LUTM} (latent unconfoundedness) and $\mathbb E[W\mid S(1),X,G=1]=\rho_1(S(1),X)$ by the definition of $\rho_1$.

	By Knight's identity \citep{knight1998},
	\begin{align*}
		&\rho_\tau\left(Y_i-q_{1,\tau}-\frac{u}{\sqrt n}\right)
		-\rho_\tau(Y_i-q_{1,\tau}) \\
		&=-\frac{u}{\sqrt n}\left[\tau-\mathbf 1(Y_i\le q_{1,\tau})\right]+\int_0^{u/\sqrt n}\left[\mathbf 1(Y_i-q_{1,\tau}\le v)-\mathbf 1(Y_i-q_{1,\tau}\le 0)\right]dv.
	\end{align*}
	Substituting this identity into the oracle sample objective and using the
	definition of $\Phi_1$ gives
	\begin{align*}
		L_n^0(u,\tau)
		=u\frac1{\sqrt n}\sum_{i=1}^n\left[\Phi_1(B_i;q_{1,\tau},q_{1,\tau},\eta_0)\right]
		+Q_{n,1}(u,\tau),
	\end{align*}
	where
	\begin{align*}
		Q_{n,1}(u,\tau)
		&=\sum_{i=1}^n\frac{G_i}{\pi}\frac{W_i}{\rho_1(S_i,X_i)}\int_0^{u/\sqrt n}\left[\mathbf 1(Y_i-q_{1,\tau}\le v)-\mathbf 1(Y_i-q_{1,\tau}\le 0)\right]dv.
	\end{align*}
	Equivalently,
	\begin{align*}
		L_n^0(u,\tau)=-uW_{n,1}(\tau)+Q_{n,1}(u,\tau),
	\end{align*}
	where
	\begin{align*}
		W_{n,1}(\tau)=-\frac1{\sqrt n}\sum_{i=1}^n\left[\Phi_1(B_i;q_{1,\tau},q_{1,\tau},\eta_0)\right].
	\end{align*}
	We show that, for every fixed $M<\infty$,
	\begin{align}
		\sup_{\tau\in\Upsilon}\sup_{|u|\le M}\left|Q_{n,1}(u,\tau)-\frac12 f_{Y(1)\vert G=1}(q_{1,\tau})u^2\right|=o_p(1).
		\label{eq:uniform-Q-expansion}
	\end{align}
	Let
	\begin{align*}
		H_{n,1}(S_i,X_i;u,\tau)
&=\mathbb E\Bigg[
		\frac{G_i}{\pi}\frac{W_i}{\rho_1(S_i,X_i)}
		\int_0^{u/\sqrt n}
		\left[\mathbf 1(Y_i-q_{1,\tau}\le v)
		-\mathbf 1(Y_i-q_{1,\tau}\le 0)\right]dv
		\;\Bigg|\;S_i,X_i\Bigg].
	\end{align*}
	Then
	\begin{align*}
		Q_{n,1}(u,\tau)
		&=\sum_{i=1}^n H_{n,1}(S_i,X_i;u,\tau) \\
		&\quad+\sum_{i=1}^n\Bigg[
		\frac{G_i}{\pi}\frac{W_i}{\rho_1(S_i,X_i)}
		\int_0^{u/\sqrt n}
		\left[\mathbf 1(Y_i-q_{1,\tau}\le v)
		-\mathbf 1(Y_i-q_{1,\tau}\le 0)\right]dv-H_{n,1}(S_i,X_i;u,\tau)\Bigg] \\
		&=:Q_{n,1}^{a}(u,\tau)+Q_{n,1}^{b}(u,\tau).
	\end{align*}

First consider $Q_{n,1}^{a}(u,\tau)$. By consistency and Assumption \ref{as:LUTM} (latent unconfoundedness),
\begin{align*}
	H_{n,1}(S_i,X_i;u,\tau)=\mathbb E\left[
	\frac{G_i}{\pi}\frac{W_i}{\rho_1(S_i,X_i)}
	\;\middle|\;S_i,X_i\right]
	\int_0^{u/\sqrt n}\Bigl[
	m_1(S_i,X_i;q_{1,\tau}+v)-m_1(S_i,X_i;q_{1,\tau})
	\Bigr]dv,
\end{align*}
where the outcome regression satisfies
\begin{align*}
m_1(s,x;y)=\Pr(Y\le y\mid G=1,W=1,S=s,X=x)=F_{Y(1)\mid G=1,S(1),X}(y\mid s,x).
\end{align*} 

By \eqref{kk0}, for every real $y$,
\begin{align*}
	&\mathbb E\Bigg[\mathbb E\left[\frac{G_i}{\pi}\frac{W_i}{\rho_1(S_i,X_i)}
	\;\middle|\;S_i,X_i\right]m_1(S_i,X_i;y)\Bigg]=\mathbb E\left[
	\frac{G_i}{\pi}\frac{W_i}{\rho_1(S_i,X_i)}
	\mathbf1(Y_i\le y)\right]=F_{Y(1)\mid G=1}(y).
\end{align*}
By the uniform Lipschitz continuity of the marginal density on the fixed enlarged quantile sets in Assumption~\ref{as:smoothness}, uniformly over $\tau\in\Upsilon$ and $|v|\leq\delta$,
\begin{align*}
	&\left|F_{Y(1)\mid G=1}(q_{1,\tau}+v)-F_{Y(1)\mid G=1}(q_{1,\tau})-f_{Y(1)\mid G=1}(q_{1,\tau})v\right|\leq \frac{L_f}{2}v^2.
\end{align*}

Accordingly, define the remainder by
\begin{align*}
	R_{1,i}(v,\tau)
	&:=m_1(S_i,X_i;q_{1,\tau}+v)
	-m_1(S_i,X_i;q_{1,\tau})
	-f_{Y(1)\mid G=1}(q_{1,\tau})v.
\end{align*}
We therefore obtain, exactly as in the original decomposition,
\begin{align*}
	&H_{n,1}(S_i,X_i;u,\tau)\\
&=\frac{u^2}{2n}\mathbb E\left[\frac{G_i}{\pi}\frac{W_i}{\rho_1(S_i,X_i)}
	\;\middle|\;S_i,X_i\right]f_{Y(1)\mid G=1}(q_{1,\tau})+\mathbb E\left[
	\frac{G_i}{\pi}\frac{W_i}{\rho_1(S_i,X_i)}
	\;\middle|\;S_i,X_i\right]
	\int_0^{u/\sqrt n}R_{1,i}(v,\tau)\,dv.
\end{align*}
The above identification result of \eqref{kk0} gives
\begin{align*}
	\mathbb E\Bigg[\mathbb E\left[\frac{G_i}{\pi}\frac{W_i}{\rho_1(S_i,X_i)}\;\middle|\;S_i,X_i\right]R_{1,i}(v,\tau)\Bigg]=F_{Y(1)\mid G=1}(q_{1,\tau}+v)
	-F_{Y(1)\mid G=1}(q_{1,\tau})
	-f_{Y(1)\mid G=1}(q_{1,\tau})v,
\end{align*}
whose absolute value can be bounded uniformly by $Cv^2$. To further control the empirical process term uniformly, consider the class of functions of $(S,X)$,
\begin{align*}
\left\{(S,X)\mapsto
\mathbb E\left[
\frac{GW}{\pi\rho_1(S,X)}\mathbf1(Y\le y)
\;\middle|\;S,X\right]:
y\text{ in a fixed compact neighborhood of }\mathcal Q_1(\Upsilon)
\right\}.
\end{align*}
For $s<t$ in that neighborhood, Jensen's inequality and the same identification further imply
\begin{align*}
	\left\|\mathbb E\left[
	\frac{GW}{\pi\rho_1(S,X)}\mathbf1(s<Y\le t)
	\;\middle|\;S,X\right]\right\|_{\mathbb P,2}^{2}&\le\mathbb E\left[
	\left(\frac{GW}{\pi\rho_1(S,X)}\right)^2
	\mathbf1(s<Y\le t)\right]
	\\
	&\le C\{F_{Y(1)\mid G=1}(t)-F_{Y(1)\mid G=1}(s)\}.
\end{align*}
As the marginal distribution function is continuous on this neighborhood, partitioning its range into increments of order $\varepsilon^2$ gives $O(\varepsilon^{-2})$ $L_2(\mathbb P)$ brackets of size $\varepsilon$. The bracketing entropy integral is finite. So, this class is $\mathbb P$-Donsker \citep[Theorem 2.5.6]{vanderVaart1998} under Assumption~\ref{as:iid}. Its stochastic equicontinuity \citep[Theorem 1.5.7]{vanderVaart1998}, the $L_2(\mathbb P)$ bound given above and the fact that $\mathbb E[\frac{GW}{\pi\rho_1(S,X)}\vert S,X]$ is uniformly bounded imply
\begin{align*}
	\sup_{\tau\in\Upsilon}\sup_{|v|\le M/\sqrt n}
	\Bigg|\frac1{\sqrt n}\sum_{i=1}^n
	\Bigg\{
	\mathbb E\left[
	\frac{G_i}{\pi}\frac{W_i}{\rho_1(S_i,X_i)}
	\middle|S_i,X_i\right]R_{1,i}(v,\tau)-
	\mathbb E\left[
	\mathbb E\left[
	\frac{G_i}{\pi}\frac{W_i}{\rho_1(S_i,X_i)}
	\middle|S_i,X_i\right]R_{1,i}(v,\tau)
	\right]\Bigg\}\Bigg|\\=o_p(1).
\end{align*}
It follows, for every fixed $M<\infty$, that
\begin{align*}
	\sup_{\tau\in\Upsilon}\sup_{|u|\le M}
	\left|\sum_{i=1}^n
	\mathbb E\left[\frac{G_i}{\pi}\frac{W_i}{\rho_1(S_i,X_i)}
	\;\middle|\;S_i,X_i\right]
	\int_0^{u/\sqrt n}R_{1,i}(v,\tau)\,dv\right|\le\frac{CM^3}{3\sqrt n}+M\,o_p(1)=o_p(1),
\end{align*}
Then, by Assumption~\ref{as:iid}, Assumption~\ref{as:overlap} and Assumption~\ref{as:smoothness}, we have
\begin{align*}
	&\sup_{\tau\in\Upsilon}\Bigg|\frac1n\sum_{i=1}^n\mathbb E\left[\frac{G_i}{\pi}\frac{W_i}{\rho_1(S_i,X_i)}\;\middle|\;S_i,X_i\right]f_{Y(1)\mid G=1}(q_{1,\tau})-f_{Y(1)\mid G=1}(q_{1,\tau})\Bigg|=o_p(1).
\end{align*}
Consequently, it follows that
\begin{equation}
	\sup_{\tau\in\Upsilon}\sup_{|u|\le M}\left|Q_{n,1}^{a}(u,\tau)
	-\frac12f_{Y(1)\mid G=1}(q_{1,\tau})u^2\right|=o_p(1).
	\label{eq:Qa-bound}
\end{equation}

Next consider $Q_{n,1}^{b}(u,\tau)$. Define
\begin{align*}
	h_{n,u,\tau}(B_i)=\frac{G_iW_i}{\pi\rho_1(S_i,X_i)}\int_0^{u/\sqrt n}\left[\mathbf 1(Y_i\le q_{1,\tau}+v)-\mathbf 1(Y_i\le q_{1,\tau})\right]dv-H_{n,1}(S_i,X_i;u,\tau).
\end{align*}
By construction, \eqref{eq:rho-identification} and Assumption~\ref{as:overlap} imply that $A(B):=\frac{GW}{\pi\rho_1(S,X)}$ is uniformly bounded and
\begin{align*}
	\mathbb E[h_{n,u,\tau}(B_i)\mid S_i,X_i]=0,~~Q_{n,1}^{b}(u,\tau)=\sum_{i=1}^nh_{n,u,\tau}(B_i),~~\sup_{\tau\in\Upsilon}\sup_{|u|\le M}|h_{n,u,\tau}(B_i)|\le CM/\sqrt n.
\end{align*}
Under Assumptions~\ref{as:overlap} and \ref{as:iid}, the bounded weighted-indicator class satisfies, for $0<\varepsilon<1$,
\begin{align*}
	&\sup_Q \mathcal N\left(\varepsilon,\left\{\frac{GW}{\pi\rho_1(S,X)}\mathbf1(Y\le y):y\in\mathcal Q_1^\delta(\Upsilon)\right\},L_2(Q)\right)\le C\varepsilon^{-C}.
\end{align*}
Thus it is $\mathbb P$-Donsker by \citet[Theorems~2.6.7 and 2.5.2]{VW96}. For $s<t$ in $\mathcal Q_1^\delta(\Upsilon)$, Assumptions~\ref{as:overlap}--\ref{as:iid} and the IPW identity~\eqref{kk0} give
\begin{align*}
	\left\|\mathbb E\left[
	\frac{GW}{\pi\rho_1(S,X)}\mathbf1(s<Y\le t)
	\;\middle|\;S,X\right]\right\|_{\mathbb P,2}^2
	&\le\mathbb E\left[\left(\frac{GW}{\pi\rho_1(S,X)}\right)^2
	\mathbf1(s<Y\le t)\right]
	\\
&\le C\big\{F_{Y(1)\mid G=1}(t)-F_{Y(1)\mid G=1}(s)\big\}.
\end{align*}
Since the conditional-expectation functions are nondecreasing in $y$, partitioning the range of $F_{Y(1)\mid G=1}$ into increments of order $\varepsilon^2$ yields $O(\varepsilon^{-2})$ $L_2(\mathbb P)$ brackets for the conditional-expectation class, so it follows that
\begin{align*}
	&\int_0^1\sqrt{\log \mathcal N_{[]}\left(\varepsilon,
		\left\{\mathbb E\left[\frac{GW}{\pi\rho_1(S,X)}\mathbf1(Y\le y)
		\;\middle|\;S,X\right]:y\in\mathcal Q_1^\delta(\Upsilon)\right\},
		L_2(\mathbb P)\right)}\,d\varepsilon<\infty.
\end{align*}
By \citet[Theorem~2.5.6]{VW96}, this class is $\mathbb P$-Donsker. Consequently, the centered class
\begin{align*}
	\mathcal D=\left\{d_y(B):=A(B)\mathbf1(Y\le y)-\mathbb E[A(B)\mathbf1(Y\le y)\mid S,X]:y\in\mathcal Q_1^\delta(\Upsilon)\right\}
\end{align*}
is $\mathbb P$-Donsker by \citet[Corollary~2.10.13]{VW96}.

Furthermore, for $n$ large enough that $M/\sqrt n\le\delta$, the IPW identity~\eqref{kk0}, Assumptions~\ref{as:unconfoundedness}--\ref{as:overlap} and uniform density bound in Assumption~\ref{as:smoothness} imply, uniformly over $\tau\in\Upsilon$ and $|v|\le M/\sqrt n$,
\begin{align*}
	&\left\|d_{q_{1,\tau}+v}-d_{q_{1,\tau}}\right\|_{\mathbb P,2}^2
	=\mathbb E\Bigg[\operatorname{Var}\Bigg(
	\frac{GW}{\pi\rho_1(S,X)}
	\big\{\mathbf1(Y\le q_{1,\tau}+v)-\mathbf1(Y\le q_{1,\tau})\big\}
	\;\Big|\;S,X\Bigg)\Bigg]\\
	&\le\mathbb E\left[\left(\frac{GW}{\pi\rho_1(S,X)}\right)^2
	\left\{\mathbf1(Y\le q_{1,\tau}+v)-\mathbf1(Y\le q_{1,\tau})\right\}^2\right]\\
	&\le C\left|F_{Y(1)\mid G=1}(q_{1,\tau}+v)-F_{Y(1)\mid G=1}(q_{1,\tau})\right|
	\le CC_f|v|\le\frac{CC_fM}{\sqrt n}.
\end{align*}
Thus $\sup_{\tau\in\Upsilon}\sup_{|v|\le M/\sqrt n}\|d_{q_{1,\tau}+v}-d_{q_{1,\tau}}\|_{\mathbb P,2}=O(n^{-1/4})=o(1)$. Since $\mathcal D$ is a fixed $\mathbb P$-Donsker function class, its asymptotic $L_2(\mathbb P)$ stochastic equicontinuity \citep[Theorem~1.5.7]{VW96}, under Assumption~\ref{as:iid}, yields
\begin{align*}
	&\sup_{\tau\in\Upsilon}\sup_{|v|\le M/\sqrt n}
	\left|\sqrt n(\mathbb E_n-\mathbb E)
	\left[d_{q_{1,\tau}+v}-d_{q_{1,\tau}}\right]\right|=o_p(1).
\end{align*}
Integrating the centered increments over $[0,u/\sqrt n]$ gives
\begin{equation}\label{eq:Qb-bound}
	\sup_{\tau\in\Upsilon}\sup_{|u|\le M}|Q_{n,1}^{b}(u,\tau)|
	\le\sqrt n\frac{M}{\sqrt n}\,o_p(1)=o_p(1).
\end{equation}
Combining \eqref{eq:Qa-bound} and \eqref{eq:Qb-bound} proves \eqref{eq:uniform-Q-expansion}.

	It follows that
	\begin{align}
		L_n^0(u,\tau)=u\frac1{\sqrt n}\sum_{i=1}^n\left[\Phi_1(B_i;q_{1,\tau},q_{1,\tau},\eta_0)\right]+\frac12f_{Y(1)\mid G=1}(q_{1,\tau})u^2+r_{n,1}(u,\tau),
		\label{eq:linearization}
	\end{align}
	where, for every fixed $M<\infty$,
	\begin{align*}
		\sup_{\tau\in\Upsilon}\sup_{|u|\le M}|r_{n,1}(u,\tau)|=o_p(1).
	\end{align*}
	Equivalently, we obtain that
	\begin{align*}
		\sup_{\tau\in\Upsilon}\sup_{|u|\le M}\left|L_n^0(u,\tau)-\left[-uW_{n,1}(\tau)+\frac12f_{Y(1)\mid G=1}(q_{1,\tau})u^2\right]\right|=o_p(1).
	\end{align*}
	Since the oracle-equivalence lemma (Lemma \ref{lem:dml-rate-oracle-equivalence}) gives
	\begin{align*}
		\sup_{\tau\in\Upsilon}\sup_{|u|\le M}|L_n(u,\tau)-L_n^0(u,\tau)|=o_p(1),
	\end{align*}
	we conclude that
	\begin{align*}
		\sup_{\tau\in\Upsilon}\sup_{|u|\le M}\left|L_n(u,\tau)-\left[-uW_{n,1}(\tau)+\frac12f_{Y(1)\mid G=1}(q_{1,\tau})u^2\right]\right|=o_p(1).
	\end{align*}
	Finally, Assumption~\ref{as:smoothness} ensures that
	\begin{align*}
		0<c\le\inf_{\tau\in\Upsilon}f_{Y(1)\vert G=1}(q_{1,\tau})\le\sup_{\tau\in\Upsilon}f_{Y(1)\vert G=1}(q_{1,\tau})\le C<\infty,
	\end{align*}
	so the limiting quadratic criterion is uniformly strictly convex. This
	completes Step~(1).
	
	\paragraph{Step (2): Total boundedness and stochastic equicontinuity of $W_{n,1}(\tau)$.}
	
	Recall that
	\begin{align*} 
		W_{n,1}(\tau)=-\frac1{\sqrt n}\sum_{i=1}^n[\Phi_1(B_i;q_{1,\tau},q_{1,\tau},\eta_0)].
	\end{align*}
	We first prove finite-dimensional convergence and stochastic
	equicontinuity. These results will imply
	$\sup_{\tau\in\Upsilon}|W_{n,1}(\tau)|=O_p(1)$. For each fixed $\tau$, the
	identification result in Theorem~\ref{thm:EIF1} gives
	\begin{align*}
		\mathbb E[\Phi_1(B;q_{1,\tau},q_{1,\tau},\eta_0)]=0. 
	\end{align*}

	Moreover, by the overlap and boundedness conditions in Assumptions~\ref{as:overlap} and \ref{as:smoothness},
	\begin{align*}
		\sup_{\tau\in\Upsilon}|\Phi_1(B;q_{1,\tau},q_{1,\tau},\eta_0)|\le C,
	\end{align*}
	almost surely. Particularly, the scores have uniformly bounded second moments. Therefore, for any finite collection $\tau_1,...,\tau_L\in\Upsilon$, the multivariate central limit theorem and the Cram\'er--Wold device under Assumption~\ref{as:iid} imply
	\begin{align*}
		&\Bigg(
		\frac1{\sqrt n}\sum_{i=1}^n[\Phi_1(B_i;q_{1,\tau_1},q_{1,\tau_1},\eta_0)],...,
		\frac1{\sqrt n}\sum_{i=1}^n[\Phi_1(B_i;q_{1,\tau_L},q_{1,\tau_L},\eta_0)]
		\Bigg)' \rightsquigarrow
		\mathcal N\left(0,\left[\Omega_{11}(\tau_\ell,\tau_r)\right]_{\ell,r=1}^L
		\right),
	\end{align*}
	where
	\begin{align*}
		\Omega_{11}(\tau_\ell,\tau_r)
		=\mathbb E\Big[\left[\Phi_1(B;q_{1,\tau_\ell},q_{1,\tau_\ell},\eta_0)\right]
		\left[\Phi_1(B;q_{1,\tau_r},q_{1,\tau_r},\eta_0)\right]\Big].
	\end{align*}
	It remains to establish the tightness through total boundedness and stochastic
	equicontinuity.
	
	Define the function class
	\begin{align*}
		\mathcal F_1=\left\{B\mapsto\Phi_1(B;q_{1,\tau},q_{1,\tau},\eta_0):\tau\in\Upsilon\right\}.
	\end{align*}
	For $f,g\in\mathcal F_1$ corresponding to $\tau_f,\tau_g\in\Upsilon$, define the metric
	\begin{align*}
		e_{\mathbb P,2}(f,g)=[\mathbb E[(f(B)-g(B))^2]]^{1/2}.
	\end{align*}
	By the definition of $\Phi_1$ and the score representation in Step~(1),
	\begin{align*}
		f(B)-g(B)
		&=\frac{G}{\pi}\frac{W}{\rho_1(S,X)}
		\Big[
		\mathbf 1(Y\le q_{1,\tau_f})
		-\mathbf 1(Y\le q_{1,\tau_g})-[m_1(S,X;q_{1,\tau_f})-m_1(S,X;q_{1,\tau_g})]
		\Big] \\
		&+\frac{G}{\pi}\Big[\overline m_1(X;q_{1,\tau_f})-\overline m_1(X;q_{1,\tau_g})-(\tau_f-\tau_g)\Big] \\
		&+\frac{1-G}{\pi}\frac{\gamma(X)}{1-\gamma(X)}\frac{W}{\varrho(X)}\Big[m_1(S,X;q_{1,\tau_f})-m_1(S,X;q_{1,\tau_g}) -[\overline m_1(X;q_{1,\tau_f})-\overline m_1(X;q_{1,\tau_g})]\Big].
	\end{align*}
	Set
\begin{align*}
 A&=\frac{GW}{\pi\rho_1(S,X)}, &
 U&=\frac G\pi, &
 C_1&=\frac{(1-G)\gamma(X)W}
 {\pi\{1-\gamma(X)\}\varrho(X)}.
\end{align*}
These weights are nonnegative and uniformly bounded under the overlap
condition. Without loss of generality, suppose $\tau_f\le\tau_g$,
and write $q=q_{1,\tau_f}$ and $q'=q_{1,\tau_g}$. Define
\begin{align*}
 \Delta I&=\mathbf 1\{Y\le q'\}-\mathbf 1\{Y\le q\},\\
 \Delta m&=m_1(S,X;q')-m_1(S,X;q),\\
 \Delta\overline m&=\overline m_1(X;q')-\overline m_1(X;q).
\end{align*}
All three increments lie in $[0,1]$. The IPW identity~\eqref{kk0},
the change-of-measure identities
\eqref{eq:observational-change-measure}--\eqref{eq:experimental-change-measure},
and the definition of $\overline m_1$ imply
\begin{align*}
 \mathbb E[A\Delta I]
 &=\mathbb E[A\Delta m]
 =\mathbb E[C_1\Delta m]\\
 &=\mathbb E[U\Delta\overline m]
 =\mathbb E[C_1\Delta\overline m]
 =\tau_g-\tau_f.
\end{align*}
Using the preceding decomposition of $f-g$, boundedness of the
weights, and $z^2\le z$ for $z\in[0,1]$, we obtain
\begin{align*}
 e_{\mathbb P,2}^2(f,g)
 &\le C\Big\{
 \mathbb E[A\Delta I]+\mathbb E[A\Delta m]
 +\mathbb E[C_1\Delta m]\\
 &\qquad\quad
 +\mathbb E[U\Delta\overline m]
 +\mathbb E[C_1\Delta\overline m]
 +|\tau_f-\tau_g|^2\Big\}\\
 &\le C|\tau_f-\tau_g|.
\end{align*}
Thus,
\begin{align*}
 e_{\mathbb P,2}(f,g)\le C|\tau_f-\tau_g|^{1/2}.
\end{align*}
Since $\Upsilon$ is compact in the Euclidean metric,
$(\mathcal F_1,e_{\mathbb P,2})$ is totally bounded.
	
	We next show stochastic equicontinuity. For $\delta>0$, define
	\begin{align*}
		\mathcal H_{1,\delta}=\left\{f-g:f,g\in\mathcal F_1,\ e_{\mathbb P,2}(f,g)<\delta\right\}.
	\end{align*}
	The class $\mathcal H_{1,\delta}$ is VC-type function class.  Moreover, by the overlap and boundedness conditions in Assumptions \ref{as:overlap} and \ref{as:smoothness}, $\mathcal H_{1,\delta}$ has an envelope $H_1$ with
	\begin{align*}
		\|H_1\|_{\mathbb P,2}<\infty.
	\end{align*}
	By construction,
	\begin{align*}
		\sup_{h\in\mathcal H_{1,\delta}}\mathbb E h^2\le\delta^2.
	\end{align*}
	A maximal inequality for VC-type classes \citep[Theorem 5.2]{chernozhukovetal2014} gives
	\begin{align*}
		\mathbb E\left[\sup_{h\in\mathcal H_{1,\delta}}\left|\sqrt n(\mathbb E_n-\mathbb E)h
		\right|\right]\le C\left\{\delta\sqrt{\log(a/\delta)}+\frac{\log(a/\delta)}{\sqrt n}\right\},
	\end{align*}
	for constants $C<\infty$ and $a<\infty$ independent of $n$ and $\delta$. Hence,
	\begin{align*}
		\lim_{\delta\downarrow0}\limsup_{n\to\infty}\mathbb E\left[\sup_{h\in\mathcal H_{1,\delta}}
		\left|\sqrt n(\mathbb E_n-\mathbb E)h\right|\right]=0.
	\end{align*}
	By Markov's inequality, for every $\varepsilon>0$,
	\begin{align*}
		\lim_{\delta\downarrow0}\limsup_{n\to\infty}\Pr\left(\sup_{\substack{f,g\in\mathcal F_1\\ e_{\mathbb P,2}(f,g)<\delta}}\left|\sqrt n(\mathbb E_n-\mathbb E)(f-g)\right|>\varepsilon\right)=0.
	\end{align*}
	This proves stochastic equicontinuity.
	
	Combining finite-dimensional convergence, total boundedness of $(\mathcal F_1,e_{\mathbb P,2})$ and the stochastic equicontinuity gives the functional central limit theorem
	\begin{align*}
		\frac1{\sqrt n}\sum_{i=1}^n
		\left[\Phi_1(B_i;q_{1,\tau},q_{1,\tau},\eta_0)\right]
		\rightsquigarrow\mathbb S_1(\tau)~~\text{in }L_\infty(\Upsilon),
	\end{align*}
	where $\mathbb S_1$ is a tight mean-zero Gaussian process with covariance
	kernel
	\begin{align*}
		\Omega_{11}(\tau,\tau')
		=\mathbb E\Big[\left[\Phi_1(B;q_{1,\tau},q_{1,\tau},\eta_0)\right]\left[\Phi_1(B;q_{1,\tau'},q_{1,\tau'},\eta_0)\right]\Big].
	\end{align*}
	Thus, we obtain that
	\begin{align*}
		W_{n,1}\rightsquigarrow-\mathbb S_1~~\text{in }L_\infty(\Upsilon),
	\end{align*}
	and, in particular, $\sup_{\tau\in\Upsilon}|W_{n,1}(\tau)|=O_p(1)$.
	
	By Step~(1), for every fixed $M<\infty$,
	\begin{align*}
		\sup_{\tau\in\Upsilon}\sup_{|u|\le M}
		\left|L_n(u,\tau)
		-\left[-uW_{n,1}(\tau)
		+\frac12f_{Y(1)\mid G=1}(q_{1,\tau})u^2\right]\right|=o_p(1).
	\end{align*}
	Moreover,
	\begin{align*}
		0<c
		\le\inf_{\tau\in\Upsilon}f_{Y(1)\mid G=1}(q_{1,\tau})
		\le\sup_{\tau\in\Upsilon}f_{Y(1)\mid G=1}(q_{1,\tau})
		\le C<\infty.
	\end{align*}
	Since $L_n(\cdot,\tau)$ is convex, $\widehat q_{1,\tau}$ is its minimizer under the local reparameterization and the uniform root-$n$ tightness of the minimizers follows from Lemma~\ref{lem:dml-rate-oracle-equivalence}, Kato's convexity lemma \citep[Theorem~1]{kato2009} yields
	\begin{align*}
		\sqrt n(\widehat q_{1,\tau}-q_{1,\tau})
		=\frac{W_{n,1}(\tau)}
		{f_{Y(1)\mid G=1}(q_{1,\tau})}
		+r_{n,1}(\tau),
	\end{align*}
	where the remainder term
	\begin{align*}
		\sup_{\tau\in\Upsilon}|r_{n,1}(\tau)|=o_p(1).
	\end{align*}
	The same argument gives
	\begin{align*}
		\sqrt n(\widehat q_{0,\tau}-q_{0,\tau})
		&=-\frac{1}{f_{Y(0)\mid G=1}(q_{0,\tau})}
		\frac1{\sqrt n}\sum_{i=1}^n
		\left[\Phi_0(B_i;q_{0,\tau},q_{0,\tau},\eta_0)\right]
		+o_p(1),
	\end{align*}
	uniformly over $\tau\in\Upsilon$. If joint weak convergence of two expansions is required, the above finite-dimensional convergence and stochastic-equicontinuity arguments are applied to the two score classes jointly, thereby retaining their cross-covariances.
	
	\paragraph{Step (3): Weak convergence and covariance kernel}
	
	Define the two-dimensional process
	\begin{align*}
		\mathbb S_n(\tau)=
		\begin{pmatrix}
			\mathbb G_n\{\Phi_1(B;q_{1,\tau},q_{1,\tau},\eta_0)\}\\
			\mathbb G_n\{\Phi_0(B;q_{0,\tau},q_{0,\tau},\eta_0)\}
		\end{pmatrix}.
	\end{align*}
	By the finite-dimensional convergence and stochastic equicontinuity shown above, we can similarly prove the joint Gaussian limit of $\mathbb S_n(\cdot)$ and obtain
	\begin{align*}
		\mathbb S_n(\tau)\rightsquigarrow\mathbb S(\tau)=
		\begin{pmatrix}
			\mathbb S_1(\tau)\\
			\mathbb S_0(\tau)
		\end{pmatrix}
		\quad
		\text{in } L_\infty(\Upsilon),
	\end{align*}
	where $\mathbb S(\cdot)$ is a mean-zero tight Gaussian process with the covariance kernel function as
	\begin{align*}
		\Omega(\tau,\tau')=\mathbb E\left[
		\begin{pmatrix}
			\Phi_1(B;q_{1,\tau},q_{1,\tau},\eta_0)\\
			\Phi_0(B;q_{0,\tau},q_{0,\tau},\eta_0)
		\end{pmatrix}
		\begin{pmatrix}
			\Phi_1(B;q_{1,\tau'},q_{1,\tau'},\eta_0),&
			\Phi_0(B;q_{0,\tau'},q_{0,\tau'},\eta_0)
		\end{pmatrix}
		\right].
	\end{align*}
	Equivalently,
	\begin{align*}
		\Omega(\tau,\tau')=
		\begin{pmatrix}
			\Omega_{11}(\tau,\tau') & \Omega_{10}(\tau,\tau')\\
			\Omega_{01}(\tau,\tau') & \Omega_{00}(\tau,\tau')
		\end{pmatrix},
	\end{align*}
	where
	\begin{align*}
		\Omega_{11}(\tau,\tau')
		&=
		\mathbb E\left[
		\{\Phi_1(B;q_{1,\tau},q_{1,\tau},\eta_0)\}
		\{\Phi_1(B;q_{1,\tau'},q_{1,\tau'},\eta_0)\}
		\right],\\
		\Omega_{10}(\tau,\tau')
		&=
		\mathbb E\left[
		\{\Phi_1(B;q_{1,\tau},q_{1,\tau},\eta_0)\}
		\{\Phi_0(B;q_{0,\tau'},q_{0,\tau'},\eta_0)\}
		\right],\\
		\Omega_{01}(\tau,\tau')
		&=
		\mathbb E\left[
		\{\Phi_0(B;q_{0,\tau},q_{0,\tau},\eta_0)\}
		\{\Phi_1(B;q_{1,\tau'},q_{1,\tau'},\eta_0)\}
		\right],\\
		\Omega_{00}(\tau,\tau')
		&=
		\mathbb E\left[
		\{\Phi_0(B;q_{0,\tau},q_{0,\tau},\eta_0)\}
		\{\Phi_0(B;q_{0,\tau'},q_{0,\tau'},\eta_0)\}
		\right].
	\end{align*}
	From the above Step (2), we obtain that
	\begin{align*}
		\sqrt n(\widehat\Delta_\tau-\Delta_\tau)
		&=
		\sqrt n(\widehat q_{1,\tau}-q_{1,\tau})-\sqrt n(\widehat q_{0,\tau}-q_{0,\tau})\\
		&=-\frac{\mathbb G_n\{\Phi_1(B;q_{1,\tau},q_{1,\tau},\eta_0)\}}{f_{Y(1)\vert G=1}(q_{1,\tau})}
		+\frac{\mathbb G_n\{\Phi_0(B;q_{0,\tau},q_{0,\tau},\eta_0)\}}{f_{Y(0)\vert G=1}(q_{0,\tau})}
		+
		o_p(1),
	\end{align*}
	uniformly over $\tau\in\Upsilon$. Define
	\begin{align*}
		\psi_\Delta^M(B;\tau)=-\frac{\Phi_1(B;q_{1,\tau},q_{1,\tau},\eta_0)}{f_{Y(1)\vert G=1}(q_{1,\tau})}+\frac{\Phi_0(B;q_{0,\tau},q_{0,\tau},\eta_0)}{f_{Y(0)\vert G=1}(q_{0,\tau})}.
	\end{align*}
	Then, it follows that
	\begin{align*}
		\sqrt n(\widehat\Delta_\tau-\Delta_\tau)=\frac1{\sqrt n}\sum_{i=1}^n
		\psi_\Delta^M(B_i;\tau)+o_p(1),
	\end{align*}
	uniformly over $\tau\in\Upsilon$. By the functional central limit theorem \citep{andrews1994empirical} for the class
	\begin{align*}
		\left\{\psi_\Delta^M(B;\tau):\tau\in\Upsilon\right\},
	\end{align*}
	we obtain
	\begin{align*}
		\sqrt n(\widehat\Delta_\tau-\Delta_\tau)\rightsquigarrow\mathcal B(\tau)
		\quad\text{in }L_\infty(\Upsilon),
	\end{align*}
	where $\mathcal B$ is a tight mean-zero Gaussian process with covariance kernel
	\begin{align*}
		\Sigma_\Delta(\tau,\tau')=\mathbb E[\psi_\Delta^M(B;\tau)\psi_\Delta^M(B;\tau')].
	\end{align*}
	Equivalently, we obtain that
	\begin{align*}
		\Sigma_\Delta(\tau,\tau')
		&=\frac{\Omega_{11}(\tau,\tau')}{f_{Y(1)\vert G=1}(q_{1,\tau})f_{Y(1)\vert G=1}(q_{1,\tau'})}+\frac{\Omega_{00}(\tau,\tau')}{f_{Y(0)\vert G=1}(q_{0,\tau})f_{Y(0)\vert G=1}(q_{0,\tau'})} \\
		&-\frac{\Omega_{10}(\tau,\tau')}{f_{Y(1)\vert G=1}(q_{1,\tau})f_{Y(0)\vert G=1}(q_{0,\tau'})}-
		\frac{\Omega_{01}(\tau,\tau')}{f_{Y(0)\vert G=1}(q_{0,\tau})f_{Y(1)\vert G=1}(q_{1,\tau'})}.
	\end{align*}
	The proof is then complete. $\blacksquare$

\bigskip
\subsection{Proofs for Section \ref{sec:bootstrap_uniform_inference}}

\noindent\textbf{Proof of Theorem \ref{thm:multiplier-bootstrap-integrated}}: Fix $w\in\{0,1\}$. Let $\{\xi_i\}_{i=1}^n$ be bootstrap weights independent of data. Define the feasible bootstrap loss by
\begin{align*} 
	\widehat{\mathcal L}_{w,\tau}^{*}(q)=\frac1n\sum_{i=1}^n\xi_i\,\widehat\ell_{w,\tau}(B_i;q),
\end{align*}
where $\widehat\ell_{w,\tau}$ denotes the loss by replacing the nuisance functions and pre-estimated quantile index in $\ell_{w,\tau}^{0}$ with their cross-fitted estimators. Define its oracle bootstrap counterpart:
\begin{align*}
	\mathcal L_{w,\tau,n}^{0,*}(q;q_{w,\tau},\eta_0)
	=
	\frac1n\sum_{i=1}^n
	\xi_i\,
	\ell_{w,\tau}^{0}(B_i;q,q_{w,\tau},\eta_0).
\end{align*}
Accordingly, define the feasible and oracle bootstrap local objectives as
\begin{align*}
	L_n^*(u,\tau)
	&=n\left[\widehat{\mathcal L}_{w,\tau}^{*}\left(q_{w,\tau}+\frac{u}{\sqrt n}\right)-\widehat{\mathcal L}_{w,\tau}^{*}(q_{w,\tau})\right],\\
	L_n^{0,*}(u,\tau)&=n\left[\mathcal L_{w,\tau,n}^{0,*}\left(q_{w,\tau}+\frac{u}{\sqrt n};
	q_{w,\tau},\eta_0\right)-\mathcal L_{w,\tau,n}^{0,*}(q_{w,\tau};q_{w,\tau},\eta_0)\right].
\end{align*}
By the same decomposition as in Lemma \ref{lem:dml-rate-oracle-equivalence}, together with the cross-fitting construction and the first-stage rate conditions, we obtain the bootstrap local oracle equivalence
\begin{align}
	&\sup_{\tau\in\Upsilon}\sup_{|u|\le M}
	\Bigg|n\left[\widehat{\mathcal L}_{w,\tau}^*\left(q_{w,\tau}+\frac{u}{\sqrt n}\right)-\widehat{\mathcal L}_{w,\tau}^*(q_{w,\tau})\right]\notag\\
	&\qquad-n\left[\mathcal L_{w,\tau,n}^{0,*}\left(q_{w,\tau}+\frac{u}{\sqrt n};
	q_{w,\tau},\eta_0\right)-\mathcal L_{w,\tau,n}^{0,*}(q_{w,\tau};q_{w,\tau},\eta_0)
	\right]\Bigg|=o_p^*(1).\label{eq:bootstrap_local_equivalence}
\end{align}
since the difference between the feasible and oracle bootstrap scores is bounded by
\begin{align*}
	&\sup_{\tau\in\Upsilon}\sup_{|v|\le M}\left|\frac{1}{\sqrt n}\sum_{k=1}^K\sum_{i\in\mathcal I_k}
	\xi_i\left[\Phi_w\left(B_i;q_{w,\tau}+\frac{v}{\sqrt n},
	\widetilde q_{w,\tau}^{(-k)},\widehat\eta^{(-k)}\right)-\Phi_w\left(B_i;q_{w,\tau}+\frac{v}{\sqrt n},
	q_{w,\tau},\eta_0\right)\right]\right|
	\\
	&\leq\sup_{\tau\in\Upsilon}\sup_{|v|\le M}\left|\frac{1}{\sqrt n}\sum_{k=1}^K\sum_{i\in\mathcal I_k}
	\left[\Phi_w\left(B_i;q_{w,\tau}+\frac{v}{\sqrt n},\widetilde q_{w,\tau}^{(-k)},\widehat\eta^{(-k)}\right)-\Phi_w\left(B_i;q_{w,\tau}+\frac{v}{\sqrt n},q_{w,\tau},\eta_0\right)\right]\right|
	\\
	&\quad+\sup_{\tau\in\Upsilon}\sup_{|v|\le M}\left|\frac{1}{\sqrt n}\sum_{k=1}^K\sum_{i\in\mathcal I_k}
	(\xi_i-1)\left[\Phi_w\left(
	B_i;q_{w,\tau}+\frac{v}{\sqrt n},\widetilde q_{w,\tau}^{(-k)},
	\widehat\eta^{(-k)}\right)-\Phi_w\left(B_i;q_{w,\tau}+\frac{v}{\sqrt n},
	q_{w,\tau},\eta_0\right)\right]\right|,
\end{align*}
where the first term is $o_p(1)$ by Lemma \ref{lem:dml-rate-oracle-equivalence}. For the second term,
conditional on each training sample, Assumption \ref{as:uniform_score_complexity} and maximal inequality \citep[Corollary 5.1]{chernozhukovetal2014} yield
\begin{align*}
O_p\left(r_n\sqrt{J_n}+\frac{J_n\log(n\vee e)}{\sqrt n}\right)=o_p(1),
\end{align*}
where $J_n=v_n\log(A_n/r_n)$. This joint-probability bound further implies $o_p^*(1)$ conditional on the full sample by Markov's inequality.
 
Next, consider the oracle bootstrap objective. Knight's identity gives
\begin{align*}
	L_n^{0,*}(u,\tau)
	&=u\frac1{\sqrt n}\sum_{i=1}^n\xi_i\left[\Phi_w(B_i;q_{w,\tau},q_{w,\tau},\eta_0)\right]\notag\\
	&+\sum_{i=1}^n\xi_i\frac{G_i}{\pi}\frac{\mathbf 1(W_i=w)}{\rho_w(S_i,X_i)}
	\int_0^{u/\sqrt n}\left[\mathbf 1(Y_i-q_{w,\tau}\le v)-\mathbf 1(Y_i-q_{w,\tau}\le0)\right]dv.
\end{align*}
in which the second term is the sum of its unweighted counterpart and
\begin{align*}
	\sum_{i=1}^n(\xi_i-1)\frac{G_i}{\pi}
	\frac{\mathbf 1(W_i=w)}{\rho_w(S_i,X_i)}
	\int_0^{u/\sqrt n}
	\left[\mathbf 1(Y_i-q_{w,\tau}\le v)
	-\mathbf 1(Y_i-q_{w,\tau}\le0)\right]dv.
\end{align*}
By the uniform quadratic approximation established in Step~(1) of Theorem \ref{thm:integratedM}, the unweighted counterpart is given by
\begin{align*}
	\frac12 f_{Y(w)\vert G=1}(q_{w,\tau})u^2+o_p(1)
\end{align*}
uniformly over $\tau\in\Upsilon$ and $|u|\le M$. For the centered multiplier component, the overlap condition (Assumption \ref{as:overlap}) gives an envelope of order $M/\sqrt n$ and the bounded conditional-density condition (Assumption \ref{as:smoothness}) gives a maximal second moment of order $M^3n^{-3/2}$ for each summand, uniformly over $\tau\in\Upsilon$ and $|u|\le M$. The indicator class is VC-type. Hence, the conditional multiplier maximal inequality \citep[Theorem 5.2]{chernozhukovetal2014} yields that the multiplier term is $o_p^*(1)$, uniformly over $\tau\in\Upsilon$ and $|u|\le M$, in probability. It follows that
\begin{align}
	L_n^{0,*}(u,\tau)=u\frac1{\sqrt n}\sum_{i=1}^n\xi_i
	\left[\Phi_w(B_i;q_{w,\tau},q_{w,\tau},\eta_0)\right]+\frac12 f_{Y(w)\vert G=1}(q_{w,\tau})u^2+o_p^*(1),
	\label{eq:linearization_bootstrap}
\end{align}
uniformly over $\tau\in\Upsilon$ and $|u|\le M$, in probability. Combining
\eqref{eq:bootstrap_local_equivalence} and
\eqref{eq:linearization_bootstrap} gives
\begin{align*}
	L_n^*(u,\tau)=u\frac1{\sqrt n}\sum_{i=1}^n\xi_i\left[\Phi_w(B_i;q_{w,\tau},q_{w,\tau},\eta_0)\right]+\frac12 f_{Y(w)\vert G=1}(q_{w,\tau})u^2+o_p^*(1),
\end{align*}
uniformly over $\tau\in\Upsilon$ and $|u|\le M$, in probability.

The joint score class for $w=1$ and $w=0$ is Donsker by the finite-dimensional convergence, total boundedness, and stochastic equicontinuity established in Steps~(2)--(3) of Theorem \ref{thm:integratedM}. Since the weights are independent of the data, have mean one and variance one, and satisfy the stated tail condition, the conditional multiplier central limit theorem gives
\begin{align*}
	\begin{pmatrix}
		\displaystyle\frac1{\sqrt n}\sum_{i=1}^n(\xi_i-1)
		[\Phi_1(B_i;q_{1,\tau},q_{1,\tau},\eta_0)]\\[6pt]
		\displaystyle\frac1{\sqrt n}\sum_{i=1}^n(\xi_i-1)
		[\Phi_0(B_i;q_{0,\tau},q_{0,\tau},\eta_0)]
	\end{pmatrix}
	\underset{\xi}{\overset{\mathbb P}{\rightsquigarrow}}
	\begin{pmatrix}
		\mathbb S_1(\tau)\\
		\mathbb S_0(\tau)
	\end{pmatrix}
	~~\text{in }L_\infty(\Upsilon),
\end{align*}
in probability. 

The conditional multiplier central limit theorem also gives conditional tightness of the linear term in the local objective. Because the multiplier weights are nonnegative, $L_n^*(\cdot,\tau)$ is convex. Moreover,
$f_{Y(w)\vert G=1}(q_{w,\tau})$ is uniformly bounded away from zero and infinity. Thus, the uniform local approximation above and the convexity lemma \citep{kato2009} imply both
\begin{align*}
	\sup_{\tau\in\Upsilon}\left|\sqrt n(\widehat q_{w,\tau}^*-q_{w,\tau})\right|=O_p^*(1)
\end{align*}
and
\begin{align*}
	\sqrt n(\widehat q_{w,\tau}^*-q_{w,\tau})=-\frac{1}{f_{Y(w)\vert G=1}(q_{w,\tau})}
	\frac1{\sqrt n}\sum_{i=1}^n\xi_i\left[\Phi_w(B_i;q_{w,\tau},q_{w,\tau},\eta_0)\right]+o_p^*(1),
\end{align*}
uniformly over $\tau\in\Upsilon$, in probability. 

The corresponding expansion for the original estimator is
\begin{align*}
	\sqrt n(\widehat q_{w,\tau}-q_{w,\tau})=-\frac{1}{f_{Y(w)\vert G=1}(q_{w,\tau})}
	\frac1{\sqrt n}\sum_{i=1}^n\left[\Phi_w(B_i;q_{w,\tau},q_{w,\tau},\eta_0)\right]+o_p(1),
\end{align*}
uniformly over $\tau\in\Upsilon$. Subtracting the last two expressions yields
\begin{align*}
	\sqrt n(\widehat q_{w,\tau}^*-\widehat q_{w,\tau})=-\frac{1}{f_{Y(w)\vert G=1}(q_{w,\tau})}
	\frac1{\sqrt n}\sum_{i=1}^n(\xi_i-1)\left[\Phi_w(B_i;q_{w,\tau},q_{w,\tau},\eta_0)\right]+o_p^*(1),
\end{align*}
uniformly over $\tau\in\Upsilon$, in probability. Therefore, it follows that
\begin{align*}
	\sqrt n(\widehat\Delta_\tau^*-\widehat\Delta_\tau)=\frac1{\sqrt n}\sum_{i=1}^n(\xi_i-1)\psi_\Delta^M(B_i;\tau)+o_p^*(1),
\end{align*}
uniformly over $\tau\in\Upsilon$, in probability.

Finally, the conditional multiplier central limit theorem applied to the
Donsker class $\{\psi_\Delta^M(\cdot;\tau):\tau\in\Upsilon\}$ gives
\begin{align*}
	\frac1{\sqrt n}\sum_{i=1}^n(\xi_i-1)\psi_\Delta^M(B_i;\tau)
	\underset{\xi}{\overset{\mathbb P}{\rightsquigarrow}}
	\mathcal B(\tau)~~\text{in }L_\infty(\Upsilon),
\end{align*}
in probability, where $\mathcal B$ is the same tight mean-zero Gaussian process as in the weak limit of the original LTQTE estimator with the covariance kernel as
\begin{align*}
	\mathbb E\left[\psi_\Delta^M(B;\tau)\psi_\Delta^M(B;\tau')\right].
\end{align*}
The conditional Slutsky theorem now implies
\begin{align*}
	\sqrt n(\widehat\Delta_\tau^*-\widehat\Delta_\tau)\underset{\xi}{\overset{\mathbb P}{\rightsquigarrow}}\mathcal B(\tau)~~\text{in }L_\infty(\Upsilon),
\end{align*}
in probability. The proof is complete. $\blacksquare$

\bigskip

\noindent\textbf{Proof of Theorem \ref{thm:pointwise_ci}.}
Following Footnote~\ref{remark1}, let $\mathrm{BL}_1$ denote the class of all functions $h:L_\infty(\Upsilon)\to[0,1]$ satisfying
\begin{align*}
	|h(f)-h(g)|\leq \sup_{\tau\in\Upsilon}|f(\tau)-g(\tau)|~~\text{for all }f,g\in L_\infty(\Upsilon).
\end{align*}
Theorem~\ref{thm:multiplier-bootstrap-integrated} implies
\begin{align}
	\sup_{h\in\mathrm{BL}_1}
	\Bigg|\mathbb E_\xi\!\left[h\!\left(\left\{\sqrt n(\widehat\Delta_\tau^*-\widehat\Delta_\tau):\tau\in\Upsilon\right\}\right)\right]-\mathbb E\!\left[h\!\left(\left\{\mathcal B(\tau):\tau\in\Upsilon\right\}\right)\right]\Bigg|\rightarrow_p\ 0,
	\label{eq:process-bl-convergence}
\end{align}
where both arguments of $h$ are the entire processes and $\mathbb E_\xi$ denotes expectation with respect to the multiplier weights conditional on $\mathcal D_n$.

Fix $0<\varepsilon<1$, $\tau\in\Upsilon$, and $z\in\mathbb R$. For every
$f\in L_\infty(\Upsilon)$, define
\begin{align*}
	\underline h_{\tau,z,\varepsilon}(f)
	&:=
	\begin{cases}
		1,
		& f(\tau)\leq z-\varepsilon,\\[2pt]
		\dfrac{z-f(\tau)}{\varepsilon},
		& z-\varepsilon<f(\tau)<z,\\[7pt]
		0,
		& f(\tau)\geq z,
	\end{cases}\\
	\overline h_{\tau,z,\varepsilon}(f)
	&:=
	\begin{cases}
		1,
		& f(\tau)\leq z,\\[2pt]
		\dfrac{z+\varepsilon-f(\tau)}{\varepsilon},
		& z<f(\tau)<z+\varepsilon,\\[7pt]
		0,
		& f(\tau)\geq z+\varepsilon.
	\end{cases}
\end{align*}
These functions satisfy
\begin{align}
	\underline h_{\tau,z,\varepsilon}(f)
	\leq \mathbf 1\{f(\tau)\leq z\}
	\leq \overline h_{\tau,z,\varepsilon}(f)
	\qquad\text{for every }f\in L_\infty(\Upsilon).
	\label{eq:indicator-bracket}
\end{align}
Moreover, both functions take values in $[0,1]$. For $h^\circ_{\tau,z,\varepsilon}\in \{\underline h_{\tau,z,\varepsilon},\overline h_{\tau,z,\varepsilon}\}$ and all $f,g\in L_\infty(\Upsilon)$,
\begin{align}
	\left|h^\circ_{\tau,z,\varepsilon}(f)-h^\circ_{\tau,z,\varepsilon}(g)\right|\leq \varepsilon^{-1}|f(\tau)-g(\tau)|\leq \varepsilon^{-1}\sup_{\widetilde\tau\in\Upsilon}|f(\widetilde\tau)-g(\widetilde\tau)|.
	\label{eq:envelope-lipschitz}
\end{align}
Because $0<\varepsilon<1$, both $\varepsilon\underline h_{\tau,z,\varepsilon}$ and $\varepsilon\overline h_{\tau,z,\varepsilon}$ take values in $[0,1]$ and are Lipschitz. Hence, they belong to $\mathrm{BL}_1$. It follows that, for either choice of $h^\circ_{\tau,z,\varepsilon}$,
\begin{align}
	&\Bigg|
	\mathbb E_\xi\!\left[h^\circ_{\tau,z,\varepsilon}\!\left(\left\{\sqrt n(\widehat\Delta_{\widetilde\tau}^*-\widehat\Delta_{\widetilde\tau}):\widetilde\tau\in\Upsilon\right\}\right)\right]-\mathbb E\!\left[h^\circ_{\tau,z,\varepsilon}\!\left(\left\{\mathcal B(\widetilde\tau):\widetilde\tau\in\Upsilon\right\}\right)\right]\Bigg|
	\notag\\
	&\leq\varepsilon^{-1}\sup_{h\in\mathrm{BL}_1}\Bigg|\mathbb E_\xi\!\left[h\!\left(\left\{\sqrt n(\widehat\Delta_{\widetilde\tau}^*-\widehat\Delta_{\widetilde\tau}):\widetilde\tau\in\Upsilon\right\}\right)\right]-\mathbb E\!\left[h\!\left(\left\{\mathcal B(\widetilde\tau):\widetilde\tau\in\Upsilon\right\}\right)\right]\Bigg|.
	\label{eq:scaled-envelope-bound}
\end{align}
By \eqref{eq:indicator-bracket} and \eqref{eq:scaled-envelope-bound},
\begin{align*}
	&\Pr\!\left(
	\sqrt n(\widehat\Delta_\tau^*-\widehat\Delta_\tau)\leq z\mid \mathcal D_n\right)
	\notag\\
	&\quad\leq\mathbb E_\xi\!\left[\overline h_{\tau,z,\varepsilon}\!\left(\left\{\sqrt n(\widehat\Delta_{\widetilde\tau}^*-\widehat\Delta_{\widetilde\tau}):\widetilde\tau\in\Upsilon\right\}\right)\right]\notag\\
	&\leq\Pr\!\left(\mathcal B(\tau)\leq z+\varepsilon\right)+\varepsilon^{-1}\sup_{h\in\mathrm{BL}_1}\Bigg|\mathbb E_\xi\!\left[h\!\left(\left\{\sqrt n(\widehat\Delta_{\widetilde\tau}^*-\widehat\Delta_{\widetilde\tau}):\widetilde\tau\in\Upsilon\right\}\right)\right]-\mathbb E\!\left[h\!\left(\left\{\mathcal B(\widetilde\tau):\widetilde\tau\in\Upsilon\right\}\right)\right]\Bigg|,
\end{align*}
where the last inequality also uses $\overline h_{\tau,z,\varepsilon}(f) \leq\mathbf 1\{f(\tau)\leq z+\varepsilon\}$. Similarly,
\begin{align*}
	&\Pr\!\left(\sqrt n(\widehat\Delta_\tau^*-\widehat\Delta_\tau)\leq z\mid \mathcal D_n\right)
	\notag\\
	&\geq\Pr\!\left(\mathcal B(\tau)\leq z-\varepsilon\right)-\varepsilon^{-1}\sup_{h\in\mathrm{BL}_1}\Bigg|\mathbb E_\xi\!\left[h\!\left(\left\{\sqrt n(\widehat\Delta_{\widetilde\tau}^*-\widehat\Delta_{\widetilde\tau}):\widetilde\tau\in\Upsilon\right\}\right)\right]-\mathbb E\!\left[h\!\left(\left\{\mathcal B(\widetilde\tau):\widetilde\tau\in\Upsilon\right\}\right)\right]\Bigg|,
\end{align*}
because $\underline h_{\tau,z,\varepsilon}(f)\geq\mathbf 1\{f(\tau)\leq z-\varepsilon\}$. Combining the above two bounds yields
\begin{align}
	&\sup_{\tau\in\Upsilon}\sup_{z\in\mathbb R}\left|\Pr\!\left(\sqrt n(\widehat\Delta_\tau^*-\widehat\Delta_\tau)\leq z\mid \mathcal D_n \right)-\Pr\!\left(\mathcal B(\tau)\leq z\right)\right|\notag\\
	&\leq\varepsilon^{-1}\sup_{h\in\mathrm{BL}_1}\Bigg|\mathbb E_\xi\!\left[h\!\left(\left\{\sqrt n(\widehat\Delta_\tau^*-\widehat\Delta_\tau):\tau\in\Upsilon\right\}\right)\right]-\mathbb E\!\left[
	h\!\left(\left\{\mathcal B(\tau):\tau\in\Upsilon\right\}\right)\right]\Bigg|+\sup_{\tau\in\Upsilon}\sup_{z\in\mathbb R}\Pr\!\left(z-\varepsilon<\mathcal B(\tau)\leq z+\varepsilon\right).
	\label{eq:first-smoothing-bound}
\end{align}
For every $\tau\in\Upsilon$, $\mathcal B(\tau)$ is Gaussian with variance $\Sigma_\Delta(\tau,\tau)$. Its density is bounded above by
\begin{align*}
	\frac{1}{\sqrt{2\pi}\,\Sigma_\Delta(\tau,\tau)^{1/2}}\leq\frac{1}{\sqrt{2\pi}\inf_{\widetilde\tau\in\Upsilon}\Sigma_\Delta(\widetilde\tau,\widetilde\tau)^{1/2}}.
\end{align*}
Therefore, the length of the interval $(z-\varepsilon,z+\varepsilon]$ gives
\begin{align}
	\sup_{\tau\in\Upsilon}\sup_{z\in\mathbb R}\Pr\!\left(z-\varepsilon<\mathcal B(\tau)\leq z+\varepsilon\right)\leq\frac{2\varepsilon}{\sqrt{2\pi}\inf_{\tau\in\Upsilon}\Sigma_\Delta(\tau,\tau)^{1/2}}.
	\label{eq:gaussian-anticoncentration}
\end{align}
For any fixed $\varepsilon$, the bounded-Lipschitz term in \eqref{eq:first-smoothing-bound} converges to zero in probability by \eqref{eq:process-bl-convergence}. Given any positive constant, first choose
$\varepsilon$ small enough and then let $n\to\infty$. This proves
\begin{align}
	\sup_{\tau\in\Upsilon}\sup_{z\in\mathbb R}
	\Bigg|\Pr\!\left(\sqrt n(\widehat\Delta_\tau^*-\widehat\Delta_\tau)\leq z\mid \mathcal D_n \right)-\Phi\!\left(\frac{z}{\Sigma_\Delta(\tau,\tau)^{1/2}}\right)\Bigg|\rightarrow_p\ 0.
	\label{eq:uniform-polya}\tag{10}
\end{align}
It remains to justify uniform inversion of the conditional distribution functions. By the definition of the conditional quantile,
\begin{align*}
	\sqrt n\{\widehat{\mathcal C}_\tau(\nu)-\widehat\Delta_\tau\}=\inf\Bigg\{z:\Pr\!\left(\sqrt n(\widehat\Delta_\tau^*-\widehat\Delta_\tau)\leq z\mid \mathcal D_n \right)\geq\nu\Bigg\}.
\end{align*}
Under \eqref{eq:nondegeneracy}, for every $\delta>0$ and $\nu\in\{0.025,0.5,0.975\}$, we have
\begin{align*}
	\inf_{\tau\in\Upsilon}\left[\Phi\!\left(z_\nu+\frac{\delta}{\Sigma_\Delta(\tau,\tau)^{1/2}}\right)-\nu\right]\geq\Phi\!\left(z_\nu+\frac{\delta}{\sup_{\tau\in\Upsilon}\Sigma_\Delta(\tau,\tau)^{1/2}}\right)-\nu>0,
\end{align*}
and
\begin{align*}
	\inf_{\tau\in\Upsilon}\left[\nu-\Phi\!\left(z_\nu-\frac{\delta}{\Sigma_\Delta(\tau,\tau)^{1/2}}\right)\right]\geq\nu-\Phi\!\left(z_\nu-\frac{\delta}{\sup_{\tau\in\Upsilon}\Sigma_\Delta(\tau,\tau)^{1/2}}\right)>0.
\end{align*}
Consequently, \eqref{eq:uniform-polya} implies that, with probability
approaching one, simultaneously for all $\tau\in\Upsilon$,
\begin{align*}
	&\Pr\!\left(\sqrt n(\widehat\Delta_\tau^*-\widehat\Delta_\tau)\leq z_\nu\Sigma_\Delta(\tau,\tau)^{1/2}-\delta\mid \mathcal D_n \right)<\nu,\\
	&\Pr\!\left(\sqrt n(\widehat\Delta_\tau^*-\widehat\Delta_\tau)\leq z_\nu\Sigma_\Delta(\tau,\tau)^{1/2}+\delta\mid \mathcal D_n \right)>\nu.
\end{align*}
The definition of the generalized inverse then gives
\begin{align}
	\sup_{\tau\in\Upsilon}\left|\sqrt n(\widehat{\mathcal C}_\tau(\nu)-\widehat\Delta_\tau)-z_\nu\Sigma_\Delta(\tau,\tau)^{1/2}\right|\rightarrow_p0
	\label{eq:uniform-quantile-consistency}
\end{align}
for $\nu\in\{0.025,0.5,0.975\}$.

Applying \eqref{eq:uniform-quantile-consistency} with $\nu=0.975$ and $\nu=0.025$, subtracting, and dividing by $z_{0.975}-z_{0.025}>0$ proves \eqref{eq:uniform-scale-consistency}. Indeed,
\begin{align*}
	\sqrt n\,\widehat\sigma_\tau
	-\Sigma_\Delta(\tau,\tau)^{1/2}
	&=
	\frac{
		\sqrt n\{\widehat{\mathcal C}_\tau(0.975)-\widehat\Delta_\tau\}
		-z_{0.975}\Sigma_\Delta(\tau,\tau)^{1/2}}
	{z_{0.975}-z_{0.025}}-
	\frac{
		\sqrt n\{\widehat{\mathcal C}_\tau(0.025)-\widehat\Delta_\tau\}
		-z_{0.025}\Sigma_\Delta(\tau,\tau)^{1/2}}
	{z_{0.975}-z_{0.025}},
\end{align*}
so the supremum of the absolute value of the left-hand side over $\tau\in\Upsilon$ converges to zero in probability. 
Furthermore, by \eqref{eq:nondegeneracy},
\begin{align*}
	   \sup_{\tau\in\Upsilon}\left|\sqrt n\,\widehat\sigma_\tau\Sigma_\Delta(\tau,\tau)^{-1/2}-1\right|
		\leq\frac{\sup_{\tau\in\Upsilon}\left|\sqrt n\,\widehat\sigma_\tau-\Sigma_\Delta(\tau,\tau)^{1/2}\right|}{\inf_{\tau\in\Upsilon}\Sigma_\Delta(\tau,\tau)^{1/2}}\rightarrow_p0,
\end{align*}
which proves \eqref{eq:uniform-scale-consistency}.

Finally, for every fixed $\tau\in\Upsilon$, Theorem~\ref{thm:integratedM} gives
\begin{align*}
	\sqrt n(\widehat\Delta_\tau-\Delta_\tau)\rightsquigarrow \mathcal N(0,\Sigma_\Delta(\tau,\tau)),
\end{align*}
whereas \eqref{eq:uniform-scale-consistency} gives
\begin{align*}
	\sqrt n\,\widehat\sigma_\tau\rightarrow_p\Sigma_\Delta(\tau,\tau)^{1/2}>0.
\end{align*}
Slutsky's theorem therefore yields
\begin{align*}
	\frac{\widehat\Delta_\tau-\Delta_\tau}{\widehat\sigma_\tau}=\frac{\sqrt n(\widehat\Delta_\tau-\Delta_\tau)}{\sqrt n\,\widehat\sigma_\tau}\rightsquigarrow \mathcal N(0,1).
\end{align*}
Under $\mathcal H_{0,\tau}$, this establishes \eqref{eq:pointwise-normality}. Since the standard normal distribution is continuous at $z_{\alpha/2}$ and $z_{1-\alpha/2}$, the definitions of pointwise test and confidence interval give \eqref{eq:pointwise-size} and \eqref{eq:pointwise-coverage}. The proof is complete. $\blacksquare$

\bigskip

\noindent\textbf{Proof of Theorem \ref{thm:uniform}}: Because $z_{0.5}=0$, \eqref{eq:uniform-quantile-consistency} gives
\begin{align}
	\sup_{\tau\in\Upsilon}\sqrt n\left|\widetilde\Delta_\tau^*-\widehat\Delta_\tau\right|\rightarrow_p0.
	\label{eq:median-centering}
\end{align}
Next, \eqref{eq:nondegeneracy} and \eqref{eq:uniform-scale-consistency} imply
\begin{align}
	\Pr\!\left(\inf_{\tau\in\Upsilon}\sqrt n\,\widehat\sigma_\tau>\frac{1}{2}\inf_{\tau\in\Upsilon}\Sigma_\Delta(\tau,\tau)^{1/2}\right)\rightarrow1.
	\label{eq:estimated-scale-positive}
\end{align}
On the event in \eqref{eq:estimated-scale-positive},
\begin{align*}
	&\sup_{\tau\in\Upsilon}\left|\frac{1}{\sqrt n\,\widehat\sigma_\tau}-\frac{1}{\Sigma_\Delta(\tau,\tau)^{1/2}}\right|\leq\frac{2}{(\inf_{\tau\in\Upsilon}\Sigma_\Delta(\tau,\tau)^{1/2})^{2}}\cdot\sup_{\tau\in\Upsilon}\left|\sqrt n\,\widehat\sigma_\tau-\Sigma_\Delta(\tau,\tau)^{1/2}\right|\rightarrow_p0.
\end{align*}
The conditional weak convergence in Theorem~\ref{thm:multiplier-bootstrap-integrated} also implies conditional asymptotic tightness. Thus, for every $\varepsilon>0$,
\begin{align}
	\lim_{M\to\infty}\limsup_{n\to\infty}\Pr\!\Bigg\{\Pr\!\left(\sup_{\tau\in\Upsilon}\sqrt n|\widehat\Delta_\tau^*-\widehat\Delta_\tau|>M\ \middle|\ \mathcal D_n\right)>\varepsilon\Bigg\}=0.
	\label{eq:conditional-tightness}
\end{align}
Since the estimated standard deviations are positive, the triangle inequality gives
\begin{align*}
	&\sup_{\tau\in\Upsilon}\left|\frac{\widehat\Delta_\tau^*-\widetilde\Delta_\tau^*}{\widehat\sigma_\tau}-\frac{\sqrt n(\widehat\Delta_\tau^*-\widehat\Delta_\tau)}{\Sigma_\Delta(\tau,\tau)^{1/2}}\right|\notag\\
	&\leq\sup_{\tau\in\Upsilon}\sqrt n|\widehat\Delta_\tau^*-\widehat\Delta_\tau|\sup_{\tau\in\Upsilon}\left|\frac{1}{\sqrt n\,\widehat\sigma_\tau}-\frac{1}{\Sigma_\Delta(\tau,\tau)^{1/2}}\right|+\frac{\sup_{\tau\in\Upsilon}\sqrt n|\widetilde\Delta_\tau^*-\widehat\Delta_\tau|}{\inf_{\tau\in\Upsilon}\sqrt n\,\widehat\sigma_\tau}.
\end{align*}
It follows from \eqref{eq:median-centering}--\eqref{eq:conditional-tightness} that, for every $\varepsilon>0$,
\begin{align*}
	\Pr\!\Bigg(\sup_{\tau\in\Upsilon}\left|\frac{\widehat\Delta_\tau^*-\widetilde\Delta_\tau^*}{\widehat\sigma_\tau}-\frac{\sqrt n(\widehat\Delta_\tau^*-\widehat\Delta_\tau)}{\Sigma_\Delta(\tau,\tau)^{1/2}}\right|>\varepsilon\ \Biggm|\ \mathcal D_n \Bigg)\rightarrow_p0.
\end{align*}
The map that divides the coordinate indexed by $\tau$ by $\Sigma_\Delta(\tau,\tau)^{1/2}$ is a bounded linear map on $L_\infty(\Upsilon)$ because of \eqref{eq:nondegeneracy}. The conditional continuous mapping and Slutsky theorems yield
\begin{align}
	\frac{\widehat\Delta_\tau^*-\widetilde\Delta_\tau^*}{\widehat\sigma_\tau}\underset{\xi}{\overset{\mathbb P}{\rightsquigarrow}}\frac{\mathcal B(\tau)}{\Sigma_\Delta(\tau,\tau)^{1/2}}~~~~\text{in }L_\infty(\Upsilon),
	\label{eq:studentized-bootstrap-process}
\end{align}
in probability.

For the original empirical process, Theorem~\ref{thm:integratedM} implies both
\begin{align*}
	\sqrt n(\widehat\Delta_\tau-\Delta_\tau)\rightsquigarrow \mathcal B(\tau),~~~~\text{in }L_\infty(\Upsilon)
\end{align*}
and $\sup_{\tau\in\Upsilon}\sqrt n|\widehat\Delta_\tau-\Delta_\tau|=O_p(1)$. Consequently,
\begin{align*}
	\sup_{\tau\in\Upsilon}\left|\frac{\widehat\Delta_\tau-\Delta_\tau}{\widehat\sigma_\tau}-\frac{\sqrt n(\widehat\Delta_\tau-\Delta_\tau)}{\Sigma_\Delta(\tau,\tau)^{1/2}}\right|\leq\sup_{\tau\in\Upsilon}
	\sqrt n|\widehat\Delta_\tau-\Delta_\tau|\cdot\sup_{\tau\in\Upsilon}\left|\frac{1}{\sqrt n\,\widehat\sigma_\tau}-\frac{1}{\Sigma_\Delta(\tau,\tau)^{1/2}}\right|\rightarrow_p0.
\end{align*}
Combining this bound with Theorem~\ref{thm:integratedM}, rather than with Theorem~\ref{thm:multiplier-bootstrap-integrated}, gives
\begin{align}
	\frac{\widehat\Delta_\tau-\Delta_\tau}{\widehat\sigma_\tau}\rightsquigarrow\frac{\mathcal B(\tau)}{\Sigma_\Delta(\tau,\tau)^{1/2}},~~~~~\text{in }L_\infty(\Upsilon).
	\label{eq:studentized-original-process}
\end{align} 

The supremum norm $\Vert\cdot\Vert_{\mathbb P,\infty}$ is Lipschitz continuous on $L_\infty(\Upsilon)$. Hence the conditional and ordinary continuous mapping theorems applied to
\eqref{eq:studentized-bootstrap-process} and
\eqref{eq:studentized-original-process} give weak convergence of the related absolute suprema. The standardized limit process is a tight, separable, centered Gaussian process with unit variances. Gaussian coupling \citep[Corollary~2.1]{chernozhukovetal2014} implies that
\begin{align*}
	\lim_{\varepsilon\downarrow0}\sup_{z\in\mathbb R}\Pr\!\left(z-\varepsilon<\sup_{\tau\in\Upsilon}\left|\frac{\mathcal B(\tau)}{\Sigma_\Delta(\tau,\tau)^{1/2}}\right|\leq z+\varepsilon\right)=0.
\end{align*}
The conditional and ordinary P\'{o}lya theorems \citep[Lemmas 2.4 and 2.5]{bucherkojadinovic2019} give
\begin{align}
	\sup_{z\in\mathbb R}
	\Bigg|\Pr\!\left(\sup_{\tau\in\Upsilon}\left|\frac{\widehat\Delta_\tau^*-\widetilde\Delta_\tau^*}{\widehat\sigma_\tau}\right|\leq z\ \middle|\ \mathcal D_n\right)-\Pr\!\left(\sup_{\tau\in\Upsilon}\left|\frac{\mathcal B(\tau)}{\Sigma_\Delta(\tau,\tau)^{1/2}}\right|\leq z\right)\Bigg|\rightarrow_p0
	\label{eq:bootstrap-sup-polya}
\end{align}
and
\begin{align}
	\sup_{z\in\mathbb R}\Bigg|\Pr\!\left(\sup_{\tau\in\Upsilon}\left|\frac{\widehat\Delta_\tau-\Delta_\tau}{\widehat\sigma_\tau}\right|\leq z\right)-\Pr\!\left(\sup_{\tau\in\Upsilon}\left|\frac{\mathcal B(\tau)}{\Sigma_\Delta(\tau,\tau)^{1/2}}\right|\leq z\right)\Bigg|\rightarrow0.
	\label{eq:original-sup-polya}
\end{align}

Since the limiting distribution function is continuous and strictly increasing at its $(1-\alpha)$-th quantile, inversion of the uniform convergence in \eqref{eq:bootstrap-sup-polya}, together with the definition of $\widehat c_{1-\alpha}$ in \eqref{eq:uniform_critical}, yields
\begin{align}
	\Bigg|\widehat c_{1-\alpha}-\inf\Bigg\{z:\Pr\!\left(\sup_{\tau\in\Upsilon}\left|\frac{\mathcal B(\tau)}{\Sigma_\Delta(\tau,\tau)^{1/2}}\right|\leq z\right)\geq1-\alpha\Bigg\}\Bigg|\rightarrow_p0.
	\label{eq:critical-value-consistency}
\end{align}
Under $\mathcal H_0$,
\begin{align*}
	T_n=\sup_{\tau\in\Upsilon}\left|\frac{\widehat\Delta_\tau-\Delta_\tau}{\widehat\sigma_\tau}\right|.
\end{align*}
Equations \eqref{eq:original-sup-polya} and \eqref{eq:critical-value-consistency} further give that
\begin{align*}
	\Pr(T_n\leq\widehat c_{1-\alpha})\rightarrow1-\alpha,
\end{align*}
proving \eqref{eq:uniform-size}. Finally, without imposing $\mathcal H_0$, the simultaneous-coverage event is
exactly
\begin{align*}
	\left\{
	\sup_{\tau\in\Upsilon}
	\left|
	\frac{\widehat\Delta_\tau-\Delta_\tau}
	{\widehat\sigma_\tau}
	\right|
	\leq\widehat c_{1-\alpha}
	\right\}.
\end{align*}
The same argument based on \eqref{eq:original-sup-polya} and \eqref{eq:critical-value-consistency} then proves \eqref{eq:uniform-coverage}. The proof is complete. $\blacksquare$

\bigskip
\subsection{Proofs for Section \ref{sec:np-first-stage-uniform}}

\noindent\textbf{Proof of Theorem \ref{thm:np-uniform-first-stage}}: 
Under Assumptions~\ref{as:iid} and \ref{as:np-sieve-logit}, Theorems~6.1 and 6.2 of \citet{belloni2017program} establish rates for the indicator-response linear-Lasso and logistic-Lasso regressions:
\begin{align*}
	&\max_{k\in[K]}\Vert\widehat\eta_{\mathrm{prop}}^{(-k)}-\eta_{\mathrm{prop}}\Vert_{\mathbb P,2}
	=O_p(r_n),\\
	&\max_{k,w}\sup_{q\in\mathcal Q_w}\Vert\widehat m_w^{(-k)}(\cdot;q)-m_w(\cdot;q)\Vert_{\mathbb P,2}=O_p(r_n).
\end{align*}
For the second-stage Lasso, conditional on the cross fitting sample, the fitted functions are fixed. 
{With the clipping rule, \eqref{eq:lasso_conditional_entropy} and \eqref{eq:lasso_path_second}, both established below, imply that the squared prediction-error class admits a uniformly bounded envelope and has covering numbers of order $O(n\varepsilon^-2)$.} Combining the established population $L_2$ rate with the conditional maximal inequality of
\citet[Corollary~5.1]{chernozhukovetal2014} yields
\begin{align*}
	\max_{k,w}\sup_{q\in\mathcal Q_w}
	\Vert\widehat m_w^{(-k)}(\cdot;q)-m_w(\cdot;q)
	\Vert_{\mathbb P_{n_{0w}^{(-k)}},2}=O_p(r_n).
\end{align*}
Treating the first-stage prediction error as another approximation error, we use the basic inequality argument of \citet[Supplementary Lemma~I.3]{belloni2017program}. The uniform score domination ensured by the penalty rule in Assumption~\ref{as:np-sieve-logit}(vii) yields uniform empirical prediction and $\ell_1$ estimation rates of $O_p(r_n)$ and $O_p(\sqrt{s_n}r_n)$, respectively. Combining these rates with the sparse eigenvalue conditions in Assumption~\ref{as:np-sieve-logit}(vi), the approximation-error bounds in Assumption \ref{as:np-sieve-logit}, clipping rule and overlap in Assumption \ref{as:overlap} gives
\begin{align*}
	\max_{k,w}\sup_{q\in\mathcal Q_w}\Vert\widehat{\overline m}_w^{(-k)}(\cdot;q)-\overline m_w(\cdot;q)\Vert_{\mathbb P,2}=O_p(r_n).
\end{align*}
The remaining proof follows from Theorems 6.1 and 6.2 of \cite{belloni2017program}. 

\emph{Proof of \eqref{eq:lasso_path_second}.}
Fix $(k,w)$. The first-stage response vector and second-stage pseudo-responses satisfy
\begin{align}
	\bigl(\mathbf1\{Y_i\le b\}:i\notin I_k,G_i=1,W_i=w\bigr)\ \mapsto\ \widehat m_w^{(-k)}(\cdot;b)\ \mapsto\ \widehat{\overline m}_w^{(-k)}(\cdot;b).
\end{align}
Under the stipulated tuning, penalty loadings and deterministic selection rules, the binary response vector uniquely determines both fitted functions. Since this vector changes only when the threshold crosses one of the $n_{1w}^{(-k)}$ observed outcomes, it takes at most $n_{1w}^{(-k)}+1$ distinct values, establishing \eqref{eq:lasso_path_second}.

\emph{Proof of \eqref{eq:lasso_conditional_entropy}.}
Write $D_w=\mathbf1\{W=w\}$, $p_1=\varrho$, $p_0=1-\varrho$, and
\begin{align*}
	A=\frac{GD_w}{\pi\rho_w},~~
	U=\frac G\pi,~~
	C_w=\frac{(1-G)\gamma D_w}{\pi(1-\gamma)p_w}.
\end{align*}
Their estimated counterparts are
\begin{align*}
	\widehat A=\frac{GD_w}{\widehat\pi^{(-k)}\widehat\rho_w^{(-k)}},~~\widehat U=\frac{G}{\widehat\pi^{(-k)}},~~\widehat C_w=\frac{(1-G)\widehat\gamma^{(-k)}D_w}{\widehat\pi^{(-k)}(1-\widehat\gamma^{(-k)})\widehat p_w^{(-k)}},
\end{align*}
where $\widehat p_1^{(-k)}=\widehat\varrho^{(-k)}$ and $\widehat p_0^{(-k)}=1-\widehat\varrho^{(-k)}$.
For simplicity, write
\begin{align*}
	m_u=m_w(\cdot;u),~~\overline m_u=\overline m_w(\cdot;u),~~\widehat m_u=\widehat m_w^{(-k)}(\cdot;u),~~\widehat{\overline m}_u=\widehat{\overline m}_w^{(-k)}(\cdot;u).
\end{align*}
Set $a=q_{w,\tau}$, $\widehat a=\widetilde q_{w,\tau}^{(-k)}$, $t=a+v/\sqrt n$ and $I_t=\mathbf1\{Y\le t\}$.

We establish the covering bound for every probability measure $Q$. Write
\begin{align*}
	\mathbf m_b=(m_w(\cdot;b),\overline m_w(\cdot;b)),~~\|\mathbf f\|_{Q,2}^2=\mathbb E_Q(f_1^2+f_2^2).
\end{align*}
Since both coordinates are conditional expectations of nested indicators, they are nondecreasing in $b$ and lie in $[0,1]$ \citep[Supplementary Lemma~L.2]{belloni2017program}. Thus, for $b<b'$,
\begin{equation*}
	\|\mathbf m_{b'}-\mathbf m_b\|_{Q,2}^2\le \mathbb E_Q[m_{b'}-m_b+\bar m_{b'}-\bar m_b].
\end{equation*}
Let $b_1<...<b_J$ index a maximal $\delta$-separated subset of $\{\mathbf m_b:b\in Q_w\}$ in $L_2(Q)$. By telescoping,
\begin{align*}
	(J-1)\delta^2\le\sum_{j=1}^{J-1}\|\mathbf m_{b_{j+1}}-\mathbf m_{b_j}\|_{Q,2}^2\le
	\mathbb E_Q[m_{b_J}-m_{b_1}+\bar m_{b_J}-\bar m_{b_1}]\le2.
\end{align*}
Maximality makes $\mathcal M_\delta$ an $L_2(Q)$
$\delta$-net for the oracle-pair class, with
\begin{align*}
	\mathcal N(\delta,\{\mathbf m_b\},L_2(Q))\le |\mathcal M_\delta|\le 1+2\delta^{-2}.
\end{align*}
Applying the same argument to the indicator class $\mathcal I=\{\mathbf 1(Y\le t):t\in\mathbb R\}$ yields an $L_2(Q)$ $\delta$-net $\mathcal I_\delta$. For the scalar class $[0,1]$, let $\mathcal T_\delta$ be a $\delta$-net under the absolute-value metric. Then,
\begin{align*}
	|\mathcal I_\delta|&\le 1+\delta^{-2},~~|\mathcal M_\delta|\le 1+2\delta^{-2},~~|\mathcal T_\delta|\le 1+\delta^{-1}.
\end{align*}
Fix $k,w,Q$ and condition on $\mathcal D_{-k}$. For a score with $p\in\mathcal P_{k,w}$, write the score identity as
\begin{align*}
	h_{n,k,w,\tau,v}= g_p + c_I I_t + \mathbf c_m^\prime \mathbf m_a + c_\tau\tau,
\end{align*}
where
\begin{align*}
	g_p&=(\widehat C_w-\widehat A)\widehat m_{\widehat a}
	+(\widehat U-\widehat C_w)\widehat{\overline m}_{\widehat a},~~c_I=\widehat A-A,~~
	\mathbf c_m=(A-C_w,C_w-U)^\prime,~~ c_\tau=U-\widehat U.
\end{align*}
Conditional on $\mathcal D_{-k}$, $g_p$ is determined by $p$, whereas $c_I,\mathbf c_m,c_\tau$ do not depend on $(\tau,v)$. Since $K$ is fixed and the training sample sizes are of order $n$, the sample proportions satisfy $\max_k|\widehat\pi^{(-k)}-\pi|=o_p(1)$. By Assumption~\ref{as:np-sieve-logit}(v), the clipping rules and this uniform consistency, there exists a deterministic $C_0\ge1$, independent of $n$, such that, on an event $\mathcal E_n$ with $\Pr(\mathcal E_n)\to1$,
	\begin{align*}
		\max_{k,w}\{\|A\|_{\mathbb P,\infty},\|U\|_{\mathbb P,\infty},\|C_w\|_{\mathbb P,\infty},
		\|\widehat A\|_{\mathbb P,\infty},\|\widehat U\|_{\mathbb P,\infty},\|\widehat C_w\|_{\mathbb P,\infty}\}\le C_0,~~\max_{k,w}\{\|c_I\|_{\mathbb P,\infty},\|\mathbf c_m\|_{\mathbb P,\infty,2},\|c_\tau\|_{\mathbb P,\infty}\}\le C_0,
	\end{align*}
	in which $\|\mathbf c_m\|_{\mathbb P,\infty,2}:=\sup_{B\in\mathcal B}\sqrt{c_{m,1}(B)^2+c_{m,2}(B)^2}$. Here $\mathcal E_n$ is the intersection of the corresponding $\mathcal D_{-k}$-measurable events over folds. Since the estimates are truncated to $[0,1]$, then
	\begin{align*}
		\Phi_w(B;t,\widehat a,\widehat\eta^{(-k)})
		&=\widehat A(I_t-\widehat m_{\widehat a})
		+\widehat U(\widehat{\overline m}_{\widehat a}-\tau)
		+\widehat C_w(\widehat m_{\widehat a}-\widehat{\overline m}_{\widehat a})
	\end{align*}
	and its oracle counterpart imply, on $\mathcal E_n$,
	\begin{align*}
		\max_{k,w}\sup_{h\in\mathcal H_{k,w}}\|h\|_{\mathbb P,\infty}\le 6C_0=:F_M.
\end{align*}

Let $\mathcal I_\delta$, $\mathcal M_\delta$ and $\mathcal T_\delta$ be the $\delta$-nets for the indicator, oracle-pair and scalar classes, respectively. For each fitted pair $p$, form the candidate centers
\begin{align*}
	g_p+c_I I^\star+\mathbf c_m^\prime\mathbf m^\star+c_\tau\tau^\star,~~(I^\star,\mathbf m^\star,\tau^\star)\in\mathcal I_\delta\times\mathcal M_\delta\times\mathcal T_\delta.
\end{align*}
To cover a given score $h$, use its own fitted pair $p$ and choose each starred term within $\delta$ of its counterpart. The resulting center $h^\star$ satisfies
\begin{align*}
	\|h-h^\star\|_{Q,2}\le C_0\bigl(\|I_t-I^\star\|_{Q,2}+\|\mathbf m_a-\mathbf m^\star\|_{Q,2}+|\tau-\tau^\star|
	\bigr)\le 3C_0\delta.
\end{align*}
By re-centering each nonempty candidate ball at a score in $\mathcal H_{n,k,w}(M)$ and applying the triangle inequality, we obtain
\begin{align*}
	\mathcal N(3C_0\delta,\mathcal H_{n,k,w}(M),L_2(Q))
	\le \#\mathcal P_{k,w}
	|\mathcal I_\delta|
	|\mathcal M_\delta|
	|\mathcal T_\delta|.
\end{align*}

The net sizes multiply within each fitted pair and the resulting cover sizes add over fitted pairs, as in
\citet[Supplementary Lemma~L.1(2)]{belloni2017program}. Taking $\delta=\varepsilon\min\{1,F_M/(6C_0)\}$ and using $\#\mathcal P_{k,w}\le n+1$ gives
\begin{align*}
	\sup_Q\mathcal N\bigl(\varepsilon F_M,\mathcal H_{n,k,w}(M),L_2(Q)
	\bigr)&\le \#\mathcal P_{k,w}(1+\delta^{-2})(1+2\delta^{-2})(1+\delta^{-1})\le C(n+1)\varepsilon^{-5},~~ 0<\varepsilon\le 1.
\end{align*}
This proves \eqref{eq:lasso_conditional_entropy}. In particular, the entropy part of Assumption~\ref{as:uniform_score_complexity} holds with $v_n=5$ and $A_n=\max\{e,[C(n+1)]^{1/5}\}$. The proof is then complete. $\blacksquare$

\bigskip

\noindent\textbf{Proof of Theorem \ref{thm:initial-ipw-quantile-rate}}:	We first show that the feasible IPW distribution estimator converges uniformly at rate $n^{-1/2}+r_n$. Define the infeasible empirical IPW distribution function
\begin{align*}
	\widehat F_w^0(q)=\mathbb E_n\left[\frac{G_iD_w\mathbf 1(Y_i\le q)}
	{\pi\rho_w(S_i,X_i)}\right]
\end{align*}
and the population IPW distribution function
\begin{align*}
	F_w(q)=\mathbb E\left[\frac{G_iD_w\mathbf 1(Y_i\le q)}{\pi\rho_w(S_i,X_i)}\right].
\end{align*}
By consistency and Assumptions~\ref{as:unconfoundedness}, \ref{as:external_validity}, and \ref{as:LUTM}, together with the positivity in Assumption~\ref{as:overlap}, the IPW identification argument underlying \eqref{eq:ipw0} and \eqref{eq:ipw1} gives, for every
$q\in\mathcal Q_w$,
\begin{align*}
	F_w(q)=F_{Y(w)\vert G=1}(q).
\end{align*}
Assumption~\ref{as:smoothness} ensures continuity at $q_{w,\tau}$; hence, by the definition of $q_{w,\tau}$, we obtain that
\begin{align*}
	F_w(q_{w,\tau})=\tau,~~w\in\{0,1\},~~\tau\in\Upsilon.
\end{align*}
For the infeasible estimator, consider the class
\begin{align*}
	\mathcal F_w=\left\{B\mapsto\frac{GD_w\mathbf 1(Y\le q)}{\pi\rho_w(S,X)}:q\in\mathcal Q_w\right\}.
\end{align*}
The threshold-indicator class is a VC-subgraph class of fixed VC dimension. Assumption~\ref{as:overlap}, together with the identifion of $\rho_w$ under Assumptions~\ref{as:unconfoundedness} and \ref{as:external_validity}, provides a bounded envelope for
$\mathcal F_w$. Therefore, under the i.i.d. sampling condition in Assumption~\ref{as:iid}, the maximal inequality of \citet[Corollary~5.1]{chernozhukovetal2014} yields
\begin{align}
	\label{eq:infeasible-ipw-cdf-rate}
	\max_{w\in\{0,1\}}\sup_{q\in\mathcal Q_w}\vert\widehat F_w^0(q)-F_w(q)\vert=O_p(n^{-1/2}).
\end{align}

We next control the difference between the feasible and infeasible IPW
distribution estimators. By Assumptions~\ref{as:overlap}
and \ref{as:np-sieve-logit}(v) and the
definition of $\widehat\rho_w$ in \eqref{eq:rho-identification},
\begin{align}
	\label{eq:rho-in-sample-lipschitz}
	|\widehat\rho_w(s,x)-\rho_w(s,x)|
	\le C\Big\{|\widehat e_w(s,x)-e_w(s,x)|
	+|\widehat h_w(x)-h_w(x)|
	+|\widehat t(x)-t(x)|\Big\}.
\end{align}
The fitted $\widehat e_w$, $\widehat h_w$ and $\widehat t$ are clipped to
$[c_*,1-c_*]$, which also contains their true counterparts and $\rho_w$.
The odds-ratio map in \eqref{eq:rho-identification} is uniformly
Lipschitz on $[c_*,1-c_*]^3$; since the final projection
$\Pi_{[c_*,1-c_*]}$ is nonexpansive, \eqref{eq:rho-in-sample-lipschitz} follows.

Under Assumptions~\ref{as:iid} and \ref{as:np-sieve-logit}(i), (ii), (v)--(ix), the five logistic regressions for $e_w$, $h_w$ and $t$ satisfy \citet[Assumption~6.2 and Algorithm~6.1]{belloni2017program}. In particular, (vii) specifies their penalty and loading rules, (ix) imposes the approximation-error and score-variance bounds. Conditional on the corresponding subsample, the observations are i.i.d. Then, their Theorem~6.2, together with the approximation-error bounds and clipping, yields
\begin{align}
	\label{eq:component-in-sample-bounds}
	&\max_{w\in\{0,1\}}\mathbb E_n\big[D_w(\widehat e_w(S,X)-e_w(S,X))^2\big]+\max_{w\in\{0,1\}}\mathbb E_n\big[D_w(\widehat h_w(X)-h_w(X))^2\big]+\mathbb E_n\big[G(\widehat t(X)-t(X))^2\big]
	=O_p(r_n^2),
\end{align}
Here the regressions for $e_w$ and $h_w$ use observations with $W=w$, while that for $t$ uses observations with $G=1$; Assumption~\ref{as:np-sieve-logit}(i) ensures that the sample sizes are of order $n$. 

By Assumption~\ref{as:overlap}, the fitted truncation $\widehat\rho_w\in[c_*,1-c_*]$ and root-$n$ rate of
$\widehat\pi$ under Assumption~\ref{as:iid}, on an event whose probability approaches one on which $\widehat\pi\ge\pi/2>0$, uniformly in $q\in\mathcal Q_w$,
\begin{align*}
	&|\widehat F_w(q)-\widehat F_w^0(q)|\\
	&\le C|\widehat\pi-\pi|
	+C\mathbb E_n\big[GD_w
	|\widehat\rho_w(S,X)-\rho_w(S,X)|\big]\\
	&\le C|\widehat\pi-\pi|
	+C\Big\{
	\big(\mathbb E_n[D_w(\widehat e_w-e_w)^2]\big)^{1/2}
	+\big(\mathbb E_n[D_w(\widehat h_w-h_w)^2]\big)^{1/2}
	+\big(\mathbb E_n[G(\widehat t-t)^2]\big)^{1/2}
	\Big\}\\
	&=O_p(n^{-1/2}+r_n),
\end{align*}
where the second inequality follows from \eqref{eq:rho-in-sample-lipschitz}, $GD_w\le D_w$, $GD_w\le G$ and Cauchy--Schwarz inequality and the last equality follows from \eqref{eq:component-in-sample-bounds}. Therefore, combining this bound with \eqref{eq:infeasible-ipw-cdf-rate} by the triangle inequality, we obtain
\begin{align}
	\label{eq:feasible-ipw-cdf-uniform-rate-main}
	\max_{w\in\{0,1\}}\sup_{q\in\mathcal Q_w}
		|\widehat F_w(q)-F_w(q)|
		=O_p(n^{-1/2}+r_n).
\end{align}

It remains to prove the uniform quantile rate. Since $q_{w,\tau}\in\mathcal Q_w$, the definition of
$\widetilde q_{w,\tau}$ gives
\begin{align*}
	|\widehat F_w(\widetilde q_{w,\tau})-\tau|\le|\widehat F_w(q_{w,\tau})-\tau
	|=|\widehat F_w(q_{w,\tau})-F_w(q_{w,\tau})|\le\max_{w\in\{0,1\}}
	\sup_{q\in\mathcal Q_w}|\widehat F_w(q)-F_w(q)|,
\end{align*}
in which the equality uses the identification in \eqref{eq:ipw0} and \eqref{eq:ipw1} and Assumption~\ref{as:smoothness}. The triangle inequality therefore implies
\begin{align}
	\label{eq:population-cdf-at-initial-q}
	&\max_{w\in\{0,1\}}\sup_{\tau\in\Upsilon}|F_w(\widetilde q_{w,\tau})-\tau|\le2\max_{w\in\{0,1\}}
	\sup_{q\in\mathcal Q_w}|\widehat F_w(q)-F_w(q)|=O_p(n^{-1/2}+r_n),
\end{align}
where the final equality follows from
\eqref{eq:feasible-ipw-cdf-uniform-rate-main}. 

Under Assumptions~\ref{as:unconfoundedness}--\ref{as:overlap}, IPW identification underlying \eqref{eq:ipw0}--\eqref{eq:ipw1} gives $F_w(q)=F_{Y(w)\mid G=1}(q)$ for $q\in\mathcal Q_w$.
By Assumption~\ref{as:smoothness}, the marginal density satisfies that $f_{Y(w)\mid G=1}(y)\ge c_f$ for every $y\in\mathcal Q_w^\delta(\Upsilon)$ and $w\in\{0,1\}$. Taking $c_F=c_f$, this identification and the monotonicity of $F_w$ imply, uniformly over $w\in\{0,1\}$, $\tau\in\Upsilon$, and $q\in\mathcal Q_w$,
\begin{align}
	\label{eq:local-strong-identification-ipw}
	|F_w(q)-F_w(q_{w,\tau})|
	&\ge \int_0^{\min\{|q-q_{w,\tau}|,\delta\}}
	f_{Y(w)\mid G=1}\big(q_{w,\tau}
	+\operatorname{sgn}(q-q_{w,\tau})t\big)\,dt \notag\\
	&\ge c_F\min\{|q-q_{w,\tau}|,\delta\}.
\end{align}
Since $F_w(q_{w,\tau})=\tau$ by Assumption~\ref{as:smoothness} and the IPW identification, and
$\widetilde q_{w,\tau}\in\mathcal Q_w$, \eqref{eq:local-strong-identification-ipw} and \eqref{eq:population-cdf-at-initial-q} yield
\begin{align*}
	&c_F\min\left\{
	\max_{w\in\{0,1\}}\sup_{\tau\in\Upsilon}
	|\widetilde q_{w,\tau}-q_{w,\tau}|,\delta\right\}\le
	\max_{w\in\{0,1\}}\sup_{\tau\in\Upsilon}
	|F_w(\widetilde q_{w,\tau})-\tau|\\
	&\qquad\le 2\max_{w\in\{0,1\}}
	\sup_{q\in\mathcal Q_w}|\widehat F_w(q)-F_w(q)|
	=O_p(n^{-1/2}+r_n)=o_p(1),
\end{align*}
where the $O_p$ bound follows from \eqref{eq:feasible-ipw-cdf-uniform-rate-main}. Therefore, for every fixed $\epsilon>0$,
\begin{align*}
	&\Pr\left(
	\max_{w\in\{0,1\}}\sup_{\tau\in\Upsilon}
	|\widetilde q_{w,\tau}-q_{w,\tau}|\ge\epsilon
	\right)\le\Pr\left(
	2\max_{w\in\{0,1\}}\sup_{q\in\mathcal Q_w}
	|\widehat F_w(q)-F_w(q)|
	\ge c_F\min\{\epsilon,\delta\}\right)\longrightarrow0.
\end{align*}
Thus, it follows that $\max_{w\in\{0,1\}}\sup_{\tau\in\Upsilon}|\widetilde q_{w,\tau}-q_{w,\tau}|=o_p(1)$.

Consequently, with probability approaching one, $|\widetilde q_{w,\tau}-q_{w,\tau}|\le\delta$ uniformly over $w\in\{0,1\}$ and $\tau\in\Upsilon$. On this event, \eqref{eq:local-strong-identification-ipw}, the identity $F_w(q_{w,\tau})=\tau$, and \eqref{eq:population-cdf-at-initial-q} give
\begin{align*}
	c_F|\widetilde q_{w,\tau}-q_{w,\tau}|\le|F_w(\widetilde q_{w,\tau})-F_w(q_{w,\tau})|=|F_w(\widetilde q_{w,\tau})-\tau|.
\end{align*}
Taking the supremum over $\tau\in\Upsilon$ and the maximum over $w\in\{0,1\}$ therefore yields
\begin{align*}
	\max_{w\in\{0,1\}}\sup_{\tau\in\Upsilon}|\widetilde q_{w,\tau}-q_{w,\tau}|=O_p(n^{-1/2}+r_n).
\end{align*}
This proves the first claim.

The same argument applies to each estimator $\widetilde q_{w,\tau}^{(-k)}$ constructed on the training sample outside $\mathcal I_k$. Because $K$ is fixed, the size of each training sample is proportional to $n$ and Assumption~\ref{as:np-sieve-logit} imposes the first-stage bounds uniformly over $1\le k\le K$, the above result hold uniformly over all the folds. Hence, it follows that
\begin{align*}
	\max_{1\le k\le K}\max_{w\in\{0,1\}}\sup_{\tau\in\Upsilon}|\widetilde q_{w,\tau}^{(-k)}-q_{w,\tau}|=O_p(n^{-1/2}+r_n).
\end{align*}
Finally, if $r_n\gtrsim n^{-1/2}$, then $n^{-1/2}+r_n\lesssim r_n$, yielding that
\begin{align*}
	\max_{1\le k\le K}\max_{w\in\{0,1\}}\sup_{\tau\in\Upsilon}|\widetilde q_{w,\tau}^{(-k)}-q_{w,\tau}|=O_p(r_n),
\end{align*}
which verifies \eqref{eq:initial-q-rate-simple} of Assumption~\ref{as:first-stage-rate}. The proof is complete. $\blacksquare$

\section{Implementation of Estimation and the Multiplier Bootstrap}
\label{sec:implementation}

This section collects the computational details for estimation and bootstrap inference.  All nuisance functions and initial quantiles used for observations in $\mathcal I_k$ are estimated from observations outside $\mathcal I_k$. The calculations below are repeated for every required value of $\tau$.

\subsection{Cross-fitted estimation}\label{sec:implementation1}

Write the nuisance estimates as
\begin{align*}
	\widehat\eta^{(-k)}=\big(\widehat m_0^{(-k)},\widehat m_1^{(-k)},\widehat{\overline m}_0^{(-k)},\widehat{\overline m}_1^{(-k)},\widehat\rho_0^{(-k)},\widehat\rho_1^{(-k)},\widehat\varrho^{(-k)},\widehat\gamma^{(-k)},
	\widehat\pi^{(-k)}\big),
\end{align*}
and let $\widetilde q_{1,\tau}^{(-k)}$ and $\widetilde q_{0,\tau}^{(-k)}$ denote the initial estimators of
$q_{1,\tau}$ and $q_{0,\tau}$, respectively. 

For $w=1$, the cross-fitted objective is
\begin{align}
	\widehat{\mathcal L}_{1,\tau}(q)
	=&\frac{1}{n}\sum_{k=1}^K\sum_{i\in\mathcal I_k}
	\Bigg[
	\frac{G_i}{\widehat\pi^{(-k)}}
	\frac{W_i}{\widehat\rho_1^{(-k)}(S_i,X_i)}
	\Big\{
	\rho_\tau(Y_i-q)
	+\big(\tau-\widehat m_1^{(-k)}
	(S_i,X_i;\widetilde q_{1,\tau}^{(-k)})\big)q
	\Big\}
	\notag\\
	&\quad+
	\frac{G_i}{\widehat\pi^{(-k)}}
	\Big\{
	\widehat{\overline m}_1^{(-k)}
	(X_i;\widetilde q_{1,\tau}^{(-k)})-\tau
	\Big\}q
	\notag\\
	&\quad+
	\frac{1-G_i}{\widehat\pi^{(-k)}}
	\frac{\widehat\gamma^{(-k)}(X_i)}
	{1-\widehat\gamma^{(-k)}(X_i)}
	\frac{W_i}{\widehat\varrho^{(-k)}(X_i)}
	\notag\\[-2pt]
	&\hspace{31mm}\times
	\Big\{
	\widehat m_1^{(-k)}
	(S_i,X_i;\widetilde q_{1,\tau}^{(-k)})
	-\widehat{\overline m}_1^{(-k)}
	(X_i;\widetilde q_{1,\tau}^{(-k)})
	\Big\}q
	\Bigg].
	\label{aeq:sampleloss1}
\end{align}
For $w=0$, the objective is
\begin{align}
	\widehat{\mathcal L}_{0,\tau}(q)
	=&\frac{1}{n}\sum_{k=1}^K\sum_{i\in\mathcal I_k}
	\Bigg[
	\frac{G_i}{\widehat\pi^{(-k)}}
	\frac{1-W_i}{\widehat\rho_0^{(-k)}(S_i,X_i)}
	\Big\{
	\rho_\tau(Y_i-q)
	+\big(\tau-\widehat m_0^{(-k)}
	(S_i,X_i;\widetilde q_{0,\tau}^{(-k)})\big)q
	\Big\}
	\notag\\
	&\quad+
	\frac{G_i}{\widehat\pi^{(-k)}}
	\Big\{
	\widehat{\overline m}_0^{(-k)}
	(X_i;\widetilde q_{0,\tau}^{(-k)})-\tau
	\Big\}q
	\notag\\
	&\quad+
	\frac{1-G_i}{\widehat\pi^{(-k)}}
	\frac{\widehat\gamma^{(-k)}(X_i)}
	{1-\widehat\gamma^{(-k)}(X_i)}
	\frac{1-W_i}{1-\widehat\varrho^{(-k)}(X_i)}
	\notag\\[-2pt]
	&\hspace{31mm}\times
	\Big\{
	\widehat m_0^{(-k)}
	(S_i,X_i;\widetilde q_{0,\tau}^{(-k)})
	-\widehat{\overline m}_0^{(-k)}
	(X_i;\widetilde q_{0,\tau}^{(-k)})
	\Big\}q
	\Bigg].
	\label{aeq:sampleloss0}
\end{align}
The estimators are
\begin{align}
	\widehat q_{1,\tau}
	&=\arg\min_{q\in\mathcal Q}\widehat{\mathcal L}_{1,\tau}(q),
	\label{aeq:qhat1-new}\\
	\widehat q_{0,\tau}
	&=\arg\min_{q\in\mathcal Q}\widehat{\mathcal L}_{0,\tau}(q),
	\label{aeq:qhat0-new}\\
	\widehat\Delta_\tau
	&=\widehat q_{1,\tau}-\widehat q_{0,\tau}.
	\label{aeq:Deltahat-new}
\end{align}
Whenever an objective has multiple minimizers, we take its smallest minimizer.

\subsection{Computation from the subgradient conditions}\label{sec:implementation2}

Both $\widehat{q}_{1,\tau}$ and $\widehat{q}_{0,\tau}$ can be computed explicitly based on subgradient conditions.  Specifically, we have $(\widehat{q}_{1,\tau},\widehat{q}_{0,\tau}) = (Y_{i_1},Y_{i_0})$ such that $G_{i_1} = G_{i_0}= 1$, $W_{i_1} = 1$, $W_{i_0} = 0$, 
\begin{align}
	& \tau\left(\sum_{k=1}^K \sum_{i\in\mathcal I_k} \frac{G_i W_i}{\widehat\pi^{(-k)}  \widehat\rho_1^{(-k)}(S_i,X_i)}\right) \notag \\
	& -\sum_{k=1}^K \sum_{i\in\mathcal I_k} 
	\begin{bmatrix}
		& \frac{G_i W_i}{\widehat\pi^{(-k)}  \widehat\rho_1^{(-k)}(S_i,X_i)} (\tau-\widehat m_1^{(-k)}(S_i,X_i;\widetilde q^{(-k)}_{1,\tau})) \notag \\
		& + \frac{G_i}{\widehat\pi^{(-k)}} \left( \widehat{\overline m}_1^{(-k)}(X_i;\widetilde q^{(-k)}_{1,\tau}) - \tau \right) \notag \\
		& + \frac{1-G_i}{\widehat\pi^{(-k)}}\frac{\widehat\gamma^{(-k)}(X_i)}{1-\widehat\gamma^{(-k)}(X_i)}\frac{W_i}{\widehat\varrho^{(-k)}(X_i)}\big\{\widehat m_1^{(-k)}(S_i,X_i;\widetilde q^{(-k)}_{1,\tau})-\widehat{\overline m}_1^{(-k)}(X_i;\widetilde q^{(-k)}_{1,\tau})\big\}	
	\end{bmatrix} 
	\notag \\
	& \geq \sum_{k=1}^K \sum_{i\in\mathcal I_k} \frac{G_i W_i}{\widehat\pi^{(-k)}  \widehat\rho_1^{(-k)}(S_i,X_i)}1\{ Y_i < Y_{i_1}\} \notag \\
	& \geq \tau\left(\sum_{k=1}^K \sum_{i\in\mathcal I_k} \frac{G_i W_i}{\widehat\pi^{(-k)}  \widehat\rho_1^{(-k)}(S_i,X_i)}\right) - \frac{1}{\widehat\pi^{(-k)}  \widehat\rho_1^{(-k)}(S_{i_1},X_{i_1})} \notag  \\
	& -\sum_{k=1}^K \sum_{i\in\mathcal I_k} 
	\begin{bmatrix}
		& \frac{G_i W_i}{\widehat\pi^{(-k)}  \widehat\rho_1^{(-k)}(S_i,X_i)} (\tau-\widehat m_1^{(-k)}(S_i,X_i;\widetilde q^{(-k)}_{1,\tau})) \\
		& + \frac{G_i}{\widehat\pi^{(-k)}} \left( \widehat{\overline m}_1^{(-k)}(X_i;\widetilde q^{(-k)}_{1,\tau}) - \tau \right) \\
		& + \frac{1-G_i}{\widehat\pi^{(-k)}}\frac{\widehat\gamma^{(-k)}(X_i)}{1-\widehat\gamma^{(-k)}(X_i)}\frac{W_i}{\widehat\varrho^{(-k)}(X_i)}\big\{\widehat m_1^{(-k)}(S_i,X_i;\widetilde q^{(-k)}_{1,\tau})-\widehat{\overline m}_1^{(-k)}(X_i;\widetilde q^{(-k)}_{1,\tau})\big\}	
	\end{bmatrix}
	\label{eq:q1_gradient}
\end{align}
and 
\begin{align}
	& \tau\left(\sum_{k=1}^K \sum_{i\in\mathcal I_k} \frac{G_i (1-W_i)}{\widehat\pi^{(-k)}  \widehat\rho_0^{(-k)}(S_i,X_i)}\right) \notag \\
	& -\sum_{k=1}^K \sum_{i\in\mathcal I_k} 
	\begin{bmatrix}
		& \frac{G_i (1-W_i)}{\widehat\pi^{(-k)}  \widehat\rho_0^{(-k)}(S_i,X_i)} (\tau-\widehat m_0^{(-k)}(S_i,X_i;\widetilde q^{(-k)}_{0,\tau})) \notag \\
		& + \frac{G_i}{\widehat\pi^{(-k)}} \left( \widehat{\overline m}_0^{(-k)}(X_i;\widetilde q^{(-k)}_{0,\tau}) - \tau \right) \notag \\
		& + \frac{1-G_i}{\widehat\pi^{(-k)}}\frac{\widehat\gamma^{(-k)}(X_i)}{1-\widehat\gamma^{(-k)}(X_i)}\frac{1-W_i}{1-\widehat\varrho^{(-k)}(X_i)}\big\{\widehat m_0^{(-k)}(S_i,X_i;\widetilde q^{(-k)}_{0,\tau})-\widehat{\overline m}_0^{(-k)}(X_i;\widetilde q^{(-k)}_{0,\tau})\big\}	
	\end{bmatrix} 
	\notag \\
	& \geq \sum_{k=1}^K \sum_{i\in\mathcal I_k} \frac{G_i (1-W_i)}{\widehat\pi^{(-k)}  \widehat\rho_0^{(-k)}(S_i,X_i)}1\{ Y_i < Y_{i_0}\} \notag \\
	& \geq \tau\left(\sum_{k=1}^K \sum_{i\in\mathcal I_k} \frac{G_i (1-W_i)}{\widehat\pi^{(-k)}  \widehat\rho_0^{(-k)}(S_i,X_i)}\right) - \frac{1}{\widehat\pi^{(-k)}  \widehat\rho_0^{(-k)}(S_{i_0},X_{i_0})} \notag  \\
	& -\sum_{k=1}^K \sum_{i\in\mathcal I_k} 
	\begin{bmatrix}
		& \frac{G_i (1-W_i)}{\widehat\pi^{(-k)}  \widehat\rho_0^{(-k)}(S_i,X_i)} (\tau-\widehat m_0^{(-k)}(S_i,X_i;\widetilde q^{(-k)}_{0,\tau})) \\
		& + \frac{G_i}{\widehat\pi^{(-k)}} \left( \widehat{\overline m}_0^{(-k)}(X_i;\widetilde q^{(-k)}_{0,\tau}) - \tau \right) \\
		& + \frac{1-G_i}{\widehat\pi^{(-k)}}\frac{\widehat\gamma^{(-k)}(X_i)}{1-\widehat\gamma^{(-k)}(X_i)}\frac{1-W_i}{1-\widehat\varrho^{(-k)}(X_i)}\big\{\widehat m_0^{(-k)}(S_i,X_i;\widetilde q^{(-k)}_{0,\tau})-\widehat{\overline m}_0^{(-k)}(X_i;\widetilde q^{(-k)}_{0,\tau})\big\}	
	\end{bmatrix}.
	\label{eq:q0_gradient}
\end{align}
We note that $(i_1,i_0)$ are uniquely defined as long as all the inequalities in \eqref{eq:q1_gradient} and \eqref{eq:q0_gradient} are strict, which is usually the case. If we have
\begin{align*}
	& \tau\left(\sum_{k=1}^K \sum_{i\in\mathcal I_k} \frac{G_i W_i}{\widehat\pi^{(-k)}  \widehat\rho_1^{(-k)}(S_i,X_i)}\right) \notag \\
	& -\sum_{k=1}^K \sum_{i\in\mathcal I_k} 
	\begin{bmatrix}
		& \frac{G_i W_i}{\widehat\pi^{(-k)}  \widehat\rho_1^{(-k)}(S_i,X_i)} (\tau-\widehat m_1^{(-k)}(S_i,X_i;\widetilde q^{(-k)}_{1,\tau})) \notag \\
		& + \frac{G_i}{\widehat\pi^{(-k)}} \left( \widehat{\overline m}_1^{(-k)}(X_i;\widetilde q^{(-k)}_{1,\tau}) - \tau \right) \notag \\
		& + \frac{1-G_i}{\widehat\pi^{(-k)}}\frac{\widehat\gamma^{(-k)}(X_i)}{1-\widehat\gamma^{(-k)}(X_i)}\frac{W_i}{\widehat\varrho^{(-k)}(X_i)}\big\{\widehat m_1^{(-k)}(S_i,X_i;\widetilde q^{(-k)}_{1,\tau})-\widehat{\overline m}_1^{(-k)}(X_i;\widetilde q^{(-k)}_{1,\tau})\big\}	
	\end{bmatrix} 
	\notag \\
	& = \sum_{k=1}^K \sum_{i\in\mathcal I_k} \frac{G_i W_i}{\widehat\pi^{(-k)}  \widehat\rho_1^{(-k)}(S_i,X_i)}1\{ Y_i \leq Y_{i_1}\} \notag \\
\end{align*}
then both $i_1$ and $i_1'$ satisfy \eqref{eq:q1_gradient}, where $i_1'$ is the index such that $G_{i_1'} = 1$, $W_{i_1'}=1$, and $Y_{i_1'}$ is the smallest observation in the treatment group that is larger than $Y_{i_1}$. In this case, we let $\widehat{q}_{1.\tau} = Y_{i_1}$.\footnote{In this case, any value that belongs to $[Y_{i_1},Y_{i_1'}]$ can be viewed as a solution. Because $|Y_{i_1}-Y_{i_1'}| = O_p(1/n)$, all choices are asymptotically equivalent.} The same remark applies to $\widehat{q}_{0,\tau}$.

Thus, $\widehat q_{1,\tau}=Y_{i_1}$ and $\widehat q_{0,\tau}=Y_{i_0}$, where $G_{i_1}=G_{i_0}=1$, $W_{i_1}=1$, $W_{i_0}=0$ and the selected outcomes are the smallest ones satisfying \eqref{eq:q1_gradient} and \eqref{eq:q0_gradient}, respectively. They can be found by sorting the corresponding observed outcomes and accumulating their inverse-probability weights. If the middle expression in either inequality is nonpositive or is at least the total check-loss weight, the minimizer is the lower or upper endpoint of $\mathcal Q$, respectively. If equality occurs at a cumulative weight, the objective is flat between two adjacent outcome values, and the smaller value is selected.

\subsection{Multiplier bootstrap}\label{sec:implementation3}

Let $\{\xi_i\}_{i=1}^n$ be nonnegative iid multiplier weights, independent of the data, satisfying
\begin{align*}
	\mathbb E[\xi_i]=1~~\text{and}~~\mathbb E[(\xi_i-1)^2]=1,
\end{align*}
and having a sub-exponential upper tail.  For instance, we can use $\xi_i\sim\operatorname{Exp}(1)$.  The nuisance functions and initial quantiles are held fixed across bootstrap draws.

For each observation $i$, let $k(i)$ denote the unique fold such that $i\in\mathcal I_{k(i)}$.
Both $\widehat{q}_{1,\tau}^*$ and $\widehat{q}_{0,\tau}^*$ can be computed explicitly based on subgradient conditions.  Specifically, we have $(\widehat{q}_{1,\tau}^*,\widehat{q}_{0,\tau}^*) = (Y_{i_1^*},Y_{i_0^*})$ such that $G_{i_1^*} = G_{i_0^*}= 1$, $W_{i_1^*} = 1$, $W_{i_0^*} = 0$, 
\begin{align}
	& \tau\left(\sum_{k=1}^K \sum_{i\in\mathcal I_k} \frac{\xi_i G_i W_i}{\widehat\pi^{(-k)}  \widehat\rho_1^{(-k)}(S_i,X_i)}\right) \notag \\
	& -\sum_{k=1}^K \sum_{i\in\mathcal I_k} 
	\begin{bmatrix}
		& \frac{\xi_i G_i W_i}{\widehat\pi^{(-k)}  \widehat\rho_1^{(-k)}(S_i,X_i)} (\tau-\widehat m_1^{(-k)}(S_i,X_i;\widetilde q^{(-k)}_{1,\tau})) \notag \\
		& + \frac{\xi_i G_i}{\widehat\pi^{(-k)}} \left( \widehat{\overline m}_1^{(-k)}(X_i;\widetilde q^{(-k)}_{1,\tau}) - \tau \right) \notag \\
		& + \frac{\xi_i(1-G_i)}{\widehat\pi^{(-k)}}\frac{\widehat\gamma^{(-k)}(X_i)}{1-\widehat\gamma^{(-k)}(X_i)}\frac{W_i}{\widehat\varrho^{(-k)}(X_i)}\big\{\widehat m_1^{(-k)}(S_i,X_i;\widetilde q^{(-k)}_{1,\tau})-\widehat{\overline m}_1^{(-k)}(X_i;\widetilde q^{(-k)}_{1,\tau})\big\}	
	\end{bmatrix} 
	\notag \\
	& \geq \sum_{k=1}^K \sum_{i\in\mathcal I_k} \frac{\xi_i G_i W_i}{\widehat\pi^{(-k)}  \widehat\rho_1^{(-k)}(S_i,X_i)}1\{ Y_i < Y_{i_1^*}\} \notag \\
	& \geq \tau\left(\sum_{k=1}^K \sum_{i\in\mathcal I_k} \frac{\xi_i G_i W_i}{\widehat\pi^{(-k)}  \widehat\rho_1^{(-k)}(S_i,X_i)}\right) - \frac{\xi_{i_1^*}}{\widehat\pi^{(-k(i_1^*))} \widehat\rho_1^{(-k(i_1^*))}(S_{i_1^*},X_{i_1^*})} \notag  \\
	& -\sum_{k=1}^K \sum_{i\in\mathcal I_k} 
	\begin{bmatrix}
		& \frac{\xi_i G_i W_i}{\widehat\pi^{(-k)}  \widehat\rho_1^{(-k)}(S_i,X_i)} (\tau-\widehat m_1^{(-k)}(S_i,X_i;\widetilde q^{(-k)}_{1,\tau})) \\
		& + \frac{\xi_i G_i}{\widehat\pi^{(-k)}} \left( \widehat{\overline m}_1^{(-k)}(X_i;\widetilde q^{(-k)}_{1,\tau}) - \tau \right) \\
		& + \frac{\xi_i(1-G_i)}{\widehat\pi^{(-k)}}\frac{\widehat\gamma^{(-k)}(X_i)}{1-\widehat\gamma^{(-k)}(X_i)}\frac{W_i}{\widehat\varrho^{(-k)}(X_i)}\big\{\widehat m_1^{(-k)}(S_i,X_i;\widetilde q^{(-k)}_{1,\tau})-\widehat{\overline m}_1^{(-k)}(X_i;\widetilde q^{(-k)}_{1,\tau})\big\}	
	\end{bmatrix},
	\label{eq:q1_gradient_boot}
\end{align}
and 
\begin{align}
	& \tau\left(\sum_{k=1}^K \sum_{i\in\mathcal I_k} \frac{\xi_i G_i (1-W_i)}{\widehat\pi^{(-k)}  \widehat\rho_0^{(-k)}(S_i,X_i)}\right) \notag \\
	& -\sum_{k=1}^K \sum_{i\in\mathcal I_k} 
	\begin{bmatrix}
		& \frac{\xi_i G_i (1-W_i)}{\widehat\pi^{(-k)}  \widehat\rho_0^{(-k)}(S_i,X_i)} (\tau-\widehat m_0^{(-k)}(S_i,X_i;\widetilde q^{(-k)}_{0,\tau})) \notag \\
		& + \frac{\xi_i G_i}{\widehat\pi^{(-k)}} \left( \widehat{\overline m}_0^{(-k)}(X_i;\widetilde q^{(-k)}_{0,\tau}) - \tau \right) \notag \\
		& + \frac{\xi_i(1-G_i)}{\widehat\pi^{(-k)}}\frac{\widehat\gamma^{(-k)}(X_i)}{1-\widehat\gamma^{(-k)}(X_i)}\frac{1-W_i}{1-\widehat\varrho^{(-k)}(X_i)}\big\{\widehat m_0^{(-k)}(S_i,X_i;\widetilde q^{(-k)}_{0,\tau})-\widehat{\overline m}_0^{(-k)}(X_i;\widetilde q^{(-k)}_{0,\tau})\big\}	
	\end{bmatrix} 
	\notag \\
	& \geq \sum_{k=1}^K \sum_{i\in\mathcal I_k} \frac{\xi_i G_i (1-W_i)}{\widehat\pi^{(-k)}  \widehat\rho_0^{(-k)}(S_i,X_i)}1\{ Y_i < Y_{i_0^*}\} \notag \\
	& \geq \tau\left(\sum_{k=1}^K \sum_{i\in\mathcal I_k} \frac{\xi_i G_i (1-W_i)}{\widehat\pi^{(-k)}  \widehat\rho_0^{(-k)}(S_i,X_i)}\right) - \frac{\xi_{i_0^*}}{\widehat\pi^{(-k(i_0^*))} \widehat\rho_0^{(-k(i_0^*))}(S_{i_0^*},X_{i_0^*})} \notag  \\
	& -\sum_{k=1}^K \sum_{i\in\mathcal I_k} 
	\begin{bmatrix}
		& \frac{\xi_i G_i (1-W_i)}{\widehat\pi^{(-k)}  \widehat\rho_0^{(-k)}(S_i,X_i)} (\tau-\widehat m_0^{(-k)}(S_i,X_i;\widetilde q^{(-k)}_{0,\tau})) \\
		& + \frac{\xi_i G_i}{\widehat\pi^{(-k)}} \left( \widehat{\overline m}_0^{(-k)}(X_i;\widetilde q^{(-k)}_{0,\tau}) - \tau \right) \\
		& + \frac{\xi_i(1-G_i)}{\widehat\pi^{(-k)}}\frac{\widehat\gamma^{(-k)}(X_i)}{1-\widehat\gamma^{(-k)}(X_i)}\frac{1-W_i}{1-\widehat\varrho^{(-k)}(X_i)}\big\{\widehat m_0^{(-k)}(S_i,X_i;\widetilde q^{(-k)}_{0,\tau})-\widehat{\overline m}_0^{(-k)}(X_i;\widetilde q^{(-k)}_{0,\tau})\big\}	
	\end{bmatrix}.
	\label{eq:q0_gradient_boot}
\end{align}
We note that $(i_1,i_0)$ are uniquely defined as long as all the inequalities in \eqref{eq:q1_gradient} and \eqref{eq:q0_gradient} are strict, which is the case if the bootstrap weights are continuously distributed.

Accordingly,
\begin{align*}
	(\widehat q_{1,\tau}^*,\widehat q_{0,\tau}^*)
	=(Y_{i_1^*},Y_{i_0^*}),
\end{align*}
where $G_{i_1^*}=G_{i_0^*}=1$, $W_{i_1^*}=1$, and $W_{i_0^*}=0$, and $Y_{i_1^*}$ and $Y_{i_0^*}$ are the smallest observed outcomes satisfying \eqref{eq:q1_gradient_boot} and \eqref{eq:q0_gradient_boot}, respectively. The multiplier $\xi_i$ enters every observation-level contribution, including the jumps in the cumulative weights. The boundary and tie-breaking rules used for the original estimator continue to apply. Repeating this procedure for $R$ independent multiplier draws yields $\{\widehat\Delta_\tau^{*,r}\}_{r=1}^R$. The nuisance functions are not re-estimated.

\end{document}